\documentclass[a4paper,12pt]{article}

\usepackage{jheppub}

\title{\boldmath A superspace approach to AdS$_3$ string theory}

\abstract{We carefully examine the Polyakov path integral for strings on $\text{AdS}_3$ in superspace, both for type II and heterotic superstrings. We construct a free-field realization of the supersymmetric $\text{SL}(2,\mathbb{R})$ WZW model which manifestly preserves worldsheet supersymmetry and use this free-field realization to construct spectrally-flowed vertex operators describing the emission of long strings in the bulk. By working directly with the moduli space of super Riemann surfaces, we exactly compute tree-level correlation functions of long strings in the `near-boundary' limit without resorting to the standard picture-changing-operator (PCO) procedure. Finally, we argue how these correlators schematically reproduce correlation functions of the conjectured boundary CFTs, and as a result provide a novel proposal for the CFT dual for heterotic superstrings in $\text{AdS}_3$.
}

\author[a]{Bob Knighton,}
\author[a,b]{Nathan McStay,}
\author[c,d]{Vit Sriprachyakul}

\affiliation[a]{Department of Applied Mathematics \& Theoretical Physics, University of Cambridge,\\
Wilberforce Road, Cambridge CB3 0WA, United Kingdom}
\affiliation[b]{International Centre for Theoretical Sciences-TIFR,\\
Shivakote, Hesaraghatta Hobli, Bengaluru North 560089, India.}
\affiliation[c]{Institut f\"ur Theoretische Physik,
ETH Z\"urich,\\
Wolfgang-Pauli-Strasse 27,
8093 Z\"urich, Switzerland}
\affiliation[d]{Department of Particle Physics and Astrophysics, Weizmann Institute of Science,\\ Rehovot 7610001, Israel}

\emailAdd{rik23@cam.ac.uk}
\emailAdd{nathan.mcstay@icts.res.in}
\emailAdd{vit.sriprachyakul@weizmann.ac.il}

\usepackage{bm} 
\usepackage[dvipsnames]{xcolor}
\definecolor{green_maf}{RGB}{28, 166, 46}
\definecolor{blue_mrg}{RGB}{12, 143, 145}
\definecolor{detail}{RGB}{110,110,110}
\usepackage{amsmath} 
\usepackage{nccmath} 
\usepackage{amssymb} 
\usepackage{amsthm} 
\usepackage{mathtools} 
\usepackage[utf8]{inputenc} 
\usepackage{braket}
\usepackage{enumerate}
\usepackage[many]{tcolorbox}
\usepackage[inline]{enumitem}
\usepackage{comment}
\usepackage{soul} 
\usepackage{tikz}
\usepackage{csquotes}
\usepackage{mathrsfs}
\usepackage{tensor}
\usepackage{dsfont}
\usepackage{slashed}

\newtcolorbox{empheqboxed}{colback=gray!30, 
 colframe=white,
 width=\textwidth,
 sharpish corners,
 top=-2mm, 
 bottom=0pt
}

\hypersetup{
		pdfencoding=unicode,
		colorlinks=true,
		urlcolor=Maroon,
		linkcolor=RoyalBlue,
		citecolor=Maroon,
		pdfstartview=FitH,
		linktocpage=true
}

\usetikzlibrary{calc,decorations.markings}
\usetikzlibrary{decorations.pathmorphing}
\usetikzlibrary{shapes,backgrounds}
\usetikzlibrary{fadings}
\usetikzlibrary{cd}
\usetikzlibrary{decorations.pathreplacing,calligraphy}
\usetikzlibrary{hobby}

\tikzset{
	partial ellipse/.style args={#1:#2:#3}{
		insert path={+ (#1:#3) arc (#1:#2:#3)}
	}
}

\tikzset{
  every overlay node/.style={
    draw=black,fill=white,rounded corners,anchor=north west,
  },
}

\tikzfading
[
  name=fade out,
  inner color=transparent!0,
  outer color=transparent!100
]

\newif\ifdetails
\detailstrue

\newcommand{\p}{\partial}

\def\be{\begin{equation}}
\def\ee{\end{equation}}

\begin{document}

\maketitle

\section{Introduction}

String theories compactified to $\text{AdS}_3$ have been a subject of interest since the early days of the AdS/CFT correspondence \cite{Maldacena:1997re,Aharony:1999ti} and are among the most well-understood from the point of view of the worldsheet CFT. Despite being a curved background, one can study analytically the worldsheet sigma model on $\text{AdS}_3$ without resorting to perturbation theory in $\alpha'$. Specifically, the worldsheet CFT of bosonic string theory on $\text{AdS}_3$ is given exactly by a Wess-Zumino-Witten (WZW) model on the (universal cover of the) group $\text{SL}(2,\mathbb{R})$ \cite{Maldacena:2000hw,Maldacena:2000kv,Maldacena:2001km}. Given that the $\text{SL}(2,\mathbb{R})$ WZW model and its Euclidean cousin, the $\mathbb{H}^+_3$ model, are among the richest and most well-studied 2D CFTs, bosonic string theory on $\text{AdS}_3$ enjoys a uniquely well-developed and powerful analytic toolkit that can be employed for the computation of partition functions and correlation functions. Correlation functions of the $\text{SL}(2,\mathbb{R})$ WZW model in particular have been studied intensively for nearly 30 years \cite{Giribet:1999ft,Ishibashi:2000fn,Maldacena:2001km,Giribet:2001ft,Giribet:2005ix,Ribault:2005ms,Giribet:2005mc,Minces:2005nb,Iguri:2007af,Giribet:2007wp,Baron:2008qf,Iguri:2009cf,Giribet:2011xf,Cagnacci:2013ufa,Giribet:2015oiy}, and there has been enormous recent progress in determining the correlation functions from first principles \cite{Dei:2019osr,Eberhardt:2019ywk,Eberhardt:2020akk,Hikida:2020kil,Dei:2021xgh,Dei:2021yom,Dei:2022pkr,Iguri:2022eat,Bufalini:2022toj,Hikida:2023jyc,Knighton:2023mhq,Knighton:2024qxd,Sriprachyakul:2024gyl}.\footnote{See \cite{Kovensky:2026usc} for a modern review on worldsheet approaches to $\text{AdS}_3/\text{CFT}_2$.}

Type II superstrings on $\text{AdS}_3\times\mathcal{N}$ supported by pure NS-NS flux admit a similar description in terms of an $\mathscr{N}=(1,1)$ supersymmetric WZW model on $\text{SL}(2,\mathbb{R})$. Given that the background is curved, naively the worldsheet sigma model in the RNS formalism will include an interaction term of the form $G_{\mu\nu}(X)\bar{\psi}\,^{\mu}\slashed\partial\psi^{\nu}$,\footnote{The full supersymmetric action will also include four-fermion terms like $R_{\mu\nu\rho\sigma}(X)\bar\psi^{\mu}\psi^{\nu}\bar{\psi}^{\rho}\psi^{\sigma}$, as well as couplings determined by the Kalb-Ramond field $B_{\mu\nu}$ in the case of backgrounds with NS-NS flux.} and thus it seems like the supersymmetric worldsheet theory will manifestly be more complicated than the bosonic one. However, as for any supersymmetric WZW model, one can actually consistently decouple the fermions from the target space coordinates, yielding a quantum theory consisting of a bosonic WZW model and a set of free fermions \cite{DiVecchia:1984nyg}. For the case of $\text{AdS}_3$, this field redefinition gives the equivalence of conformal field theories:
\begin{equation}\label{eq:intro-sl2-decomposition}
\mathfrak{sl}(2,\mathbb{R})^{(1)}_k\cong\mathfrak{sl}(2,\mathbb{R})_{k+2}\oplus(3\text{ free fermions})\,.
\end{equation}
For compactifications like $\mathcal{N}=\text{S}^3\times\mathbb{T}^4$ and $\mathcal{N}=\text{S}^3\times\text{S}^3\times\text{S}^1$, a similar equivalence exists on each factor, and the full ten-dimensional sigma model can be rewritten as a bosonic sigma model and 10 decoupled free fermions. The consequence of this isomorphism is that the worldsheet CFT of superstrings on $\text{AdS}_3$ is in principle no more complicated than that of bosonic strings and a few free fermions.

In practice, however, things are much less simple. String theory is more than just a worldsheet CFT -- one must gauge the superconformal symmetry on the worldsheet (BRST quantization) and sum over all possible spin structures (GSO projection) to define a consistent superstring theory. It is in the former procedure where the difference between bosonic and supersymmetric $\text{AdS}_3$ string theory becomes most apparent. Unlike in flat space, where worldsheet supersymmetry simply exchanges worldsheet bosons and fermions, the supersymmetry transformations for the CFT in equation \eqref{eq:intro-sl2-decomposition} take the (schematic) form
\begin{equation}
\delta\mathcal{J}^a=\psi^a\,,\quad\delta\psi^a=\mathcal{J}^a+f\indices{^a_b_c}\psi^b\psi^c\,,
\end{equation}
where $\psi^a$ are the free fermions, $\mathcal{J}^a$ are the affine currents in the (bosonic) $\text{SL}(2,\mathbb{R})$ WZW model, and $f\indices{^a_b_c}$ are the $\mathfrak{sl}(2,\mathbb{R})$ structure constants. The quadratic term in the transformation of $\psi^a$ makes it much more difficult to identify states that are invariant under supersymmetry, and more crucially makes the physical state conditions on the worldsheet extremely cumbersome to work with.

Things take an even worse turn when one attempts to compute correlation functions. In bosonic string theory, string amplitudes are simply obtained by computing correlation functions of physical operators and integrating them over the moduli space $\mathcal{M}_{g,n}$ of (bosonic) Riemann surfaces. For tree-level ($g=0$) correlators of operators in the vacuum of the compact spacetime $\mathcal{N}$, this boils down to computing correlation functions of the form
\begin{equation}
\Braket{V_1(0)V_2(1)V_3(\infty)\prod_{i=4}^{n}\int\mathrm{d}^2z_iV_i(z_i)}
\end{equation}
in the $\text{SL}(2,\mathbb{R})$ WZW model. For the supersymmetric theory, however, things are complicated by the fact that each vertex operator has an infinite number of equivalent descriptions, differing by their `picture number' \cite{Friedan:1985ge}. The canonical form of a vertex operator has picture $-1$ in the NS sector and $-1/2$ in the Ramond sector.
However, in order to give nonzero values to scattering amplitudes of superstrings, the total picture number in a correlation function must be equal to $-2$ at tree level. For example, a non-vanishing tree-level amplitude in the NS sector would take the form
\begin{equation}
\Braket{V_1^{(-1)}(0)V_{2}^{(-1)}(1)V_3^{(0)}(\infty)\prod_{i=4}^{n}\int\mathrm{d}^2z_i\,V_i^{(0)}(z_i)}\,.
\end{equation}
The picture-0 operators are obtained from the canonical ones via the action of \textit{picture-changing operators} (PCOs). Even in superstring theories with a flat target, picture-changing yields vertex operators that are extremely cumbersome to work with, and this effect is only compounded in curved backgrounds like $\text{AdS}_3$. As a result, only a handful of string correlators in supersymmetric $\text{AdS}_3$ backgrounds have been calculated in the RNS formalism \cite{Giribet:2007wp,Dabholkar:2007ey,Iguri:2022pbp,Iguri:2023khc,Sriprachyakul:2024gyl,Yu:2024kxr,Yu:2025qnw}.\footnote{In \cite{Sriprachyakul:2024gyl}, the near-boundary limit was used in a slightly different way than what we will be doing subsequently in this article. More precisely, in \cite{Sriprachyakul:2024gyl}, the near-boundary limit was adopted for the decoupled bosonic theory in the decomposition \eqref{eq:intro-sl2-decomposition}. The same application of the near-boundary limit was also useful in the discussion of the DDF operators dual to the dilaton field in the spacetime CFT \cite{Sriprachyakul:2024xih}.}

\paragraph{A manifestly supersymmetric approach:} In this paper, we describe an approach to $\text{AdS}_3$ superstring theory that circumvents the cumbersome features described above.
\begin{enumerate}

    \item In order to maintain manifest worldsheet supersymmetry, we do not decouple the fermions from the bosons in the $\text{SL}(2,\mathbb{R})$ sigma model. The price of doing this is that one ends up with a theory of interacting bosons and fermions. We will find that this is more than a reasonable price to pay for manifest worldsheet supersymmetry.

    \item We work explicitly in superspace. This not only makes worldsheet supersymmetry even more manifest, but also allows us to circumvent the picture-changing procedure by instead integrating over the moduli space $\mathfrak{M}_{g,n}$ of \textit{super Riemann surfaces} \cite{Witten:2012bh}. At tree level, assuming all vertex operators are in the NS sector, this amounts to computing the correlation function
    \begin{equation}\label{eq:intro-m0n,integral}
    \Braket{\mathscr{V}_1(0;0)\mathscr{V}_2(1;0)\int\mathrm{d}^2\theta_3\,\mathscr{V}_3(\infty;\theta_3)\prod_{i=4}^{n}\int\mathrm{d}^2z_i\,\mathrm{d}^2\theta_i\,\mathscr{V}_i(z_i;\theta_i)}\,.
    \end{equation}
    While evaluating this integral is formally equivalent to the picture-changing procedure, we will find that the integral form in equation \eqref{eq:intro-m0n,integral} is much easier to work with.

\end{enumerate}
The superspace formulation of RNS superstring theory has a long history \cite{Martinec:1983um,DHoker:1986xro,DHoker:1987rxo,Atick:1987rk,DHoker:1988pdl,Verlinde:1988tx,DHoker:1989hhv,DHoker:2001jaf,DHoker:2002hof,Witten:2012ga,Witten:2012bh,Donagi:2013dua}, but to date has only seen applications to strings propagating in flat spacetimes or in the context of topological string theory \cite{Dijkgraaf:1990qw,Jia:2016jlo,Jia:2016rdn}, and has instead largely been superseded by the picture-changing approach. As we will demonstrate throughout the present work, the superspace formulation is particularly well-suited for studying a certain class of correlators in $\text{AdS}_3$.

\begin{figure}
\centering
\begin{tikzpicture}[scale = 0.75, rotate = 5]
\draw[thick] (0,0) circle (3);
\path[draw, closed=true, thick, black, opacity = 0.7] (3,0) to[out = 135, in = -45] (2.25,1.5) to[out = 135, in = -45] (0.2,2.98) to[out = -135, in = 45] (-2,1.8) to[out = -135, in = 45] (-3,0) to[out = -45, in = 135] (-1.7,-2) to[out = -45, in = 135]  (-0.2,-2.98) to[out = 45, in = -135] (2,-1.8) to[out = 45, in = -135] (3,0);
\path[fill, closed=true, thick, CornflowerBlue, opacity = 0.05] (3,0) to[out = 135, in = -45] (2.25,1.5) to[out = 135, in = -45] (0.2,2.98) to[out = -135, in = 45] (-2,1.8) to[out = -135, in = 45] (-3,0) to[out = -45, in = 135] (-1.7,-2) to[out = -45, in = 135]  (-0.2,-2.98) to[out = 45, in = -135] (2,-1.8) to[out = 45, in = -135] (3,0);
\path[draw, closed=true, thick, black, opacity = 0.7] (3,0) to[out = 135, in = -45] (2,2) to[out = 135, in = -45] (0.2,2.98) to[out = -135, in = 45] (-2,2) to[out = -135, in = 45] (-3,0) to[out = -45, in = 135] (-2,-2) to[out = -45, in = 135]  (-0.2,-2.98) to[out = 45, in = -135] (2,-2) to[out = 45, in = -135] (3,0);
\path[fill, closed=true, thick, CornflowerBlue, opacity = 0.05] (3,0) to[out = 135, in = -45] (2,2) to[out = 135, in = -45] (0.2,2.98) to[out = -135, in = 45] (-2,2) to[out = -135, in = 45] (-3,0) to[out = -45, in = 135] (-2,-2) to[out = -45, in = 135]  (-0.2,-2.98) to[out = 45, in = -135] (2,-2) to[out = 45, in = -135] (3,0);
\path[draw, closed=true, thick, black, opacity = 0.7] (3,0) to[out = 135, in = -45] (1.5,2.25) to[out = 135, in = -45] (0.2,2.98) to[out = -135, in = 45] (-1.8,2) to[out = -135, in = 45] (-3,0) to[out = -45, in = 135] (-2,-1.7) to[out = -45, in = 135]  (-0.2,-2.98) to[out = 45, in = -135] (1.8,-2) to[out = 45, in = -135] (3,0);
\path[fill, closed=true, thick, CornflowerBlue, opacity = 0.05] (3,0) to[out = 135, in = -45] (1.5,2.25) to[out = 135, in = -45] (0.2,2.98) to[out = -135, in = 45] (-1.8,2) to[out = -135, in = 45] (-3,0) to[out = -45, in = 135] (-2,-1.7) to[out = -45, in = 135]  (-0.2,-2.98) to[out = 45, in = -135] (1.8,-2) to[out = 45, in = -135] (3,0);
\node[right] at (3,0) {$x_1$};
\node[left] at (2.9,0) {$z_1$};
\node[above] at (0.2,3) {$x_2$};
\node[below] at (0.2,2.9) {$z_2$};
\node[left] at (-3,0) {$x_3$};
\node[right] at (-2.9,0) {$z_3$};
\node[below] at (-0.2,-3) {$x_4$};
\node[above] at (-0.2,-2.9) {$z_4$};
\end{tikzpicture}
\caption{A worldsheet instanton can move arbitrarily close to the boundary of $\text{AdS}_3$ at a finite cost of energy.}
\label{fig:worldsheet-instanton}
\end{figure}
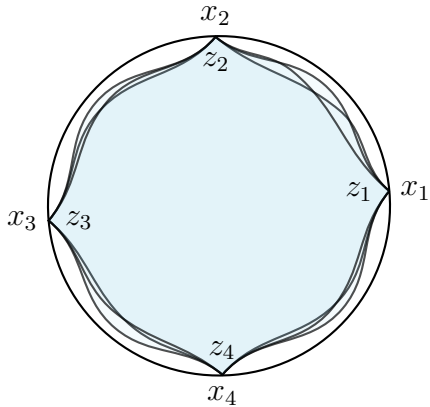

\paragraph{Computing near-boundary correlators:} The vertex operators whose correlators we will be interested in correspond to the emission of long strings from the boundary of $\text{AdS}_3$. They depend on various quantum numbers, including the position $x_i$ on the boundary where they are emitted, the winding $w_i$ around the boundary point, the boundary conformal weight $h_i$, and the $\text{SL}(2,\mathbb{R})$ spin $j_i$, which determines the center-of-mass momentum in the radial direction. 
The resulting superstring amplitudes, defined by the integrated correlator \eqref{eq:intro-m0n,integral}, are analytic functions $\mathscr{A}(j_1,\ldots,j_n)$ of the spins $j_i$. It has long been known \cite{Maldacena:2001km} that these amplitudes have poles that occur when the sums of the spins approach certain critical values, namely
\begin{equation}\label{eq:intro-pole-locations}
\sum_{i=1}^{n}j_i=\frac{L^2}{2\alpha'}(n-m-2)+1\,,\quad m\in\mathbb{Z}_{\geq 0}\,,
\end{equation}
where $L$ is the AdS radius. These poles arise physically due to the existence of `worldsheet instantons' \cite{Seiberg:1999xz,Maldacena:2001km} -- finite-action string configurations that live arbitrarily close to the boundary of $\text{AdS}_3$ (see Figure \ref{fig:worldsheet-instanton}).

Determining the full amplitude $\mathscr{A}(j_1,\ldots,j_n)$ is currently out of reach and, indeed, has only been computed in the bosonic case for 3- and 4-point functions \cite{Dei:2021xgh,Dei:2021yom,Dei:2022pkr}. However, the residues of these amplitudes at the poles \eqref{eq:intro-pole-locations} have recently been calculated in bosonic string theory in full generality at tree level \cite{Knighton:2023mhq,Hikida:2023jyc,Knighton:2024qxd} by computing the integral over all worldsheet configurations which lie arbitrarily close to the conformal boundary. Since only worldsheet instantons have a finite action near the boundary of $\text{AdS}_3$, this subsector of the path integral is the path integral of the worldsheet instantons (thus why it reproduces the poles in equation \eqref{eq:intro-pole-locations}). Amazingly, this subsector of the path integral \textit{localizes} on the moduli space $\mathcal{M}_{g,n}$, and only receives contributions from worldsheets whose moduli are tuned such that the worldsheet holomorphically covers the boundary of $\text{AdS}_3$ \cite{Maldacena:2001km,Eberhardt:2019ywk}.

The main technical result of this paper is that this feature generalizes to superstrings in a remarkably simple way. The path integral of superstring worldsheet instantons only receives contributions from worldsheets for which there exists a map
\begin{equation}
\Gamma:\Sigma\to\partial\text{AdS}_3
\end{equation}
satisfying a particular supersymmetric generalization of holomorphicity. As we will argue in the main text, the near-boundary correlation functions of NS-sector vertex operators can be written in closed form as an integral over the space of all such maps, which at tree level can be constructed very explicitly. The end result, equation \eqref{eq:final worldsheet path integral result}, is a compact expression for the residues of the poles in worldsheet correlators, which can be readily compared to computations in the dual CFT.

\vspace{0.5cm}

\hrule

\vspace{0.5cm}

This paper is organized as follows. In Section \ref{sec:superstrings}, we introduce the type II $\text{AdS}_3$ sigma model in superspace, taking care to organize our field content in such a way as to maintain manifest worldsheet supersymmetry at every step. In Section \ref{sec:physical-spectrum}, we solve the physical state conditions on the worldsheet and write down superspace vertex operators in their canonical pictures. In Section \ref{sec:localization}, we set up the worldsheet path integral, and show that the sub-sector of long strings in the path integral localizes to the supermoduli space of worldsheet instantons. In Section \ref{sec:heterotic}, we repeat the analysis of Sections \ref{sec:superstrings}, \ref{sec:physical-spectrum}, and \ref{sec:localization} for heterotic strings, obtaining closed-form expressions for the near-boundary correlation functions on the heterotic worldsheet. Finally, we conclude in Section \ref{sec:discussion} with a discussion of the holographic CFTs dual to these backgrounds, as well as a discussion of potential future work. We also include three technical appendices for the convenience of the reader.

\vspace{0.5cm}

\paragraph{Note:} This paper is the first in a series of two on superstring theory in $\text{AdS}_3$. The second paper will be devoted to a detailed study of the dual CFTs to type II and heterotic strings in $\text{AdS}_3$ \cite{Knighton:2026xxx}.

\section{\boldmath The worldsheet SCFT}\label{sec:superstrings}

As mentioned in the introduction, the main purpose of this article is the study of superstrings in spacetimes of the form $\text{AdS}_3\times\mathcal{N}$ supported by pure NS-NS flux. As we will formulate the theory in superspace, we will now introduce our superspace conventions. A precise definition of the worldsheet theory involves the formalism of super Riemann surfaces, which we review in Appendix \ref{app:srs}. For the purposes of this section, however, we will only need to know the local description of the worldsheet, which we briefly review below. In this section and throughout the rest of the paper we follow the conventions of \cite{Witten:2012ga}.

The superstring worldsheet $\Sigma$ is locally a complex supermanifold with local holomorphic coordinates $(z|\theta)$. The choice of a superconformal structure is equivalent to a choice of superderivative $D$ (as well as its right-moving counterpart $\bar{D}$), and a choice of superconformal coordinates is one in which $D,\bar{D}$ take the canonical form
\begin{equation}\label{eq:canonical-D}
D=\partial_{\theta}+\theta\partial\,,\quad \bar{D}=\partial_{\bar\theta}+\bar\theta\bar\partial\,,\quad D^2=\partial\,,\quad \bar{D}^2=\bar{\partial}\,,
\end{equation}
with $\partial\equiv\partial_z$ and $\bar{\partial}\equiv\partial_{\bar{z}}$. The superconformal structure is invariant under global supersymmetry transformations which act on the holomorphic coordinates as
\begin{equation}
z\to z+\varepsilon\theta\,,\quad\theta\to\theta-\varepsilon\,.
\end{equation}

We will frequently refer to the pair of coordinates $(z|\theta)$ by the symbol $\boldsymbol{z}$. Given two points $\boldsymbol{z}_1,\boldsymbol{z}_2$ on the worldsheet, there are two natural supertranslation-invariant intervals between them, which we will refer to as the bosonic and fermionic intervals:
\begin{equation}\label{eq:intervals}
I_b(\boldsymbol{z}_1,\boldsymbol{z}_2)=z_1-z_2-\theta_1\theta_2\,,\quad I_f(\boldsymbol{z}_1,\boldsymbol{z}_2)=\theta_1-\theta_2\,.
\end{equation}
Integration over the odd components of superspace is defined such that only the top component of the integrand survives, i.e.\footnote{This is opposite to the convention used in Polchinksi \cite{Polchinski:1998rr}.}
\begin{equation}
\int_{\Sigma}\mathrm{d}^{2|2}\boldsymbol{z}\,f(z,\bar{z})\theta\bar\theta=\int_{\Sigma_{\text{red}}}\mathrm{d}^2z\,f(z,\bar z)\,.
\end{equation}
Here, $\Sigma_{\text{red}}$ is the underlying bosonic Riemann surface $\Sigma_{\text{red}}\subset\Sigma$, frequently referred to as the `reduced space' of $\Sigma$.\footnote{Strictly speaking, integrating over the odd coordinates of $\Sigma$ to obtain an integral over $\Sigma_{\text{red}}$ is only valid globally when $\Sigma$ is a `split' super Riemann surface --- see Appendix \ref{app:srs}.}

A superfield $\Phi(z,\bar{z}|\theta,\bar\theta)$ of weight $(h,\bar{h})$ on the worldsheet admits an expansion
\begin{equation}
\Phi=\phi+\theta\lambda+\bar\theta\bar\lambda+\theta\bar\theta F\,,
\end{equation}
where $\phi$ has weight $(h,\bar{h})$, $\lambda$ has weight $(h+1/2,\bar{h})$, $\bar{\lambda}$ has weight $(h,\bar{h}+1/2)$, and $F$ has weight $(h+1/2,\bar{h}+1/2)$. By a \textit{chiral} superfield, we mean a superfield $\Gamma$ satisfying $\bar{D}\Gamma=0$. Such a field thus has an expansion
\begin{equation}
\Gamma=\gamma(z)+\theta\psi(z)\,,
\end{equation}
where $\gamma$ and $\psi$ are holomorphic on $\Sigma_{\text{red}}$. Antichiral superfields are analogously annihilated by $D$. We call a superfield $\Phi$ \textit{harmonic} if it satisfies the wave equation $D\bar{D}\Phi=0$, which is equivalent to the usual wave and Dirac equations on its components:
\begin{equation}
\partial\overline{\partial}\phi=0\,,\quad \bar\partial\lambda=0\,,\quad\partial\bar\lambda=0\,,\quad F=0\,.
\end{equation}

\subsection[The AdS\texorpdfstring{$_3$}{3} sigma model]{\boldmath The AdS\texorpdfstring{$_3$}{3} sigma model}

Here, we will carefully write down the worldsheet sigma model for strings in $\text{AdS}_3\times\mathcal{N}$. The derivation closely follows the well-known descriptions of bosonic strings in $\text{AdS}_3$ \cite{Giveon:1998ns} adapted to superspace. Throughout this section, we will frequently ignore the compactification manifold $\mathcal{N}$, focusing entirely on the $\text{AdS}_3$ factor.

Type II superstrings propagating in a background with metric $G_{\mu\nu}$ and Kalb-Ramond field $B_{\mu\nu}$ are described by the supersymmetric Polyakov action
\begin{equation}
S_{\text{II}}=\frac{1}{2\pi\alpha'}\int\mathrm{d}^{2|2}\boldsymbol{z}\left(G_{\mu\nu}(X)+B_{\mu\nu}(X)\right)DX^{\mu}\bar{D}X^{\nu}\,,
\end{equation}
where $X^{\mu}(z,\bar{z}|\theta,\bar\theta)$
is the superfield of the worldsheet embedding coordinates into the target space. In the case of $\text{AdS}_3$, we adopt Poincar\'e coordinates $(\phi,\gamma,\bar\gamma)$ for which the target metric and Kalb-Ramond field take the form
\begin{equation}
\begin{gathered}
\mathrm{d}s^2=L^2\left(\mathrm{d}\phi^2+e^{2\phi}\mathrm{d}\gamma\mathrm{d}\bar\gamma\right)\,,\\
B=-L^2e^{2\phi}\mathrm{d}\gamma\wedge\mathrm{d}\bar\gamma\,.
\end{gathered}
\end{equation}
For notational convenience, we refer to the worldsheet superfields in these coordinates as $(\Phi,\Gamma,\bar\Gamma)$ and will maintain their lowercase versions for their bottom components. Here, $L$ is the AdS radius. The worldsheet action in these coordinates is thus\footnote{We emphasize that the fields $\Gamma,\bar\Gamma$ in the equation just below are full superfields, in the sense that they have expansions of the form $\Gamma=\gamma+\theta\psi^\gamma+\bar\theta{\tilde{\psi}}^\gamma+\theta\bar\theta F^\gamma$ and similarly for $\bar\Gamma$, see also \cite{Polchinski:1998rr}. However, the fields $\Gamma,\bar\Gamma$ in equation \eqref{eq:superspace action with auxiliary fields} are (anti-)chiral superfields, meaning $\Gamma=\gamma+\theta\psi^\gamma,\bar\Gamma=\bar\gamma+\bar\theta\psi^{\bar\gamma}$. In other words, there is a nontrivial repackaging of the superfield involved when going from equation \eqref{eq:superspace action} to equation \eqref{eq:superspace action with auxiliary fields}. We will explain this in detail in Appendix \ref{app:measure}.}
\begin{equation}
S_{\text{II}}=\frac{L^2}{2\pi\alpha'}\int\mathrm{d}^{2|2}\boldsymbol{z}\left(D\Phi\overline{D}\Phi+e^{2\Phi}D\bar\Gamma\bar{D}\Gamma\right)\,.
\label{eq:superspace action}
\end{equation}
As we will be interested in the near-boundary geometry of the bulk, which lies at $\Phi\to\infty$, we will find it convenient to work in a first-order formalism by introducing an anticommuting chiral superfield $\Omega$ of weight $(1/2,0)$ and an analogous antichiral superfield $\bar\Omega$ of weight $(0,1/2)$ as Lagrange multipliers and writing the action as
\begin{equation}
S_{\text{II}}=\frac{1}{2\pi}\int\mathrm{d}^{2|2}\boldsymbol{z}\left(kD\Phi\bar{D}\Phi+\Omega\bar{D}\Gamma-\bar{\Omega}D\bar\Gamma-\mu\Omega\bar\Omega\,e^{-2\Phi}\right)\,,
\label{eq:superspace action with auxiliary fields}
\end{equation}
where we have introduced the `level' $k=L^2/\alpha'$ and a parameter $\mu$, which can be set to any desired value via a shift $\Phi\to\Phi+\Phi_0$, and so will be kept arbitrary.

\paragraph{Decoupling the fields:} Before we can move on to quantization, there is one last subtlety which needs to be taken care of, namely the path integral measure. The natural measure of the sigma model couples the fields $\Phi,\Gamma,\bar\Gamma$, and schematically takes the form
\begin{equation}
\mathcal{D}(e^{\Phi}\Gamma)\mathcal{D}(e^{\Phi}\bar{\Gamma})\mathcal{D}\Phi\,,
\label{eq:the measure for susic sl(2,R)}
\end{equation}
and thus $\Phi,\Gamma,\bar\Gamma$ cannot be quantized like free fields. We can, however, decouple the fields in the path integral measure at the expense of introducing a Jacobian which we derive in Appendix \ref{app:measure}:
\begin{equation}
\mathcal{D}(e^{\Phi}\Gamma)\mathcal{D}(e^{\Phi}\bar{\Gamma})\mathcal{D}\Phi=\mathcal{D}\Gamma\,\mathcal{D}\bar\Gamma\,\mathcal{D}\Phi\,\exp\left(\frac{1}{4\pi}\int\mathrm{d}^{2|2}\boldsymbol{z}\,\mathcal{R}\Phi\right)\,.
\label{eq:change of variables in the sl(2,R) measure}
\end{equation}
Here, $\mathcal{R}$ is the curvature superfield on $\Sigma$ whose top component is the usual Ricci scalar, which in superconformal coordinates is simply $\mathcal{R}=\theta\bar\theta R$.\footnote{The full superfield $\mathcal{R}$ includes derivatives of the worldsheet gravitino, which vanish in the superconformal gauge, as well as an auxiliary field, which we integrate out. See Appendix C of \cite{Fan:2021wsb} for detailed expressions.} Introducing this Jacobian into the action and rescaling $\Phi$, we finally land on the first-order action
\begin{equation}\label{eq:superspace-action}
S_{\text{II}}=\frac{1}{2\pi}\int\mathrm{d}^{2|2}\boldsymbol{z}\,\left(\frac{1}{2}D\Phi\bar{D}\Phi-\frac{Q}{4}\mathcal{R}\Phi+\Omega\bar{D}\Gamma-\bar{\Omega}D\bar{\Gamma}-\mu\Omega\bar\Omega\,e^{-Q\Phi}\right)\,,
\end{equation}
where $Q=\sqrt{2/k}$ is the background charge of $\Phi$.

\paragraph{The reduced action:} For the sake of completeness, we will also write down the `reduced' action obtained by integrating over the odd coordinates of $\Sigma$. To this end, we expand our superfields
\begin{equation}
\Phi=\phi+\theta\lambda+\bar\theta\bar\lambda+\theta\bar\theta F\,,\quad \Omega=\omega+\theta\beta\,,\quad\Gamma=\gamma+\theta\psi\,,
\end{equation}
and similarly for $\bar\Omega$ and $\bar\Gamma$. Plugging these expansions into the action \eqref{eq:superspace-action} gives the reduced action
\begin{equation}\label{eq:reduced-action}
\begin{split}
S_{\text{II}}=\frac{1}{2\pi}\int_{\Sigma_{\text{red}}}\mathrm{d}^2z\,\bigg(&\frac{1}{2}\partial\phi\overline{\partial}\phi-\frac{Q}{4}R\phi-\frac{1}{2}\lambda\overline{\partial}\lambda-\frac{1}{2}\bar\lambda\partial\bar\lambda+\beta\overline{\partial}\gamma+\bar\beta\partial\bar\gamma\\
&\hspace{1.5cm}-\omega\overline{\partial}\psi-\bar\omega\partial\bar\psi-\mu(\beta+Q\omega\lambda)(\bar\beta+Q\bar\omega\bar\lambda)e^{-Q\phi}\bigg)\,.
\end{split}
\end{equation}
Note that we have integrated out the auxilliary field $F$ since it plays no role in the dynamics of the theory. Since the above action was constructed in superspace, it is manifestly supersymmetric. Specifically, it is invariant under the left-moving supersymmetry transformation\footnote{Actually, due to the background charge for $\phi$, the supersymmetry transformation of $\lambda$ is slightly more complicated. The supersymmetry transformations in \eqref{eq:sus_trans_local_frame} only hold in a Weyl frame where $R=0$ locally. The more general transformation is determined by the OPE with the supercurrent, $G$, in \eqref{eq:free-field-stress-tensor}.}
\begin{equation}\label{eq:sus_trans_local_frame}
\begin{split}
\delta\phi=\varepsilon\lambda\,,&\quad\delta\lambda=-\varepsilon\partial\phi\,,\\
\delta\gamma=\varepsilon\psi\,,&\quad\delta\psi=-\varepsilon\partial\gamma\,,\\
\delta\omega=\varepsilon\beta\,,&\quad\delta\beta=-\varepsilon\partial\omega\,.
\end{split}
\end{equation}
The part of the action containing only the bosonic fields $(\phi,\gamma,\bar\gamma,\beta,\bar\beta)$ is the action of the usual Wakimoto representation for bosonic string theory in $\text{AdS}_3$.

\subsection{Quantization}\label{sec:quantization}

In the regime $\Phi\gg 1$, the action \eqref{eq:superspace-action} describes a system of nearly free fields: a free linear dilaton $\Phi$ and a first-order system $\Omega,\Gamma$. The OPEs of these fields are easily derived from the Green's functions for the operators $D\bar{D}$ and $\bar{D}$, and they read
\begin{equation}\label{eq:ii-superspace-opes}
\begin{gathered}
\Phi(\boldsymbol{z}_1)\Phi(\boldsymbol{z}_2)\sim-\log|z_1-z_2-\theta_1\theta_2|^2\,,\quad\Omega(\boldsymbol{z}_1)\Gamma(\boldsymbol{z}_2)\sim-\frac{\theta_1-\theta_2}{z_1-z_2}\,.
\end{gathered}
\end{equation}
Alternatively, one can derive the OPEs of the component fields $\phi,\lambda,\gamma,\beta,\psi,\omega$ from the action \eqref{eq:reduced-action}, which read
\begin{equation}\label{eq:free-field-opes}
\begin{gathered}
\phi(z_1)\phi(z_2)\sim-\log|z_1-z_2|^2\,,\quad\lambda(z_1)\lambda(z_2)\sim-\frac{1}{z_1-z_2}\\
\beta(z_1)\gamma(z_2)\sim-\frac{1}{z_1-z_2}\,,\quad\omega(z_1)\psi(z_2)\sim-\frac{1}{z_1-z_2}\,,
\end{gathered}
\end{equation}
which is equivalent to the superfield OPEs above.

The worldsheet stress tensor and supercurrent, which generate superconformal transformations, are given by 
\begin{equation}\label{eq:free-field-stress-tensor}
\begin{gathered}
T=-\frac{1}{2}(\partial\phi)^2-\frac{Q}{2}\partial^2\phi+\frac{1}{2}\lambda\partial\lambda-\beta\partial\gamma+\frac{1}{2}\omega\partial\psi-\frac{1}{2}\partial\omega \psi\,,\\
G=-\lambda\partial\phi-Q\partial\lambda-\omega\partial\gamma-\beta\psi\,.
\end{gathered}
\end{equation}
Alternatively, they can be combined into a single superfield
\begin{equation}\label{eq:ii-superspace-stress-tensor}
\mathscr{T}=\frac{1}{2}G+\theta T=-\frac{1}{2}D\Phi D^2\Phi-\frac{Q}{2}D^3\Phi-\frac{1}{2}\Omega D^2\Gamma-\frac{1}{2}D\Omega D\Gamma\,.
\end{equation}
Together, they generate the $\mathscr{N}=1$ superconformal algebra with central charge
\begin{equation}
c(\text{AdS}_3)=\frac{9}{2}+3Q^2=\frac{9}{2}+\frac{6}{k}\,,
\end{equation}
which is readily checked from the third-order pole in the $GG$ OPE.

\paragraph{\boldmath Relation to the $\text{SL}(2,\mathbb{R})$ WZW model:}

An alternative starting point for the study of superstrings in $\text{AdS}_3$ is the supersymmetric WZW model $\mathfrak{sl}(2,\mathbb{R})_k^{(1)}$. Let us briefly make contact with how our worldsheet action is related to the WZW model description.

First, we remind the reader of some generalities of supersymmetric current algebras, following the conventions in Appendix A of \cite{Ferreira:2017pgt}. The supersymmetric WZW model on a group $G$ with level $k$, written $\mathfrak{g}_k^{(1)}$, is a field theory possessing a supersymmetric extension of the usual $\mathfrak{g}_k$ affine symmetry. The currents $j^a$ of the theory take values in the Lie algebra $\mathfrak{g}$ and satisfy the OPEs:
\begin{equation}
j^a(z)j^b(w)\sim\frac{k\kappa^{ab}}{(z-w)^2}+\frac{f\indices{^a^b_c}j^c(w)}{z-w}\,,
\end{equation}
where $f\indices{^a^b_c}$ are the $\mathfrak{g}$ structure constants and $\kappa^{ab}$ is the (inverse) Killing form. In addition, the worldsheet theory contains $\text{dim}(\mathfrak{g})$ fermions $\psi^a$ which transform in the adjoint representation:
\begin{equation}
j^a(z)\psi^b(w)\sim\frac{f\indices{^a^b_c}\psi^c(w)}{z-w}\,,\quad \psi^a(z)\psi^b(w)\sim\frac{k\kappa^{ab}}{z-w}\,.
\end{equation}
In superspace, we can combine the currents and fermions into a single chiral superfield
\begin{equation}
J^a(z|\theta)=\psi^a(z)+\theta j^a(z)\,,
\end{equation}
which satisfies the superspace OPEs
\begin{equation}
J^a(\boldsymbol{z}_1)J^b(\boldsymbol{z}_2)\sim\frac{k\kappa^{ab}}{z_1-z_2-\theta_1\theta_2}+\frac{(\theta_1-\theta_2)f\indices{^a^b_c}J^c(\boldsymbol{z}_2)}{z_1-z_2-\theta_1\theta_2}\,.
\end{equation}
There is also a natural supersymmetric extension of the Sugawara construction which expresses the stress tensor superfield $\mathscr{T}$ in terms of the currents. Explicitly,\footnote{Note that, unlike in the case of bosonic WZW models, there is no shift in the level $k$.}
\begin{equation}\label{eq:super-sugawara}
\mathscr{T}=\frac{1}{2k}\kappa_{ab}(J^aDJ^b)\,.
\end{equation}
Expanding in components, we find
\begin{equation}
T=\frac{1}{2k}\kappa_{ab}\left((j^aj^b)-(\psi^a\partial\psi^b)\right)\,,\quad G=\frac{1}{k}\kappa_{ab}(\psi^aj^b)\,.
\end{equation}
The central charge can be read off by computing, for example, the third-order pole in the $GG$ OPE, and works out to be
\begin{equation}
c(\mathfrak{g}_k^{(1)})=\left(\frac{3}{2}-\frac{h^{\vee}}{k}\right)\text{dim}(\mathfrak{g}_k^{(1)})\,,
\end{equation}
where $h^{\vee}$ is the dual Coxeter number of $\mathfrak{g}$ ($h^{\vee}=-2$ for $\mathfrak{g}=\mathfrak{sl}(2,\mathbb{R})$).

Returning to the case of $\mathfrak{sl}(2,\mathbb{R})_k^{(1)}$, we can choose the Cartan basis $a=\{+,-,3\}$ for the Lie algebra. It is not difficult to see, then, that the worldsheet fields $\Omega,\Gamma,\Phi$ generate a free-field realization of $\mathfrak{sl}(2,\mathbb{R})_k^{(1)}$. Explicitly, the currents
\begin{equation}
J^+=\Omega\,,\quad J^3=(\Omega\Gamma)-\frac{1}{Q}D\Phi\,,\quad J^-=(\Omega(\Gamma\Gamma))-\frac{2}{Q}\Gamma D\Phi-kD\Gamma\,.
\end{equation}
satisfy the $\mathfrak{sl}(2,\mathbb{R})_k^{(1)}$ OPEs
\begin{equation}
\begin{split}
J^3(\boldsymbol{z}_1)J^{\pm}(\boldsymbol{z}_2)&\sim\pm\frac{(\theta_1-\theta_2)J^{\pm}(\boldsymbol{z}_2)}{z_1-z_2-\theta_1\theta_2}\,,\\
J^3(\boldsymbol{z}_1)J^3(\boldsymbol{z}_2)&\sim-\frac{k}{2(z_1-z_2-\theta_1\theta_2)}\,,\\
J^+(\boldsymbol{z}_1)J^-(\boldsymbol{z}_2)&\sim\frac{k}{z_1-z_2-\theta_1\theta_2}-2\frac{(\theta_1-\theta_2)J^3(\boldsymbol{z}_2)}{z_1-z_2-\theta_1\theta_2}\,.
\end{split}
\end{equation}
For completeness, we write the bosonic currents and adjoint fermions in terms of the reduced worldsheet fields:
\begin{equation}
\begin{gathered}
j^+=\beta\,,\quad j^3=(\beta\gamma)-(\omega\psi)-\frac{1}{Q}\partial\phi\,,\\
j^-=(\beta(\gamma\gamma))-2\gamma(\omega\psi)-\frac{2}{Q}\gamma\partial\phi+\frac{2}{Q}\psi\lambda-k\partial\gamma\,,\\
\psi^+=\omega\,,\quad\psi^3=\omega\gamma-\frac{1}{Q}\lambda\,,\quad \psi^-=\omega(\gamma\gamma)-\frac{2}{Q}\gamma\lambda-k\psi\,.
\end{gathered}
\end{equation}

We thus see that our worldsheet action provides a free-field realization of the $\mathfrak{sl}(2,\mathbb{R})^{(1)}_k$ WZW model. It is in fact the most natural supersymmetric generalization of the Wakimoto representation for the bosonic $\mathfrak{sl}(2,\mathbb{R})_k$ WZW model \cite{Wakimoto:1986gf}. We should also note that this particular free field realization has appeared previously in the literature in the context of the lightcone-gauge quantization of superstrings in $\text{AdS}_3$, see Section 3 of \cite{Yu:1998qw}.

\subsection{Representations and spectral flow}\label{sec:representations}

Since the worldsheet CFT is equivalent to the supersymmetric $\text{SL}(2,\mathbb{R})$ WZW model, its states fall into representations of the supersymmetric current algebra $\mathfrak{sl}(2,\mathbb{R})_k^{(1)}$. Analogously to the bosonic $\text{SL}(2,\mathbb{R})$ WZW model, these representations come in two types: highest-weight and spectrally-flowed representations \cite{Maldacena:2000hw}. The spectrum of the supersymmetric theory is usually deduced via the isomorphism
\begin{equation}
\mathfrak{sl}(2,\mathbb{R})_k^{(1)}\cong\mathfrak{sl}(2,\mathbb{R})_{k+2}\oplus(3\text{ free fermions})\,,
\end{equation}
allowing one to study the representations of the supersymmetric theory in terms of representations of $\mathfrak{sl}(2,\mathbb{R})_{k+2}$ and the mode algebra of three free fermions. However, this decomposition breaks manifest worldsheet supersymmetry, and so we will not use it in what follows. 

Instead, we will directly study the representations of $\mathfrak{sl}(2,\mathbb{R})_k^{(1)}$ by specifying OPEs in superspace. Consider a chiral superfield $\Psi(\boldsymbol{z})$ in the NS sector. We say that $\Psi$ is highest-weight if
\begin{equation}
J^a(\boldsymbol{z}_1)\Psi(\boldsymbol{z}_2)\sim\mathcal{O}\left(\frac{\theta_1-\theta_2}{z_1-z_2-\theta_1\theta_2}\right)\,.
\end{equation}
In terms of the mode algebra, this means that the bottom component of $\Psi$ is annihilated by $j^a_n$ for $n\geq 1$ and by $\psi^a_r$ for $r\geq 1/2$. Since these states are not annihilated by $j^a_0$, they automatically fall into representations of the global $\mathfrak{sl}(2,\mathbb{R})$ algebra. We can take a particular basis of such states labeled by their $j^3_0$ eigenvalue $m$ and their $\mathfrak{sl}(2,\mathbb{R})$ spin. Such a state satisfies
\begin{equation}
j^3_0\ket{m,j}=m\ket{m,j}\,,\quad j^{\pm}_0\ket{m,j}=(m\pm j)\ket{m\pm 1,j}\,.
\end{equation}
Let us call the associated vertex operator $\mathscr{V}_{m,j}$. It satisfies the OPEs
\begin{equation}\label{eq:superspace-highest-weight}
\begin{split}
J^3(\boldsymbol{z}_1)\mathscr{V}_{m,j}(\boldsymbol{z}_2)&\sim\frac{m(\theta_1-\theta_2)\mathscr{V}_{m,j}(\boldsymbol{z}_2)}{z_1-z_2-\theta_1\theta_2}\,,\\
J^\pm(\boldsymbol{z}_1)\mathscr{V}_{m,j}(\boldsymbol{z}_2)&\sim\frac{(m\pm j)(\theta_1-\theta_2)\mathscr{V}_{m\pm 1,j}(\boldsymbol{z}_2)}{z_1-z_2-\theta_1\theta_2}\,.
\end{split}
\end{equation}
The conformal weight of the operator $\mathscr{V}_{m,j}$ (i.e. the conformal weight of the bottom component $V_{m,j}$) can be read off from the Sugawara construction, which gives
\begin{equation}
L_0V_{m,j}=\frac{j(1-j)}{k}V_{m,j}\,.
\end{equation}

Using the free-field realization of $\mathfrak{sl}(2,\mathbb{R})_k^{(1)}$, we can write an explicit expression for the vertex operators $\mathscr{V}_{m,j}$ in the limit $\Phi\to\infty$. Indeed, it is straightforward to check that the operator\footnote{Here, we implicitly mean the left-moving half of $\Phi$.}
\begin{equation}
\mathscr{V}_{m,j}=e^{-Qj\Phi}\Gamma^{-m-j}
\end{equation}
satisfies all of the same properties as the highest-weight states just described.

\paragraph{Spectral flow:} Just as in the bosonic $\text{SL}(2,\mathbb{R})$ WZW model, the supersymmetric current algebra $\mathfrak{sl}(2,\mathbb{R})_k^{(1)}$ admits a spectral flow automorphism which allows us to construct new, non-highest-weight representations from the highest-weight ones discussed above. At the level of the modes of $j^a$ and $\psi^a$, $w$ units of spectral flow acts as
\begin{equation}
\begin{gathered}
\sigma^{w}(j^3_n)=j^3_n+\frac{kw}{2}\delta_{n,0}\,,\quad \sigma^w(j^{\pm}_n)=j^{\pm}_{n\mp w}\,,\\
\sigma^w(\psi^3_r)=\psi^3_r\,,\quad \sigma^w(\psi^{\pm}_r)=\psi^{\pm}_{r\mp w}\,.
\end{gathered}
\end{equation}
Composing $\sigma^w$ with a representation of $\mathfrak{sl}(2,\mathbb{R})_k^{(1)}$ yields a new representation. Such representations are no longer highest-weight, since, for example, $j^+_n$ will only annihilate states for $n> w$. Given a state $\ket{m,j}$, its spectrally-flowed image will not be highest-weight. It will still be annihilated by the positive modes of $\psi^3$ and $j^3$, but for the other generators it will satisfy
\begin{equation}
\begin{gathered}
j^+_n\ket{m,j}^{w}=0\,,\,\,\, n>w\,,\quad j^-_n\ket{m,j}^{w}=0\,,\,\,\,n>-w\,,\\
\psi^+_r\ket{m,j}^{w}=0\,,\,\,\,r>w\,,\quad \psi^-_r\ket{m,n}^{w}=0\,,\,\,\,r>-w\,.
\end{gathered}
\end{equation}
Put another way, the leading order contributions to the OPEs between the currents and the vertex operators $\mathscr{V}_{m,j}^{w}$ are
\begin{equation}
\begin{split}
J^3(\boldsymbol{z}_1)\mathscr{V}_{m,j}^{w}(\boldsymbol{z}_2)&\sim\frac{(m+\tfrac{kw}{2})(\theta_1-\theta_2)\mathscr{V}_{m,j}^{w}(\boldsymbol{z}_2)}{z_1-z_2-\theta_1\theta_2}\,,\\
J^{\pm}(\boldsymbol{z}_1)\mathscr{V}_{m,j}^{w}(\boldsymbol{z}_2)&\sim\frac{(m\pm j)(\theta_1-\theta_2)\mathscr{V}^{w}_{m\pm 1,j}(\boldsymbol{z}_2)}{(z_1-z_2-\theta_1\theta_2)^{\pm w+1}}\,.
\end{split}
\end{equation}
The worldsheet conformal weight of $\ket{m,j}^{w}$ can be worked out from the Sugawara construction:
\begin{equation}\label{eq:spectrally-flowed-weight}
\Delta=\frac{j(1-j)}{k}-hw+\frac{kw^2}{4}\,,
\end{equation}
where $h=m+kw/2$.

Spectral flow can also be understood at the level of the free field realization. It is easiest to express the action in terms of the reduced fields, for which it acts as
\begin{equation}
\begin{gathered}
\sigma^{w}(\gamma_n)=\gamma_{n+w}\,,\quad\sigma^w(\beta_n)=\beta_{n-w}\,,\quad\sigma_w((\partial\phi)_n)=(\partial\phi)_n-\frac{w}{Q}\delta_{n,0}\,,\\
\sigma^w(\psi_r)=\psi_{r+w}\,,\quad\sigma^w(\omega_r)=\omega_{r-w}\,,\quad\sigma^w(\lambda_r)=\lambda_{r}\,.
\end{gathered}
\end{equation}
Spectrally-flowed vertex operators can be written nicely in terms of the free field realization. Since it acts on the bosonic free fields $\gamma,\beta,\phi$ in precisely the same way as in the bosonic WZW model, we can borrow the results of \cite{Knighton:2023mhq}, for which the bosonic part of the spectrally-flowed operator in the Wakimoto representation was given, which takes the form
\begin{equation}
e^{(w/Q-Qj)\phi}\left(\frac{\partial^w\gamma}{w!}\right)^{-m-j}\delta_w(\gamma)\,,
\end{equation}
where $\delta_w(\gamma)$ is a shorthand for the distributional operator 
\begin{equation}\label{eq:gamma_delta_function}
\delta_w(\gamma)=\delta(\gamma)\delta(\partial\gamma)\cdots\delta(\partial^{w-1}\gamma)\,.
\end{equation}
As for the fermionic part, a natural candidate is simply the delta function 
\begin{equation}\label{eq:psi_delta_function}
\delta_w(\psi)=\psi\,\partial\psi\,\cdots\,\partial^{w-1}\psi\,,
\end{equation}
which is annihilated by $\psi_r$ for $r>-w$. Putting these two pieces together, we arrive at a proposal for the bottom component of the spectrally-flowed vertex operator:\footnote{Note that the expression below in the $x$-basis is simply equation (3.22) of \cite{Sriprachyakul:2024gyl} with the identification $\psi_{\text{here}}=(\psi^--2\gamma\psi^3+\gamma^2\psi^+)_{\text{there}}$. One can similarly derive the relation for $\lambda$, but since we do not need such an expression, we will not write it down.}
\begin{equation}\label{eq:vertex-operator-bottom-component}
V_{m,j}^{w}=e^{(w/Q-Qj)\phi}\left(\frac{\partial^w\gamma}{w!}\right)^{-m-j}\delta_w(\psi)\delta_w(\gamma)\,.
\end{equation}
Indeed, it is readily checked that this field satisfies the correct OPEs with the $\mathfrak{sl}(2,\mathbb{R})_k^{(1)}$ currents $j^a,\psi^a$, has spacetime conformal weight $h=m+kw/2$ and worldsheet conformal weight given by \eqref{eq:spectrally-flowed-weight}.

Passing back to a manifestly supersymmetric setting, we can promote \eqref{eq:vertex-operator-bottom-component} by promoting the free fields to their superfield counterparts. The only subtle piece is the product $\delta_w(\psi)\delta_w(\gamma)$. The natural object to replace this with is the supersymmetric delta function operator
\begin{equation}\label{eq:Superspace_delta_function}
\delta_{w|w}(\Gamma)=\delta(\Gamma)\delta(D\Gamma)\delta(D^2\Gamma)\cdots\delta(D^{2w-1}\Gamma)\,.
\end{equation}
The subscript is to remind us that the delta function imposes $w$ commuting and $w$ anticommuting constraints on the field $\Gamma$. With this in mind, we can write down the spectrally-flowed superspace vertex operator
\begin{equation}\label{eq:NS_vo_superspace}
\mathscr{V}^{w}_{m,j}=e^{(w/Q-Qj)\Phi}\left(\frac{D^{2w}\Gamma}{w!}\right)^{-m-j}\delta_{w|w}(\Gamma)\,.
\end{equation}

\paragraph{\boldmath The $x$-basis:} The vertex operators written down so far describe a string being emitted in the infinite past with winding number $w$. For holographic applications, we want to be able to describe strings which are emitted from arbitrary points $x$ on the boundary of $\text{AdS}_3$. Since translations on the conformal boundary are generated by $j^+_0=\beta_0$, this leads us to define the $x$-basis vertex operators
\begin{equation}
\mathscr{V}_{m,j}^{w}(\boldsymbol{z};x)=e^{x\beta_0}\mathscr{V}_{m,j}^{w}(\boldsymbol{z})e^{-x\beta_0}\,.
\end{equation}
Only $\Gamma$ has a non-zero OPE with $\beta$, for which $e^{\beta_0x}\Gamma\,e^{-\beta_0x}=\Gamma-x$, such that the $x$-basis vertex operator is
\begin{equation}\label{eq:superspace-ns-vertex}
\mathscr{V}^{w}_{m,j}(\boldsymbol{z};x)=e^{(w/Q-Qj)\Phi}(\boldsymbol{z})\left(\frac{D^{2w}\Gamma(\boldsymbol{z})}{w!}\right)^{-m-j}\delta_{w|w}(\Gamma(\boldsymbol{z})-x)
\end{equation}
for $w\geq 0$.\footnote{The $x$-basis allows us to restrict to $w\geq 0$, since the automorphism which maps $x\to-1/x$ also maps $w\to-w$.}

\subsection{The Ramond sector}\label{sec:ramond_sector}

Although we will exclusively consider correlation functions of vertex operators in the NS sector in Section \ref{sec:localization}, our discussion would be incomplete without a description of the spectrally-flowed operators in the Ramond sector. We will present both the `standard' description in terms of spin fields as well as a natural superspace description in terms of vertex operators inserted along Ramond punctures on the worldsheet. See Appendix \ref{app:srs} for a brief review Ramond vertex operators in superspace.

\paragraph{The standard description:} The usual description of Ramond-sector vertex operators for $\rm{AdS}_3$ is based on the introduction of spin fields for the free fermions $\lambda,\psi,\omega$. In the Ramond sector, these fields have zero modes that obey the three-dimensional Dirac algebra
\begin{equation}\label{eq:AdS3_fermion_zero_modes}
\{\lambda_0,\lambda_0\}=-1\,,\quad\{\omega_0,\psi_0\}=-1\,,
\end{equation}
with all other anticommutators vanishing. In the full string theory on $\rm{AdS}_3 \times \mathcal{N}$, we also have fermions living in the compact CFT, $\mathcal{N}$. In Section \ref{sec:Ramond_phys_states}, we will assume that this compact CFT contributes 7 free fermions and we will explicitly include contributions to the vertex operator from this compact sector. Here, we will make a more mild assumption that there exists a free fermion, $\tilde{\lambda}$, in $\mathcal{N}$ that can be paired with $\lambda$ to construct a spin field. Assuming the zero modes of $\tilde{\lambda}$ satisfy
\begin{equation}\label{eq:lambda_tilde}
    \{\tilde{\lambda}_0,\tilde{\lambda}_0\} = -1,
\end{equation}
the Ramond ground states for AdS$_3$ coupled to $\tilde{\lambda}$ furnish a four-dimensional representation of the algebra generated by \eqref{eq:AdS3_fermion_zero_modes} and \eqref{eq:lambda_tilde}.

In order to construct a spin field for $\lambda$ and $\tilde{\lambda}$, we define the bosonization
\begin{equation}\label{eq:kappa_definition}
\lambda = \frac{i}{\sqrt{2}}(e^{i\kappa} + e^{-i\kappa})\,, \qquad \tilde{\lambda} = \frac{1}{\sqrt{2}}(e^{i\kappa} - e^{-i\kappa})\,,
\end{equation}
from which the spin field $S_{\kappa}^{\epsilon} = e^{\frac{i\epsilon\kappa}{2}}$ can be defined. Similarly, one can bosonize the fermions $\psi$ and $\omega$ via
\begin{equation}
    \psi = e^{iH}\,, \qquad \omega = -e^{-iH}\,.
\end{equation}
The associated spin field labeling the Ramond vacuum for this system is $S_{H}^{\epsilon} = e^{\frac{i\epsilon H}{2}}$. Unflowed Ramond-sector ground states for $\rm{AdS}_3$ coupled to a free fermion then take the form
\begin{equation}
V_{m,j}^{(R,\epsilon_1,\epsilon_2)}=e^{-Qj\phi}\gamma^{-m-j-\frac{\epsilon_1}{2}}S_{H}^{\epsilon_1} S_{\kappa}^{\epsilon_2}\,.
\end{equation}
The state $V_{m,j}^{(R,\epsilon_1,\epsilon_2)}$ has $j^3_0$ eigenvalue $m$ and worldsheet conformal weight
\begin{equation}
\Delta\left(V_{m,j}^{(R,\epsilon_1,\epsilon_2)}\right)=\frac{j(1-j)}{k}+\frac{1}{4}\,.
\end{equation}

Spectrally-flowed vertex operators are constructed in essentially the same way as in the NS sector. Let $|\epsilon_1,\epsilon_2\rangle$ represent the Ramond vacuum state corresponding to the product of spin fields $S_H^{\epsilon_1}S_{\kappa}^{\epsilon_2}$. Since spectral flow shifts the mode number of the fermions $\psi$ and $\omega$, we have
\begin{equation}
\ket{\tfrac{1}{2},\epsilon_2}^{w}=\psi_{-w+1}\cdots\psi_{-1}\psi_0\ket{\tfrac{1}{2},\epsilon_2}\,,\quad\ket{-\tfrac{1}{2},\epsilon_2}^{w}=\psi_{-w}\cdots\psi_{-1}\psi_0\ket{\tfrac{1}{2},\epsilon_2}\,,
\end{equation}
for $w>0$, where we define $|\epsilon_1,\epsilon_2\rangle$ such that $|-\tfrac{1}{2},\epsilon_2\rangle = \psi_0|\tfrac{1}{2},\epsilon_2\rangle$. Thus, spectrally-flowed vertex operators of Ramond ground states are given by
\begin{equation}\label{eq:Ramond_vo_reduced_space}
V^{w,(R,\epsilon_1,\epsilon_2)}_{m,j}=e^{(w/Q-Qj)\phi}\left(\frac{\partial^w\gamma}{w!}\right)^{-m-j-\frac{\epsilon_1}{2}}\delta_{w}(\gamma)e^{i(w+\frac{\epsilon_1}{2})H} S_{\kappa}^{\epsilon_2}\,.
\end{equation}
This has spacetime conformal weight $h = m+kw/2$ and worldsheet conformal weight
\begin{equation}
\Delta\left(V^{w,(R,\epsilon_1,\epsilon_2)}_{m,j}\right)=\frac{j(1-j)}{k}-hw+\frac{kw^2}{4} + \frac{1}{4}\,.
\end{equation}

\paragraph{In superspace:} In the language of super Riemann surfaces, the existence of a Ramond vertex operator signals a degeneration in the superconformal structure of the worldsheet. Specifically, a Ramond-sector vertex operator is naturally associated to a \textit{Ramond divisor} $\mathscr{F}$, a submanifold $\mathscr{F}\subset\Sigma$ of dimension $0|1$ on which the superderivative degenerates, i.e.
\begin{equation}
D^2=0\text{ on }\mathscr{F}\,.
\end{equation}
From this perspective, a Ramond vertex operator is not just a state inserted at a point on the worldsheet, but rather a change in the geometry of the worldsheet itself that cannot be compensated by a globally-defined (super)conformal transformation.

Let us take the vertex operator to sit on the divisor $z=z_i$. Then we can always choose coordinates for which
\begin{equation}
D=\partial_{\theta}+(z-z_i)\theta\partial_z\,.
\end{equation}
Away from $z=z_i$, we can perform a change of coordinates to bring $D$ back into the canonical form \eqref{eq:canonical-D} for a superconformal structure away from a degeneration. However, this change of coordinates will always have a square-root branch cut originating at $z=z_i$. Specifically, in a coordinate frame $(z|\theta_{\text{mv}})$ where $\theta_{\text{mv}} = (z-z_i)^{1/2}\theta$, the superconformal structure takes the form
\begin{equation}
    D_{\text{mv}} = \p_{\theta_{\text{mv}}} + \theta_{\text{mv}}\p_z\,,
\end{equation}
in agreement with \eqref{eq:canonical-D}. The subscript `mv' is to signify that $\theta_{\text{mv}}$ is multi-valued in a neighborhood of the Ramond divisor.

As outlined, $\theta=(z-z_i)^{-1/2}\theta_{\text{mv}}$ is the canonical choice for defining a single-valued coordinate frame near the Ramond divisor. Nevertheless, in what follows, it will prove helpful to consider two choices of single-valued coordinate frames. We define
\begin{equation}
    \theta_{\text{sv}}^{\epsilon} = (z-z_i)^{\epsilon/2}\theta_{\text{mv}}
\end{equation}
for $\epsilon=\pm$, such that $\theta_{\text{sv}}^-$ coincides with the canonical choice. Through the chain rule, one finds that\footnote{Note that, on each side of equation \eqref{eq:superconformal_structure_transformation}, we are implicitly changing which variables are held constant in the partial derivatives as is typical in the chain rule. The precise definition is that $D_{\text{mv}} = \p_{\theta_{\text{mv}}}|_z + \theta_{\text{mv}}\p_z|_{\theta_{\text{mv}}}$ and $D^{\epsilon}_{\text{sv}} = \p_{\theta^{\epsilon}_{\text{sv}}}|_z + \theta_{\text{sv}}^{\epsilon}\p_z|_{\theta^{\epsilon}_{\text{sv}}}$, such that $\p_z$ takes a different meaning on each side.}
\begin{equation}\label{eq:superconformal_structure_transformation}
    D_{\text{mv}} = (z-z_i)^{\epsilon/2}D_{\text{sv}}^{\epsilon} = (z-z_i)^{\epsilon/2}(\p_{\theta_{\text{sv}}^{\epsilon}} + \theta_{\text{sv}}^{\epsilon} (z-z_i)^{-\epsilon}\p_z)\,,
\end{equation}
such that $D_{\text{sv}}^{\epsilon}$ define conformally equivalent superconformal structures. It is evident that $(D^{\epsilon}_{\text{sv}})^2 = (z-z_i)^{-\epsilon}\p_z$, which vanishes on the Ramond divisor for the canonical choice $\epsilon=-$. Away from the Ramond divisor, we have defined $\Gamma = \gamma + \theta_{\text{mv}}\psi$. Thus, in the single-valued local coordinate frame it is helpful to define the combination
\begin{equation}\label{eq:psi_hat}
    \hat{\psi}^{\epsilon} = (z-z_i)^{-\epsilon/2}\psi\,,
\end{equation}
such that $\Gamma = \gamma + \theta_{\text{sv}}^{\epsilon}\hat{\psi}^{\epsilon}$.\footnote{For the canonical choice of $\epsilon=-$, one can interpret \eqref{eq:psi_hat} as follows \cite{Witten:2012bh}: if one wishes to define their worldsheet CFT on the sphere, then $\psi$ is the natural field to use, whilst $\hat{\psi}^-$ is the appropriate definition of the fermionic field on the cylinder. Put more simply, the difference between $\psi$ and $\hat{\psi}^-$ is the well-known fact that Ramond sector fermions are periodic on the cylinder, but anti-periodic on the sphere \cite{Polchinski:1998rr}. Since we want to view our worldsheet CFT as being defined on the sphere, we will view $\psi$ as the fundamental field in our path integral and simply define $\hat{\psi}^{\epsilon}$ via \eqref{eq:psi_hat}, not viewing it as a globally-defined field in the CFT.}

How, then, should we use this machinery to promote \eqref{eq:Ramond_vo_reduced_space} to superspace and express it in terms of superfields? In the NS sector, the delta function $\delta_w(\psi) = e^{iwH}$ imposed that $\psi(z) \sim (z-z_i)^w$ near an NS puncture and this information was neatly repackaged in superspace through the delta function $\delta_{w|w}(\Gamma)$ --- see equation \eqref{eq:NS_vo_superspace}. For the analogous statement in the Ramond sector, we first note that
\begin{equation}\label{eq:Ramond_delta_functions}
\lim_{z\to z_i}(z-z_i)^{-\frac{\epsilon w}{2}}e^{iwH(z)}e^{\frac{i\epsilon H(z_i)}{2}}=e^{i(w+\frac{\epsilon}{2})H(z_i)}\,.
\end{equation}
Next, it is easy to see that
\begin{equation}
(z-z_i)^{-\frac{\epsilon w}{2}}e^{iwH(z)}=\prod_{i=0}^{w-1}\partial_z^i((z-z_i)^{-\frac{\epsilon}{2}}\psi(z))=\delta_w(\hat{\psi}^{\epsilon}(z))\,.
\end{equation}
Thus, we shall define $e^{i(w+\frac{\epsilon}{2})H(z_i)} = \delta_w(\hat{\psi}^{\epsilon})S^{\epsilon}_H(z_i)$ where we implicitly define the right hand side through a radial ordering. Now we can rewrite \eqref{eq:Ramond_vo_reduced_space} as
\begin{equation}
V^{w,(R,\epsilon_1,\epsilon_2)}_{m,j}=e^{(w/Q-Qj)\phi}\left(\frac{\partial^w\gamma}{w!}\right)^{-m-j-\frac{\epsilon_1}{2}}\delta_{w}(\gamma)\delta_w(\hat{\psi}^{\epsilon_1})S^{\epsilon_1}_HS_{\kappa}^{\epsilon_2}\,.
\end{equation}
We shall express the lift to superspace via
\begin{equation}\label{eq:Ramond_vo_superspace}
\mathscr{V}^{w,(R,\epsilon_1,\epsilon_2)}_{m,j}=e^{(w/Q-Qj)\Phi}\left(\frac{\p^{w}\Gamma}{w!}\right)^{-m-j-\frac{\epsilon_1}{2}}\delta_{w|w}^{\epsilon_1}(\Gamma)S^{\epsilon_1}_H\mathcal{S}^{\epsilon_2}_{\kappa}\,,
\end{equation}
which in the $x$-basis becomes
\begin{equation}
\mathscr{V}^{w,(R,\epsilon_1,\epsilon_2)}_{m,j}(x,\boldsymbol{z})=e^{(w/Q-Qj)\Phi}\left(\frac{\p^{w}\Gamma}{w!}\right)^{-m-j-\frac{\epsilon_1}{2}}\delta_{w|w}^{\epsilon_1}(\Gamma-x)S^{\epsilon_1}_H\mathcal{S}^{\epsilon_2}_{\kappa}(\boldsymbol{z})\,.
\end{equation}
We define the constituents of \eqref{eq:Ramond_vo_superspace} in the following way.

\begin{itemize}

\item $\delta_{w|w}^{\epsilon_1}(\Gamma)$:
In simple analogy with \eqref{eq:Superspace_delta_function} from the NS sector, we define
\begin{equation}\label{eq:Gamma_delta_function_Ramond}
\delta_{w|w}^{\epsilon_1}(\Gamma) = \delta(\Gamma)\delta(D_{\text{sv}}^{\epsilon_1}\Gamma)\delta((D_{\text{sv}}^{\epsilon_1})^2\Gamma)\cdots\delta((D_{\text{sv}}^{\epsilon_1})^{2w-1}\Gamma)\, . 
\end{equation}
Despite the fact that $(D_{\text{sv}}^{\epsilon_1})^2 = (z-z_i)^{-\epsilon_1}\p_z$, this expression remains well-defined even in the limit $z\to z_i$ where we intend to insert the vertex operator. To see this, we first note that
\begin{equation}
    \begin{aligned}
        \delta_{w|w}^{\epsilon_1}(\Gamma) &= \delta(\Gamma)\delta(D_{\text{sv}}^{\epsilon_1}\Gamma)\delta\left( (z-z_i)^{-\epsilon_1} \p\Gamma \right) \delta\left( (z-z_i)^{-\epsilon_1}\p D_{\text{sv}}^{\epsilon_1}\Gamma \right)\cdots\\
        &\qquad \cdots \delta\left( \left((z-z_i)^{-\epsilon_1}\p\right)^{w-1}\Gamma \right) \delta\left( \left((z-z_i)^{-\epsilon_1}\p\right)^{w-1} D_{\text{sv}}^{\epsilon_1}\Gamma \right)\,.
    \end{aligned}
\end{equation}
Since half of these delta functions are bosonic and half are fermionic, we may pull out the leading factor of $(z-z_i)^{-\epsilon_1}$ from each term,
\begin{equation}
    \begin{aligned}
        \delta_{w|w}^{\epsilon_1}(\Gamma) &= \delta(\Gamma)\delta(D_{\text{sv}}^{\epsilon_1}\Gamma)\delta\left( \p\Gamma \right) \delta\left( \p D_{\text{sv}}^{\epsilon_1}\Gamma \right)\cdots\\
        &\qquad \cdots \delta\left( \p\left((z-z_i)^{-\epsilon_1}\p\right)^{w-2}\Gamma \right) \delta\left(\p \left((z-z_i)^{-\epsilon_1}\p\right)^{w-2} D_{\text{sv}}^{\epsilon_1}\Gamma \right)\,.
    \end{aligned}
\end{equation}
Next, it is a simple matter of proof by induction to show that all other factors of $(z-z_i)^{-\epsilon_1}$ may be pulled out and that they cancel between the bosonic and fermionic delta functions. Indeed, if any of the derivatives act on one of the factors of $(z-z_i)^{-\epsilon_1}$, then a lower order derivative acts on $\Gamma$ or $D_{\text{sv}}^{\epsilon_1}\Gamma$ which already vanishes by virtue of the delta function insertions with fewer derivatives. Therefore,
\begin{equation}
        \delta_{w|w}^{\epsilon_1}(\Gamma) = \prod_{i=0}^{w-1}\delta(\p^i\Gamma)\delta(\p^iD_{\text{sv}}^{\epsilon_1}\Gamma)\,,
\end{equation}
proving the claimed result. Similarly, we have replaced the expression $D^{2w}\Gamma$ from \eqref{eq:NS_vo_superspace} by $\p^w\Gamma$ in \eqref{eq:Ramond_vo_superspace} to avoid similar complications from the degeneration of the superconformal structure.

\item $e^{(w/Q-Qj)\Phi}\mathcal{S}_{\kappa}^{\epsilon_2}$: Focusing on the chiral part $\Phi =\phi + \theta_{\text{mv}}\lambda$ of the full superfield,
\begin{equation}
    e^{(\frac{w}{Q}-Qj)\Phi} = e^{(\frac{w}{Q}-Qj)\phi}\left(1 + \left(\frac{w}{Q}-Qj\right)\theta_{\text{mv}}\lambda\right)\,.
\end{equation}
We then take the normal ordered product with $S_{\kappa}^{\epsilon_2} = e^{i\epsilon_2\kappa/2}$,
\begin{equation}
    :e^{(\frac{w}{Q}-Qj)\Phi}S_{\kappa}^{\epsilon_2}: = e^{(\frac{w}{Q}-Qj)\phi}\left(S_{\kappa}^{\epsilon_2} + \frac{i}{\sqrt{2}}\left(\frac{w}{Q}-Qj\right)\theta_{\text{sv}} S_{\kappa}^{-\epsilon_2}\right)\,,
\end{equation}
where we use the OPE
\begin{equation}
    \lambda(z) S_{\kappa}^{\epsilon_2}(z_i) = \frac{i}{\sqrt{2}}\frac{1}{(z-z_i)^{1/2}}S_{\kappa}^{-\epsilon_2}(z_i) + \dots
\end{equation}
from \eqref{eq:kappa_definition} and the behaviour of $\theta_{\text{mv}} = (z-z_i)^{1/2}\theta_{\text{sv}}$. Lastly, we note that the term proportional to $\theta_{\text{sv}}$ should be equal to the action of $G_0$ on the part independent of $\theta_{\text{sv}}$ (see Section \ref{sec:Ramond_phys_states} and Appendix \ref{app:srs}). The background charge for $\phi$ induces an anomalous superconformal transformation on $\lambda$ and therefore, $S^{\epsilon_2}_{\kappa}$ transforms non-trivially under $G_0$. This leads us to define\footnote{In principle, one could also include a term $\theta_{\text{sv}}(G_C)_0S^{\epsilon_2}_{\kappa}$ in this definition for the action of the supercurrent on the free fermion $\tilde{\lambda}$ that we have coupled to AdS$_3$ in this section. This would be necessary if, in analogy with $\lambda$, $\tilde{\lambda}$ also has an anomolous superconformal transformation. Since $\psi$ and $\omega$ transform as superconformal primaries, no such correction is required for $S^{\epsilon_1}_H$ in \eqref{eq:Ramond_vo_superspace}.}
\begin{equation}
    \mathcal{S}_{\kappa}^{\epsilon_2} = S_{\kappa}^{\epsilon_2} + \theta_{\text{sv}}G_0S_{\kappa}^{\epsilon_2} = S_{\kappa}^{\epsilon_2} + \frac{iQ}{2\sqrt{2}}\theta_{\text{sv}}S^{-\epsilon_2}_{\kappa}\,.
\end{equation}
Hence, in \eqref{eq:Ramond_vo_superspace} we have defined
\begin{equation}\label{eq:phi_lambda_superfields}
    e^{(w/Q-Qj)\Phi}\mathcal{S}_{\kappa}^{\epsilon_2} = e^{(\frac{w}{Q}-Qj)\phi}\left(S_{\kappa}^{\epsilon_2} + \frac{i}{\sqrt{2}}\left(\frac{w}{Q}-Q\left(j - \frac{1}{2}\right)\right)\theta_{\text{sv}} S_{\kappa}^{-\epsilon_2}\right)\,.
\end{equation}

\end{itemize}

\subsection{The screening operator}

An important final detail needed in defining the worldsheet CFT is a specific type of screening operator
which plays a central role in bosonic string theory on $\text{AdS}_3$. It was identified in \cite{Dei:2023ivl,Knighton:2023mhq,Hikida:2023jyc,Knighton:2024qxd} as an essential ingredient in defining a worldsheet sigma model whose correlators give sensible answers. In the present context, we will need a suitable supersymmetric generalization.

We first review the bosonic construction of the screening operator. In bosonic string theory, the coordinates $\gamma,\bar\gamma$ in the worldsheet sigma model are identified with the boundary of $\rm AdS_3$. In Euclidean signature, the conformal boundary of $\text{AdS}_3$ is the 2-sphere, and $(\gamma,\bar\gamma)$ are standard stereographic coordinates. In the worldsheet sigma model, however, $\gamma$ is a free field near the conformal boundary and hence should be finite in any correlation function, provided $\gamma$ has no poles in its OPEs with any of the vertex operators. This means that the north pole of the conformal boundary is completely missing from the worldsheet description. On a technical level, this is because Poincar\'e coordinates $(\phi,\gamma,\bar\gamma)$ do not completely cover the conformal compactification of $\text{AdS}_3$ (since it misses the point $\gamma=\infty$), and in order to describe the whole spacetime one needs to work on multiple coordinate charts. Nevertheless, there is a much simpler way to re-introduce the north pole of the conformal boundary -- we simply allow $\gamma$ to diverge, and the simplest way to achieve this is by adding in (by hand) an operator whose OPE with $\gamma$ has a simple pole. Furthermore, we should require that any operator that induces the divergence should not destroy the existing symmetry of the theory (specifically conformal and $\mathfrak{sl}(2,\mathbb{R})_k$ affine symmetry in the bosonic case). These requirements then allow us to determine such an operator as done in \cite{Knighton:2023mhq}. This operator, called $D$, is a primary of conformal weight $(1,0)$ whose OPEs with the $\mathfrak{sl}(2,\mathbb{R})_k$ currents produce total derivatives and which, crutially, has a simple pole in its OPE with $\gamma$. There is a unique such operator, and it lives in the $w=-1$ spectrally-flowed sector. Specifically,
\begin{equation}
D=V^{w=-1}_{m=-\frac{k}{2},j=\frac{k-2}{2}}=\left(\oint\gamma\right)^{-(k-1)}\delta(\beta)e^{-2\phi/Q}\,.
\end{equation}
With the screening operator $D$ defined, the worldsheet theory is modified by deforming the action
\begin{equation}
S\to S-p\int_{\Sigma}D\bar{D}\,,
\end{equation}
where $p$ is a chemical potential that counts the number of poles of $\gamma$. This deformation was shown in \cite{Dei:2023ivl,Knighton:2023mhq,Knighton:2024qxd,Hikida:2023jyc} to be the correct prescription for obtaining worldsheet correlation functions that precisely match those of the dual CFT.\footnote{There are many similar (but inequivalent) proposals that can be found in the $\text{AdS}_3$ literature, see \cite{Gerasimov:1990fi,Giribet:2000fy,Giribet:2001ft,Giveon:2019gfk,Eberhardt:2019ywk,McStay:2023thk} and references therein for an incomplete list.} More recently, it was also shown to be important in understanding the relationship between the dual CFT OPEs and the worldsheet OPEs \cite{Sriprachyakul:2025ubx}.

The supersymmetric generalization of this operator should induce a simple pole in the worldsheet field $\Gamma$, while simultaneously preserving all worldsheet symmetries. In particular, it must be a superconformal primary $\mathscr{D}$ of weight $(\frac{1}{2},0)$ in the $w=-1$ spectrally flowed sector and it must be a singlet with respect to the $\mathfrak{sl}(2,\mathbb{R})^{(1)}_k$ algebra (in the sense that the OPEs of $J^a$ with $\mathscr{D}$ yield total derivatives in superspace). Just as in the bosonic case, there is a unique operator satisfying these properties which was identified in \cite{Sriprachyakul:2024gyl}. The operator is
\begin{equation}
\mathscr{D}=\mathscr{V}^{w=-1}_{m=-\frac{k}{2},j=-\frac{k}{2}}\,.
\end{equation}
Near the conformal boundary, this can be written in terms of free fields as
\begin{equation}
\mathscr{D}=e^{-2\Phi/Q}\left(\text{Res}\,\Gamma\right)^{-k}\delta(\Omega)\delta(D\Omega)\,,
\end{equation}
where the residue is defined as a contour integral around the insertion point $(\lambda,\eta)$:
\begin{equation}
\mathop{\text{Res}}_{(\lambda,\eta)}\,\Gamma:=\oint_{\lambda}\mathrm{d}z\int \mathrm{d}\theta(\theta-\eta)\Gamma(\boldsymbol{z})\,.
\end{equation}
The odd integral picks out the coefficient of the term $1/(z-\lambda-\theta\eta)$ in a local expansion around $(\lambda,\eta)$, and thus reads off the residue of the bottom component of $\Gamma$. As in the bosonic case, we can use this field to deform the action
\begin{equation}
S\to S-p\int_{\Sigma}\mathrm{d}^2\lambda\,\mathrm{d}^2\eta\,\mathscr{D}(\lambda,\eta)\bar{\mathscr{D}}(\bar\lambda,\bar\eta)\,.
\end{equation}
Later, we will show that the insertion of this field generates the right equations of motion for the superfield $\Gamma$, inducing an arbitrary number of simple poles.

\section{The physical spectrum}\label{sec:physical-spectrum}

In the previous section we introduced the worldsheet superconformal sigma model on $\text{AdS}_3$. In order to promote this theory to a \textit{bona fide} string theory, we need to tensor the theory with an $\mathscr{N}=(1,1)$ compact CFT which we call $\mathcal{N}$. Typically, we think of $\mathcal{N}$ as a 7-dimensional compact manifold such that the resulting spacetime $\text{AdS}_3\times\mathcal{N}$ is a valid classical supergravity background. However, $\mathcal{N}$ may also be some non-geometric CFT, so long as it admits an $\mathscr{N}=(1,1)$ superconformal algebra with central charge
\begin{equation}
c(\mathcal{N})=15-c(\text{AdS}_3)=\frac{21}{2}+\frac{6}{k}\,.
\end{equation}
Physical states in the theory are then superconformal primaries $\mathscr{V}$ of weight $(\frac{1}{2},\frac{1}{2})$ for the NS sector and weight $(\frac{5}{8},\frac{5}{8})$ for the R sector which survive the GSO projection. In this section, we will write down low-lying physical states in the theory, focusing on those operators built from spectrally-flowed primaries in the $\text{AdS}_3$ factor and ground states in the $\mathcal{N}$ factor.

\subsection{The NS sector}

To begin, we consider vertex operators in the NS sector. For spectrally-flowed states in the $\text{AdS}_3$ factor, these were constructed in Section \ref{sec:representations}. The precise operators $\mathscr{V}_{m,j}^{w}$ are superconformal primaries of conformal weight
\begin{equation}
\Delta=\frac{j(1-j)}{k}-hw+\frac{kw^2}{4}\,.
\end{equation}
If we tensor this with the NS-sector ground state from $\mathcal{N}$, the condition $\Delta=\frac{1}{2}$ determines $h$ in terms of the other quantum numbers:
\begin{equation}
h=\frac{j(1-j)}{kw}-\frac{1}{2w}+\frac{kw}{4}\,.
\end{equation}
Thus, there is one such state in the NS sector for each value of $j$.

\paragraph{The GSO projection:} Imposing the GSO projection is most easily done by setting $\theta=0$. The spectrally-flowed vertex operator $\mathscr{V}^{w}_{m,j}$ is built from $w$ fermions, specifically the combination
\begin{equation}
\psi\partial\psi\cdots\partial^{w-1}\psi\,.
\end{equation}
Thus, for these operators we have $(-1)^F=(-1)^w$. The GSO projection $(-1)^F=-1$ then restricts $w$ to be odd.

For even $w$, we can still satisfy the GSO projection by considering excited states, i.e. by acting with one fermion operator. The most general nontrivial ansatz found by acting with the fermions $\psi,\omega,\lambda$ is\footnote{See Section 2.3.2 of \cite{Yu:2024kxr} for a similar discussion.}
\begin{equation}
\alpha^{(\psi)}\psi_{-w-\frac{1}{2}}\ket{\mathscr{V}^{w}_{m+1,j}}+\alpha^{(\omega)}\omega_{w-\frac{1}{2}}\ket{\mathscr{V}^{w}_{m-1,j}}+\alpha^{(\lambda)}\lambda_{-\frac{1}{2}}\ket{\mathscr{V}^{w}_{m,j}}\,,
\end{equation}
since these are the first nontrivial modes which don't annihilate the spectrally-flowed ground state. The states must have the same $\mathfrak{sl}(2,\mathbb{R})$ spin to lie in the same representation, and the $j^3$ quantum numbers are chosen so each state has the same $L_0$ quantum number. The physical state condition $G_{1/2}=0$ can be read off from the anticommutators
\begin{equation}
\{G_{1/2},\psi_{-w-1/2}\}=(\partial\gamma)_{-w}\,,\quad\{G_{1/2},\omega_{w-1/2}\}=\beta_{w}\,,\quad\{G_{1/2},\lambda_{-1/2}\}=(\partial\phi)_{0}-Q\,,
\end{equation}
as well as the relations
\begin{equation}
\begin{gathered}
(\partial\gamma)_{-w}\ket{\mathscr{V}^{w}_{m,j}}=w\ket{\mathscr{V}^{w}_{m-1,j}}\,,\quad\beta_w\ket{\mathscr{V}_{m,j}^{w}}=(m+j)\ket{\mathscr{V}_{m+1,j}^{w}}\,,\\(\partial\phi)_0\ket{\mathscr{V}^{w}_{m,j}}=(Qj-w/Q)\ket{\mathscr{V}^{w}_{m,j}}\,,
\end{gathered}
\end{equation}
which together yield the constraint equations
\begin{equation}\label{eq:ferm-descendant-constraint}
w\alpha^{(\psi)}+(m+j-1)\alpha^{(\omega)}+(Qj-Q-w/Q)\alpha^{(\lambda)}=0\,.
\end{equation}
This constraint has a two-dimensional set of solutions; however, one solution corresponds to the spurious state $G_{-\frac{1}{2}}\ket{\mathscr{V}_{m,j}^{w}}$, which has
\begin{equation}
\alpha^{(\psi)}_{\text{spur}}=-(m+j)\,,\quad\alpha^{(\omega)}_{\text{spur}}=-w\,,\quad\alpha^{(\lambda)}_{\text{spur}}=-(Qj-w/Q)\,.
\end{equation}
Physical states are thus in one-to-one correspondence with solutions to \eqref{eq:ferm-descendant-constraint} modulo the spurious solution. A convenient choice is
\begin{equation}
\alpha^{(\psi)}=(m+j-1)\,,\quad\alpha^{(\omega)}=-w\,,\quad\alpha^{(\lambda)}=0\,.
\end{equation}
Similarly to the ground states, the $L_0=1/2$ condition determines $h$ to be
\begin{equation}
h=\frac{j(1-j)}{kw}+\frac{kw}{4}\,.
\end{equation}
Beyond these excited states, there will also be excited states in the NS sector of the compact CFT $\mathcal{N}$. If such a state has conformal weight $\Delta_{\mathcal{N}}$, is annihilated by $G_{1/2}$, and has fermion number $F=1$ in the compact direction, then it survives the GSO projection provided $w$ is even. Such a state survives the physical state condition provided that
\begin{equation}
h=\frac{j(1-j)}{kw}+\frac{kw}{4}+\frac{1}{w}\left(\Delta_{\mathcal{N}}-\frac{1}{2}\right)\,.
\end{equation}
For example, if $\mathcal{N}=\text{S}^3\times\mathbb{T}^4$ or $\text{S}^3\times\text{S}^3\times\text{S}^1$, there are seven first-excited states of weight $\Delta_{\mathcal{N}}=1/2$.\footnote{For discussions on how these states match with the low-lying states of the dual CFT, see for example \cite{Eberhardt:2019qcl,Yu:2024kxr}. We will also explore the correspondence further in \cite{Knighton:2026xxx}.}

\subsection{The Ramond sector}\label{sec:Ramond_phys_states}

In Section \ref{sec:ramond_sector}, we wrote down vertex operators for spectrally flowed states on AdS$_3$ coupled to a free fermion. We now interpret this free fermion as part of the CFT $\mathcal{N}$. If $\mathcal{N}$ contains 7 free fermions, then the Ramond vacua of the full string theory will contain labels for the vacua of these fermions. Thus, before imposing physical state conditions, Ramond vertex operators for spectrally flowed ground states of the string theory take the form
\begin{equation}\label{eq:Ramond_vo_superspace_full}
\mathscr{V}^{w,(R,\underline{\epsilon})}_{m,j}=e^{(w/Q-Qj)\Phi}\left(\frac{\p^{w}\Gamma}{w!}\right)^{-m-j-\frac{\epsilon_1}{2}}\delta_{w|w}^{\epsilon_1}(\Gamma)S^{\epsilon_1}_H\mathcal{S}^{\epsilon_2}_{\kappa} \mathscr{V}_C^{(R,\epsilon_3,\epsilon_4,\epsilon_5)} \,,
\end{equation}
where we have defined $\underline{\epsilon}=(\epsilon_1,\epsilon_2,\epsilon_3,\epsilon_4,\epsilon_5)$. This has conformal weight
\begin{equation}
\Delta=\frac{j(1-j)}{k}-hw+\frac{kw^2}{4} + \frac{5}{8}\,.
\end{equation}
Then the physical state condition $\Delta = \frac{5}{8}$ (valid in either picture $P=-1/2$ or $P = -3/2$) determines the spacetime conformal dimension
\begin{equation}
    h = \frac{j(1-j)}{kw}+\frac{kw}{4}\,.
\end{equation}
Below, we will see that there are precisely eight physical states of this conformal dimension for each $w\in \mathbb{N}$.

\paragraph{A choice of picture:} As is reviewed in Appendix \ref{app:srs}, there are two canonical choices for the picture number of a Ramond vertex operator in superspace. The most common choice is $P=-1/2$, which associates the vertex operator to the whole Ramond divisor, $\mathscr{F}$, on which the superconformal structure degenerates. If the divisor is located at $z=z_i$, then the vertex operator is labeled only by $z_i$ on the worldsheet and does not depend on a fermionic coordinate. However, for $P=-3/2$, we insert the vertex operator at a specific point $(z_i|\theta_i)$ on the divisor. Integration over the divisor is equivalent to the action of $G_0$, which maps this vertex operator to the $P=-1/2$ case. For RNS superstring calculations on the reduced space, one integrates out all Grassman odd coordinates and moduli and hence, the $P=-1/2$ expressions naturally appear in that context.

For our interests, we wish to express the Ramond vertex operators in terms of superfields that depend on a fermionic coordinate and hence, picture $P=-3/2$ is more natural for us. We claim that $e^{-3\varphi/2}\mathscr{V}_{m,j}^{w,(R,\underline{\epsilon})}$ is the correct form of physical vertex operators. Indeed, for $P=-3/2$, we require that $G_r=0$ for all $r>0$ which is easily verified for $\mathscr{V}_{m,j}^{w,(R,\underline{\epsilon})}$ as defined in \eqref{eq:Ramond_vo_superspace_full}. By contrast, the $P=-1/2$ physical vertex operators require that $G_r=0$ for all $r\geq 0$. Defining
\begin{equation}
V^{w,(R,\underline{\epsilon})}_{m,j}(z_i)=e^{(w/Q-Qj)\phi}\left(\frac{\partial^w\gamma}{w!}\right)^{-m-j-\frac{\epsilon_1}{2}}\delta_{w}(\gamma)e^{i(w+\frac{\epsilon_1}{2})H} S_{\kappa}^{\epsilon_2} V_C^{(R,\epsilon_3,\epsilon_4,\epsilon_5)} (z_i) \,,
\end{equation}
the general form of spectrally flowed physical vertex operators in $P=-1/2$ is\footnote{\label{footnote:cocycle} In this calculation, there is a cocycle factor that multiplies the factor of $i$ in \eqref{eq:phi_lambda_superfields} leading to the coefficient $(-1)^w\epsilon_1\epsilon_2$. Cocycle factors also played an important role in the computations of \cite{Yu:2024kxr}. In particular, see Section 2.3 of \cite{Yu:2024kxr} for a similar expression for the physical vertex operators of type II superstring theory on $\rm{AdS}_3 \times \rm{S}^3 \times \rm{T}^4$.}
\begin{equation}\label{eq:picture_-1/2_physical_vo}
    \begin{aligned}
        G_0 V_{m,j}^{w,(R,\underline{\epsilon})} &= (-1)^w\frac{\epsilon_1\epsilon_2}{\sqrt{2}}\left[ \frac{w}{Q}-Q\left(j-\frac{1}{2}\right)\right]V^{w,(R,\tilde{\underline{\epsilon}})}_{m,j}  \\
        &+\frac{1}{2}\left[ \left(w-\left(m+j-\frac{1}{2}\right)\right)+\epsilon_1\left(w+\left(m+j-\frac{1}{2}\right)\right) \right]V^{w,(R,\hat{\underline{\epsilon}})}_{m,j} \,,
    \end{aligned}
\end{equation}
where $\tilde{\epsilon} = (\epsilon_1,-\epsilon_2,\epsilon_3,\epsilon_4,\epsilon_5)$ and $\hat{\underline{\epsilon}} = (-\epsilon_1,\epsilon_2,\epsilon_3,\epsilon_4,\epsilon_5)$. This is encoded in the $P=-3/2$ expression via
\begin{equation}\label{eq:P=-3/2_expansion}
    \mathscr{V}_{m,j}^{w,(\text{R},\underline{\epsilon})}(z_i,\theta_{i}) 
        = V_{m,j}^{w,(\text{R},\underline{\epsilon})}(z_i) + \theta_{i} G_0 {V}_{m,j}^{w,(\text{R},\underline{\epsilon})}(z_i) \,,
\end{equation}
where $\theta_i$ is defined in the coordinate chart we denoted $(z|\theta_{\text{sv}}^-)$ in Section \ref{sec:ramond_sector}. One can readily verify that this is consistent with the Taylor expansion of the superfields in \eqref{eq:Ramond_vo_superspace_full} and we do this explicitly in Appendix \ref{app:Ramond_sector}. In so doing, we verify that \eqref{eq:Ramond_vo_superspace_full} is indeed the correct expression for physical Ramond-sector vertex operators.

\paragraph{The GSO projection:} There are two choices of GSO projection in the R sector
\begin{equation}
(-1)^F=
\begin{cases}
-1\,,&\text{for GSO}\,,\\
+1\,,&\text{for GSO}'\,.
\end{cases}
\end{equation}
We shall set our conventions such that $(-1)^F$ acting on the spin fields in \eqref{eq:Ramond_vo_superspace_full} gives $(-1)^F=\epsilon_1\epsilon_2\epsilon_3\epsilon_4\epsilon_5$. Then, the GSO projection on the full expression \eqref{eq:Ramond_vo_superspace_full} imposes
\begin{equation}
\begin{split}
(-1)^{w}\epsilon_1\epsilon_2\epsilon_3\epsilon_4\epsilon_5&=-1\,,\quad \text{for GSO}\,,\\
(-1)^{w}\epsilon_1\epsilon_2\epsilon_3\epsilon_4\epsilon_5&=+1\,,\quad \text{for GSO'}\,.
\end{split}
\end{equation}
Thus, in either case, only half of the Ramond states survive the GSO projection.

\paragraph{The BRST cohomology:} Thus far, it appears that we have 16 physical vertex operators of the form \eqref{eq:Ramond_vo_superspace_full} for each $w$ that satisfy the GSO projection. However, we have only checked that these operators are BRST-closed. We show in Appendix \ref{app:Ramond_sector} that only eight of these vertex operators are independent in BRST cohomology. Indeed, we show that $e^{-3\varphi/2}\mathscr{V}_{m,j}^{w,(R,\underline{\epsilon})} (z_i|\theta_i)$ with $\underline{\epsilon}=(\epsilon_1,\epsilon_2,\epsilon_3,\epsilon_4,\epsilon_5)$ is equivalent up to BRST-exact terms to a sum over operators of the same form with $\epsilon_1\mapsto -\epsilon_1$ and a similar sign flip for one of the other fermionic vacua, such that one may set $\epsilon_1 = -$ without loss of generality. This holds for any background of the form $\rm{AdS}_3 \times \mathcal{N}$.

\section{Correlation functions and localization}\label{sec:localization}

We now turn our attention to the definition and computation of the worldsheet path integral. We are particularly interested in the sub-sector of the worldsheet theory in which all dynamics take place near the conformal boundary $\phi\to\infty$. At the level of the path integral, this essentially boils down to dropping the interaction term in the worldsheet action, and instead working with the free field, `near-boundary' action
\begin{equation}
S_{\text{free}}=\frac{1}{2\pi}\int\mathrm{d}^{2|2}\boldsymbol{z}\,\left(\frac{1}{2}D\Phi\bar{D}\Phi-\frac{Q}{4}\mathcal{R}\Phi+\Omega\bar{D}\Gamma+\bar{\Omega}D\bar{\Gamma}\right)\,.
\end{equation}
We are interested in correlators of spectrally-flowed vertex operators evaluated using the near-boundary action. Keeping in mind that we need to include the screening operator $\mathscr{D}$ to define a sensible path integral, such correlators take the form
\begin{equation}\label{eq:general-amplitude}
\mathcal{A}_{g,n}=\sum_{N=0}^{\infty}\frac{p^N}{N!}\int_{\mathfrak{M}_{g,n}}\mathrm{d}\mu\,\Braket{\left(\int\mathscr{D}\bar{\mathscr{D}}\right)^N\prod_{i=1}^{n}\mathscr{V}_{m_i,j_i}^{w_i}(\boldsymbol{z}_i;x_i)\bar{\mathscr{V}}_{\bar{m}_i,j_i}^{w_i}(\bar{\boldsymbol{z}}_i;x_i)}\,.
\end{equation}
Here, $\mathfrak{M}_{g,n}$ is the moduli space of super Riemann surfaces with $n$ NS punctures of complex dimension
\begin{equation}
\dim_{\mathbb{C}}\mathfrak{M}_{g,n}=(n+3g-3)|(n+2g-2)\,.
\end{equation}
and $\mathrm{d}\mu$ is the measure on this moduli space induced from the superconformal ghost system. For the simple case of tree-level ($g=0$) amplitudes, the moduli space is an integral over $n-3$ bosonic coordinates and $n-2$ fermionic coordinates. Concretely, the global $\text{PSL}(2|1,\mathbb{C})$ symmetry on the sphere $\mathbb{CP}^{1|1}$ allows us to fix three bosonic coordinates (say $z_1,z_2,z_3$) and two fermionic coordinates (say $\theta_1,\theta_2$), while the remaining coordinates $z_4,\ldots,z_n$ and $\theta_3,\ldots,\theta_n$ are integrated. The resulting measure on $\mathfrak{M}_{0,n}$ is derived from the superconformal ghost system:
\begin{equation}
\mathrm{d}\mu=\left\langle\delta_{1|1}^{(2)}(C(\boldsymbol{z}_1))\delta_{1|1}^{(2)}(C(\boldsymbol{z}_2))\delta^{(2)}(C(\boldsymbol{z}_{3}))\right\rangle\mathrm{d}^2\theta_3\prod_{i=4}^{n}\mathrm{d}^{2|2}\boldsymbol{z}_i\,.
\end{equation}
Here $c$ and $\hat{\gamma}$ are the usual superconformal ghosts, and we have defined $C = c + \theta\hat{\gamma}$. We use the same delta-function notation that we used to define spectrally-flowed vertex operators in Section \ref{sec:superstrings}, which we define again here for convenience:
\begin{equation}
\delta^{(2)}(C)=\delta(C)\delta(\bar{C})\,,\quad\delta^{(2)}_{1|1}(C)=\delta(C)\delta(DC)\delta(\bar{C})\delta(\bar{D}\bar{C})\,.
\end{equation}
The ghost correlator can be evaluated by expanding in $\theta$ (see Appendix \ref{app:srs}), allowing us to explicitly write down the measure on $\mathfrak{M}_{0,n}$ as 
\begin{equation}\label{eq:m0n-measure}
\mathrm{d}\mu=\left|I_b(\boldsymbol{z}_1,\boldsymbol{z}_3)I_b(\boldsymbol{z}_2,\boldsymbol{z}_3)\right|^2\mathrm{d}^2\theta_3\prod_{i=4}^{n}\mathrm{d}^{2|2}\boldsymbol{z}_i\,.
\end{equation}
The bosonic superspace interval $I_b$ is defined in equation \eqref{eq:intervals}. It is also readily shown that this measure makes the integral \eqref{eq:general-amplitude} invariant under global superconformal ($\text{PSL}(2|1,\mathbb{C})$) transformations.

The goal for the rest of the section will be to evaluate the correlator \eqref{eq:general-amplitude}. In the next subsection we will show that the path integral of the $\Omega,\Gamma$ system as well as the integral over $\mathfrak{M}_{g,n}$ \textit{localizes} to a finite-dimensional super-moduli space. The dimension of this moduli space depends on the precise choices for the $\text{SL}(2,\mathbb{R})$ spins $j$, and in particular the path integral reduces to a discrete sum for a very special choice of the spins. Using this localization property, we will explicitly determine the correlation function \eqref{eq:general-amplitude} up to an in principle calculable Jacobian.

\subsection{Localization of the path integral}\label{sec:localization-argument}

Before we calculate the path integral, it will be instructive to examine it schematically. We take our vertex operators to be those given in \eqref{eq:superspace-ns-vertex} for $w_i$ odd. For a fixed value of $N$ and the locations $\boldsymbol{\lambda}_a$ of the screening operators, the matter part of the path integral factorizes into a correlation function of the scalar $\Phi$ and a correlation function of the $\Omega,\Gamma$ system. It turns out that these correlators are only nonzero provided certain conditions are met. Since we are not fully calculating the path integral, we will keep the genus of the worldsheet arbitrary for this discussion.

For the $\Phi$ correlator, things are simple enough. Due to the shift symmetry\footnote{Of course, this shift symmetry is only approximate in the full $\text{AdS}_3$ spacetime, but becomes exact at the conformal boundary $\Phi\to\infty$.} $\Phi\to\Phi+\Phi_0$, the correlator
\begin{equation}\label{eq:phi-correlator}
\Braket{\prod_{a=1}^{N}e^{-2\Phi/Q}(\boldsymbol{\lambda}_a)\prod_{i=1}^{n}e^{(w_i/Q-Qj_i)\Phi}(\boldsymbol{z}_i)}
\end{equation}
vanishes unless the momentum conservation
\begin{equation}\label{eq:momentum-conservation}
-\frac{2N}{Q}+\sum_{i=1}^{n}\left(\frac{w_i}{Q}-Qj_i\right)=Q(g-1)
\end{equation}
is satisfied. The right-hand-side arises from the background charge of $\Phi$. A consequence of this is that the value of $N$ is actually fixed in terms of the quantum numbers of the matter vertex operators.

Now we turn our attention to the $\Omega,\Gamma$ correlator, which is more subtle. For the sake of notational simplicity we will focus only on the left-moving part of the correlator, which is
\begin{equation}
\left\langle\prod_{a=1}^{N}(\mathop{\text{Res}}\Gamma)^{-k}\delta_{{1|1}}(\Omega)(\boldsymbol{\lambda_a})\prod_{i=1}^{n}\left(\frac{D^{2w_i}\Gamma(\boldsymbol{z}_i)}{w_i!}\right)^{-m_i-j_i}\delta_{w_i|w_i}(\Gamma(\boldsymbol{z}_i)-x_i)\right\rangle\,.
\end{equation}
We begin by writing the delta function $\delta_{{1|1}}(\Omega(\boldsymbol{\lambda}_a))$ in terms of its Fourier transform:
\begin{equation}
\delta_{{1|1}}(\Omega(\boldsymbol{\lambda}_a))=\int\mathrm{d}\mathfrak{c}_a\mathrm{d}\mathfrak{d}_a\,\exp\left(i\mathfrak{c}_aD\Omega(\boldsymbol{\lambda}_a)-i\mathfrak{d}_a\Omega(\boldsymbol{\lambda}_a)\right)\,,
\end{equation}
where $\mathfrak{c}_a$ (resp. $\mathfrak{d}_a$) are commuting (resp. anticommuting) Lagrange multipliers. We can think of these delta functions, then, as modifying the action. That is, multiplying $\prod_{a=1}^{N}\delta_{{1|1}}(\Omega(\boldsymbol{\lambda}_a))$ by $e^{-S[\Omega,\Gamma]}$ gives (for fixed $\mathfrak{c}_a,\mathfrak{d}_a$)
\begin{equation}
\exp\left(-\frac{1}{2\pi}\int\mathrm{d}^{2|2}\boldsymbol{z}\,\Omega\left(\bar{D}\Gamma-2\pi i\sum_{a=1}^{N}\left(\mathfrak{d}_a\delta^{(2)}(\boldsymbol{z},\boldsymbol{\lambda}_a)+\mathfrak{\boldsymbol{c}_a}D\delta^{(2)}(\boldsymbol{z},\boldsymbol{\lambda}_a)\right)\right)\right)\,.
\end{equation}
Thus, we have now removed all $\Omega$ dependence of the correlator by redefining the action. Integrating out $\Omega$ then inserts a functional delta function that imposes the equations of motion
\begin{equation}
\bar{D}\Gamma=2\pi i\sum_{a=1}^{N}\left(\mathfrak{d}_a\delta^{(2)}(\boldsymbol{z},\boldsymbol{\lambda}_a)+\mathfrak{\boldsymbol{c}_a}D\delta^{(2)}(\boldsymbol{z},\boldsymbol{\lambda}_a)\right)\,.
\end{equation}
Away from the points $\mathfrak{\lambda}_a$, this is the condition that $\Gamma$ is a holomorphic chiral superfield. Near $\boldsymbol{\lambda}_a$, however, this condition demands that $\Gamma$ has a pole of the form
\begin{equation}\label{eq:gamma-simple-pole}
\Gamma\sim\frac{\mathfrak{c}_a}{z-\lambda_a-\theta\eta_a}+\frac{\mathfrak{d}_a(\theta-\eta_a)}{z-\lambda_a-\theta\eta_a}+\cdots\,.
\end{equation}
That is, the delta functions $\delta_2(\Omega)$ in the path integral impose that $\Gamma$ is a holomorphic function on superspace with a prescribed set of $N$ poles at the locations of the screening operators $\mathscr{D}$, and the residues at these poles are determined by the Lagrange multipliers $\mathfrak{c}_a,\mathfrak{d}_a$.

The result of integrating out $\Omega$ is, thus, that the $\Gamma,\Omega$ path integral can be written schematically in the form
\begin{equation}
\int_{\mathfrak{F}_N(\boldsymbol{\lambda}_a)}\prod_{a=1}^{N}(\mathfrak{c}_a)^{-k}\prod_{i=1}^{n}\left(\frac{D^{2w_i}\Gamma(\boldsymbol{z}_i)}{w_i!}\right)^{-m_i-j_i}\delta_{w_i|w_i}(\Gamma(\boldsymbol{z}_i)-x_i)\,,
\end{equation}
where $\mathfrak{F}_N(\boldsymbol{\lambda}_a)$ is the space of chiral superfields which have poles of the form \eqref{eq:gamma-simple-pole} at the specified points $\boldsymbol{\lambda}_a$. On the sphere, assuming that none of the poles are at the point at infinity, such a function always admits an expansion
\begin{equation}\label{eq:gamma-g0-expansion}
\Gamma(\boldsymbol{z})=\mathfrak{a}+\sum_{a=1}^{N}\frac{\mathfrak{c}_a}{z-\lambda_a-\theta\eta_a}+\sum_{a=1}^{N}\frac{\mathfrak{d}_a(\theta-\eta_a)}{z-\lambda_a-\theta\eta_a}
\end{equation}
for some commuting constant $\mathfrak{a}$. Since there are $N+1$ commuting degrees of freedom ($\mathfrak{a}$, $\mathfrak{c}_a$) and $N$ anticommuting ones ($\mathfrak{d}_a$), we are integrating over a moduli space of dimension
\begin{equation}
\text{dim}\,\mathfrak{F}_N(\boldsymbol{\lambda}_a)=(N+1)|N\,,\quad g=0\,.
\end{equation}
When the worldsheet has a higher genus, the counting is slightly modified to\footnote{This can be worked out by expanding $\Gamma=\gamma+\theta\psi$ and setting $\eta_a=0$. The condition \eqref{eq:gamma-simple-pole} is equivalent to $\gamma$ having $N$ simple poles and $\psi$ having $N$ simple poles at $z=\lambda_a$. The Riemann Roch formula then tells us that the number of such functions $\gamma$ is $N+g-1$. For $\psi$, we need to take into account the fact that it takes values in the spin bundle $K^{1/2}$, and the Riemann Roch formula then gives $N$ degrees of freedom for $\psi$.}
\begin{equation}
\text{dim}\,\mathfrak{F}_N(\boldsymbol{\lambda}_a)=(N+1-g)|N\,.
\end{equation}

Since it will be useful to us when calculating correlation functions, we also note that we can parametrize $\mathfrak{F}_N(\boldsymbol{\lambda}_a)$ in a slightly different fashion for $g=0$. In particular, we consider the derivative $\partial\Gamma$. Since this is a meromorphic one-form on $\mathbb{CP}^{1|1}$ with double poles at $\boldsymbol{z}=\boldsymbol{\lambda}_a$, we can express it as a rational function:
\begin{equation}\label{eq:gamma-general-rational}
\partial\Gamma(\boldsymbol{z})=\mathfrak{C}\,\frac{\prod_{a=1}^{2N-2}I_b(\boldsymbol{z},\boldsymbol{\mu_a})}{\prod_{a=1}^{N}I_b(\boldsymbol{z},\boldsymbol{\lambda}_a)^2}\,.
\end{equation}
Written in this form, for fixed $\boldsymbol{\lambda_a}$, $\Gamma$ seemingly has $2N|(2N-1)$ degrees of freedom: the zeroes $\boldsymbol{\mu}_a$, the prefactor $\mathfrak{C}$, and two constants of integration (corresponding to the $\mathcal{O}(1)$ and $\mathcal{O}(\theta)$ terms in the Taylor expansion around $\boldsymbol{z}=0$). However, since $\partial\Gamma$ is a total derivative of a single-valued function, its local expansion near a pole must take the form
\begin{equation}
\partial\Gamma(\boldsymbol{z})\sim-\frac{\mathfrak{c}_a}{(z-\lambda_a-\theta\eta_a)^2}-\frac{\mathfrak{d}_a(\theta-\eta_a)}{(z-\lambda_a-\theta\eta_a)^2}+\text{regular}\,.
\label{eq:pole expansion of Gamma}
\end{equation}
In particular, $\partial\Gamma$ can not have poles like $1/z$ or $\theta/z$. This places $N|N$ constraints on the zeroes $\boldsymbol{\mu}_a$ of $\partial\Gamma$. However, these constraints are not independent, since the sums of the residues of \textit{any} meromorphic one-form automatically vanishes. Thus, we are left with $(N+1)|N$ degrees of freedom, as advertised.

\paragraph{Integrating over the poles:} So far, we have considered the path integral for fixed $N$ and $\boldsymbol{\lambda}_a$. In the end, we will integrate over the $\boldsymbol{\lambda}_a$, after computing the correlator of the scalar $\Phi$. Schematically, then, for fixed $N$ we must compute the integral
\begin{equation}
\int_{\mathfrak{F}_N}\prod_{a=1}^{N}(\mathfrak{c}_a)^{-k}\prod_{i=1}^{n}\left(\frac{D^{2w_i}\Gamma(\boldsymbol{z}_i)}{w_i!}\right)^{-m_i-j_i}\delta_{w_i|w_i}(\Gamma(\boldsymbol{z}_i)-x_i)\braket{\cdots}_{\Phi}\,,
\end{equation}
where $\braket{\cdots}_{\Phi}$ is the correlator in \eqref{eq:phi-correlator} and $\mathfrak{F}_N$ is the space obtained from $\mathfrak{F}_N(\boldsymbol{\lambda}_a)$ by allowing the points $\boldsymbol{\lambda}_a$ to vary. This is an integral of dimension
\begin{equation}
\text{dim}\,\mathfrak{F}_N=(2N+1-g)|2N\,.
\end{equation}

\paragraph{Integrating over the moduli space:} So far we have written down a correlation function in a worldsheet CFT. In order to construct the string amplitude, we must integrate over the moduli space $\mathfrak{M}_{g,n}$ of super Riemann surfaces. Note that the space $\mathfrak{F}_N$ depends on the moduli of the worldsheet (since the worldsheet moduli define the superconformal structure $\overline{D}$ which defines the notion of a chiral superfield), and so we must first integrate over $\mathfrak{F}_N$ before integrating over $\mathfrak{M}_{g,n}$. We can combine these integrals into a single integral over a new moduli space, which we call $\mathfrak{M}_{g,n}(\mathbb{CP}^{1},N)$ (for reasons we will explain below). This space has the structure of a vector bundle
\begin{equation}
\begin{tikzcd}
\mathfrak{F}_N \arrow[r] & \mathfrak{M}_{g,n}(\mathbb{CP}^{1},N) \arrow[d]\\
 & \mathfrak{M}_{g,n}
\end{tikzcd}
\end{equation}
and has dimension
\begin{equation}\label{eq:mgncp1n-dimension}
\begin{split}
\text{dim}\,\mathfrak{M}_{g,n}(\mathbb{CP}^{1},N)&=\text{dim}\,\mathfrak{M}_{g,n}+\text{dim}\,\mathfrak{F}_N\\
&=(2N+n+2g-2)|(2N+n+2g-2)\,.
\end{split}
\end{equation}

\paragraph{Including the delta functions:} The upshot of the above discussion is that the path integral over fields $\Gamma$ and the worldsheet moduli localizes to a space of finite dimension, and we are left with the path integral
\begin{equation}
\int_{\mathfrak{M}_{g,n}(\mathbb{CP}^{1},N)}\prod_{a=1}^{N}(\mathfrak{c}_a)^{-k}\prod_{i=1}^{n}\left(\frac{D^{2w_i}\Gamma(\boldsymbol{z}_i)}{w_i!}\right)^{-m_i-j_i}\delta_{w_i|w_i}(\Gamma(\boldsymbol{z}_i)-x_i)\braket{\cdots}_{\Phi}\,.
\end{equation}
The obvious next step is to use the delta functions $\delta_{w_i|w_i}(\Gamma-x_i)$ to perform some of the remaining integrals.

The effect of the delta function $\delta_{w_i|w_i}(\Gamma-x_i)$ is to impose a local constraint on the field $\Gamma$, namely that its Taylor expansion around $\boldsymbol{z}_i$ takes the form
\begin{equation}
\Gamma(\boldsymbol{z})\sim x_i+\mathfrak{a}_i(z-z_i-\theta\theta_i)^{w_i}+\cdots\,,
\label{eq:Gamma expansion}
\end{equation}
where $\mathfrak{a}_i$ is some commuting constant. In terms of the components, this is equivalent to demanding
\begin{equation}\label{eq:ai-definition}
\gamma(z)\sim x_i+\mathfrak{a}_i(z-z_i)^{w_i}+\cdots\,,\quad\psi(z)\sim-w_i\mathfrak{a}_i\theta_i(z-z_i)^{w_i-1}+\cdots\,.
\end{equation}
This imposes $w_i$ commuting constraints on the Taylor series coefficients of $\gamma$ and $w_i$ anticommuting constraints on the coefficients of $\psi$. Imposing these constraints at each point $\boldsymbol{z}_i$ imposes $\sum_{i}w_i|w_i$ constraints, and thus localizes the worldsheet path integral even further to a space of dimension
\begin{equation}
\text{dim}\,\mathfrak{M}_{g,n}(\mathbb{CP}^1,N)-\sum_{i=1}^{n}w_i|w_i=m|m\,,
\end{equation}
where
\begin{equation}\label{eq:number-of-extra-branch-points}
m=2N+2g-2-\sum_{i=1}^{n}(w_i-1)\,.
\end{equation}
The meaning of the number $m$ can be understood as follows. At genus $g=0$, the derivative $\partial\Gamma$ generically takes the form \eqref{eq:gamma-general-rational}. The delta functions essentially tell us that the $2N-2$ zeroes $\boldsymbol{\mu}_a$ are not all distinct: the points $\boldsymbol{z}_i$ are zeroes of multiplicity $w_i-1$. Thus,
\begin{equation}\label{eq:gamma-rational-localized}
\partial\Gamma(\boldsymbol{z})=\mathfrak{C}\,\frac{\prod_{i=1}^{n}(z-z_i-\theta\theta_i)^{w_i-1}}{\prod_{a=1}^{N}(z-\lambda_a-\theta\eta_a)^2}\prod_{\ell=1}^{m}(z-\zeta_\ell-\theta\rho_{\ell})\,,
\end{equation}
where $\boldsymbol{\zeta}_\ell=(\zeta_\ell,\rho_\ell)$ are the remaining, undetermined, zeroes of $\partial\Gamma$. These are points where
\begin{equation}
D^2\Gamma(\boldsymbol{\zeta}_\ell)=D^3\Gamma(\boldsymbol{\zeta}_\ell)=0\,.
\end{equation}
It turns out that this interpretation of the extra $m|m$ degrees of freedom works even at higher-genus, i.e. the remaining degrees of freedom are the extra zeroes of the one-form $\partial\Gamma$ that are not determined by the delta functions. Near these points, $\Gamma$ has the local behavior
\begin{equation}
\Gamma(\boldsymbol{z})=\Gamma(\boldsymbol{\zeta}_{\ell})+(\theta-\rho_{\ell})D\Gamma(\boldsymbol{\zeta}_\ell)+\mathfrak{b}_{\ell}(z-\zeta_\ell-\theta\rho_{\ell})^2+\cdots
\end{equation}
for some commuting constants $\mathfrak{b}_{\ell}$. The values $\Gamma(\boldsymbol{\zeta}_\ell)$ and $D\Gamma(\boldsymbol{\zeta}_{\ell})$ are precisely the $(m|m)$-dimensional set of unspecified degrees of freedom.

Recalling that the momentum-conserving delta function determines the value of $N$ in terms of the $\text{SL}(2,\mathbb{R})$ spins $j_i$, we can use equation \eqref{eq:momentum-conservation} to fully determine the dimension $m|m$ of the path integral. The answer turns out to depend only on the spins $j_i$ and not on the spectral flows $w_i$:
\begin{equation}\label{eq:general-j-constraint}
m=-Q^2\left(\sum_{i=1}^{n}j_i-\frac{k}{2}(n+2g-2)+(g-1)\right)\,.
\end{equation}

To recap, the worldsheet path integral in the $\Phi\to\infty$ limit vanishes unless $m$ is a non-negative integer, in which case it \textit{localizes} to an $m|m$-dimensional integral. The localization of the worldsheet path integral is completely analogous to the one found for the bosonic $\text{SL}(2,\mathbb{R})$ WZW model, originally discovered in \cite{Eberhardt:2019ywk} and explained from a path integral perspective in \cite{Knighton:2023mhq,Knighton:2024qxd}. A particularly simple situation is when the spins are chosen such that
\begin{equation}\label{eq:m=0-j-constraint}
\sum_{i=1}^{n}j_i=\frac{k}{2}(n+2g-2)-(g-1)\,,
\end{equation}
in which case $m=0$ and the path integral reduces to a \textit{sum} over a finite set of worldsheets and fields $\Gamma$. This happens in particular when $k=1$ and $j_i=1/2$ (the situation relevant for `tensionless' string theory on $\text{AdS}_3\times\text{S}^3\times\mathbb{T}^4$ \cite{Eberhardt:2018ouy}) or more generally when $k=1$ and $j_i=1/2+ip_i$ with $\sum p_i=0$ (the situation relevant for `tensionless' strings on $\text{AdS}_3\times\text{S}^3\times\text{S}^3\times\text{S}^1$ \cite{Gaberdiel:2024dva,Eberhardt:2025sbi}).

\subsection{What are we integrating over (physically)?} 

Let us briefly take stock of what we have done so far. Starting with a worldsheet action for superstrings propagating in $\text{AdS}_3$, we have shown that, focusing entirely on worldsheet configurations that live sufficiently close to the conformal boundary ($\Phi\to\infty$), the Polyakov path integral localizes to a finite-dimensional space, and that the answer is nonzero provided that the external spins $j_i$ are chosen so that the right-hand-side of equation \eqref{eq:general-j-constraint} is a nonnegative integer, in which case the path integral is delta-function divergent due to the integration over the zero mode of $\Phi$.

Physically, the string configurations that we are integrating over are precisely those for which the string action remains finite at large radius. Semiclassically, such worldsheet configurations exist because two forces act on a long string -- the string tension and the electrical coupling to the Kalb-Ramond $B$-field. These forces are individually divergent in the $\Phi\to\infty$ but precisely cancel when the string winds the asymptotic boundary $w>0$ times, with the orientation determined by the sign of $B$. At the level of the path integral, such configurations correspond to those for which the worldsheet wraps the boundary through a smooth, orientation-preserving map. These worldsheet configurations are the supersymmetric analogues of the `worldsheet instantons' of \cite{Maldacena:2001km}.

A given worldsheet instanton is determined by two bosonic degrees of freedom -- the shape with which it winds the boundary (the field $\Gamma$) and its radial profile (the field $\Phi$). Near the conformal boundary, $\Phi$ looks like a free scalar with some background charge, and in particular the vertex operators $\mathscr{V}^{w}_{m,j}$ look like momentum wavepackets which scatter off of each other from the asymptotic boundary. The approximate translation symmetry $\Phi\to\Phi+\Phi_0$ near the asymptotic boundary gives rise to the divergences in the path integral when the `momenta' $j_i$ satisfy \eqref{eq:general-j-constraint}.

The remaining degrees of freedom are given by the field $\Gamma$. The instanton constraint -- the constraint that the action is finite as $\Phi\to\infty$ -- demands $\bar{D}\Gamma=0$. For a given winding number $N$ of the worldsheet around the boundary, the space of solutions to this equation is precisely the space $\mathfrak{F}_N$. Including the integral over the worldsheet moduli, the physical moduli space of worldsheet instantons is then $\mathfrak{M}_{g,n}(\mathbb{CP}^1,N)$, which describes the space of possible shapes of a long-superstring worldsheet winding the asymptotic boundary $N$ times.

\subsection{What are we integrating over (mathematically)?}\label{sec:super-holomprphic-curves}

The moduli space $\mathfrak{M}_{g,n}(\mathbb{CP}^1,N)$ was defined as a fiber bundle over the moduli space of super Riemann surfaces $\mathfrak{M}_{g,n}$ whose fiber, $\mathfrak{F}_N$, is the space of chiral superfields $\Gamma$ with $N$ poles of the form \eqref{eq:gamma-simple-pole}. It turns out that this moduli space has a very natural geometric interpretation, and has been studied in the mathematical literature under the name of the `moduli space of super-holomorphic curves' \cite{Kessler:2019wpz}. As an aside from our main discussion, let us briefly comment on what this object is. Readers interested only in the calculation of correlation functions should feel free to skip directly to Section \ref{sec:evaluation}.

Let $X$ be an ordinary complex manifold and $(\Sigma,\mathcal{D})$ a super Riemann surface with a superconformal structure $\mathcal{D}\subset T\Sigma$ (see Appendix \ref{app:srs}). A \textit{super-holomorphic curve} $\Phi:\Sigma\to X$ is a differentiable map such that
\begin{equation}
\overline{D}_J\Phi\equiv\frac{1}{2}(\mathrm{d}\Phi+J\mathrm{d}\Phi I)\Big|_{\mathcal{D}}=0\,.
\end{equation}
Here, $I$ is the complex structure on $\Sigma$, $J$ is the complex structure on $X$, and $\mathrm{d}\Phi:T\Sigma\to TX$ is the natural map sending vector fields on $\Sigma$ to vector fields on $X$.\footnote{The original literature uses the term `super $J$-holomorphic curve' to emphasize the dependence on the (almost) complex structure on the target. Since there is never any ambiguity in which complex structure we're referring to, we will simply use the term `super-holomorphic curve'.}

To understand this condition, it is useful to write everything out in components. If $X$ has complex coordinates $(x^{i},x^{\bar\imath})$, and $\Sigma$ has standard complex coordinates $(z,\bar{z}|\theta,\bar{\theta})$, then the components of $\Phi$ can be expanded in superspace
\begin{equation}
\begin{split}
\Phi^i&=x^i+\theta\psi^i+\bar\theta\bar\psi^{i}+\theta\bar\theta F^{i}\,,\\
\Phi^{\bar\imath}&=x^{\bar\imath}+\theta\psi^{\bar\imath}+\bar\theta\bar\psi^{\bar\imath}+\theta\bar\theta F^{\bar\imath}\,,
\end{split}
\end{equation}
where $x$ is an ordinary map from $\Sigma_{\text{red}}$ to $X$, and
\begin{equation}
\begin{gathered}
\psi\in\Gamma\left(\Pi K^{1/2}\otimes x^*\text{T}X\right)\,,\quad\bar\psi\in\Gamma\left(\Pi\bar{K}^{1/2}\otimes x^*\text{T}X\right)\\
F\in\Gamma\left(K^{1/2}\otimes\bar{K}^{1/2}\otimes x^*\text{T}X\right)\,,
\end{gathered}
\end{equation}
where $K$ is the canonical bundle on $\Sigma_{\text{red}}$ and $\Pi$ denote parity reversal (i.e. sections of $\Pi K^{1/2}$ are anticommuting worldsheet fields of weight $(1/2,0)$). Using the standard superconformal structure $D=\partial_{\theta}+\theta\partial_z$, the condition $\overline{D}_J\Phi=0$ can be written as (see Corollary 2.5.3 of \cite{Kessler:2019wpz})
\begin{equation}
F^i=F^{\bar\imath}=\psi^{\bar\imath}=\bar\psi^{i}=\overline{\partial}x^i=\overline{\partial}\psi^i=\partial x^{\bar\imath}=\partial\bar\psi^{\bar\imath}=0\,,
\end{equation}
i.e. $\Phi^i$ is a chiral superfield and $\Phi^{\bar\imath}$ is an antichiral superfield.

The moduli space $\mathfrak{M}_{g,n}(X,\beta)$ of super-holomorphic curves into $X$ is specified by the data of all pairs of marked super-Riemann surfaces $(\Sigma,\mathcal{D};\boldsymbol{z}_1,\ldots,\boldsymbol{z}_n)$ and super-holomorphic curves $\Phi:\Sigma\to X$ such that the image of the bosonic body $\Sigma_{\text{red}}$ is in the homology class $\beta\in\text{H}_2(X,\mathbb{Z})$. For $n+2g-2>0$, this moduli space is a complex supermanifold\footnote{Since some points in $\mathfrak{M}_{g,n}(X,\beta)$ have automorphisms, it is actually a super-orbifold, or, more accurately, a complex superstack.} of dimension\footnote{The (virtual) dimension can be calculated as follows: the even directions in the tangent space $T_{(\Sigma,\Phi)}\mathfrak{M}_{g,n}(X,\beta)$ are spanned by $\delta x^i\in\text{H}^0(\Sigma,x^*TX)$ and the $n+3g-3$ even Beltrami differentials, while the odd directions are spanned by $\delta\psi^i\in\text{H}^0(\Sigma,x^*TX\otimes K^{1/2})$ and the $n+2g-2$ odd Beltrami differentials (i.e. gravitino zero modes). The dimension is then found by calculating the dimensions of these cohomology groups using the Hirzebruch-Riemann-Roch theorem \cite{Kessler:2019wpz}.}
\begin{equation}
\text{dim}\,\mathfrak{M}_{g,n}(X,\beta)=\left(n+(3-d)(g-1)+\int_{\beta}c_1(TX)\right)\bigg|\left(n+2g-2+\int_{\beta}c_1(TX)\right)\,,
\end{equation}
where $d$ is the complex dimension of $X$. The reduced space of $\mathfrak{M}_{g,n}(X,\beta)$ is found by setting all odd moduli (i.e. the odd moduli in $\mathfrak{M}_{g,n}$ and the odd components $\psi^i,\bar\psi^{\bar\imath}$ of $\Phi$) to zero, which yields the isomorphism
\begin{equation}
\mathfrak{M}_{g,n}(X,\beta)_{\text{red}}\cong\mathcal{SM}_{g,n}(X,\beta)\,,
\end{equation}
where $\mathcal{SM}_{g,n}(X,\beta)$ is the moduli space of holomorphic maps from ordinary Riemann surfaces with spin structure to $X$.

In the context of $\text{AdS}_3$ string theory, the super Riemann surface $\Sigma$ is the worldsheet and the ordinary complex manifold $X$ is the $\text{AdS}_3$ boundary, in our case the 2-sphere: $X\cong\mathbb{CP}^1$. Since $\text{H}_2(X,\mathbb{Z})\cong\mathbb{Z}$, the homology class $\beta$ is simply an integer $N$ -- the number of times $\Sigma_{\text{red}}$ `wraps' $X$. Holomorphic maps must preserve orientation, and so $N>0$. Using the fact that the total curvature on the 2-sphere is $\int_{\mathbb{CP}^1}c_1(T\mathbb{CP}^1)=2$, we can readily compute the dimension of the moduli space of super-holomorphic curves into $\mathbb{CP}^1$ of degree $N$, and we find
\begin{equation}
\text{dim}\,\mathfrak{M}_{g,n}(\mathbb{CP}^1,N)=(n+2g-2+2N)|(n+2g-2+2N)\,,
\end{equation}
which matches precisely with the counting in \eqref{eq:mgncp1n-dimension}.

To put everything into the language of the fundamental fields of the $\text{AdS}_3$ path integral, let $x,\bar{x}$ be the complex coordinates on $X$. Then the components $\Phi^x$ and $\Phi^{\bar{x}}$ are nothing more than our worldsheet fields:
\begin{equation}
\Phi^x=\Gamma=\gamma+\theta\psi\,,\quad\Phi^{\bar{x}}=\bar{\Gamma}=\bar{\gamma}+\bar\theta\bar\psi\,.
\end{equation}
The statement that $\Phi$ is a super-holomorphic map from $\Sigma$ to $\mathbb{CP}^1$ of degree $N$ is nothing more than the statement that $\Gamma$ is a meromorphic function on $\Sigma$ with $N$ poles. Indeed, such a pole corresponds to $\Gamma$ mapping a point $\boldsymbol{\lambda}$ to the point at infinity on $\mathbb{CP}^1$. Consider a new coordinate patch $\tilde{\Gamma}=1/\Gamma$ on $\mathbb{CP}^1$. The statement that $\Phi$ maps $\boldsymbol{\lambda}=(\lambda,\eta)$ to the north pole means that $\tilde{\Gamma}$ has a zero at $\boldsymbol{\lambda}$. Generically, this will be a simple zero, and so
\begin{equation}
\tilde{\Gamma}(\boldsymbol{z})\sim\mathcal{O}(z-\lambda-\theta\eta)
\end{equation}
as $\boldsymbol{z}\to\boldsymbol{\lambda}$.
Expressed in terms of $\Gamma$, this reads
\begin{equation}
\Gamma(\boldsymbol{z})\sim\mathcal{O}\left(\frac{1}{z-\lambda-\theta\eta}\right)\,,
\end{equation}
which is exactly the condition provided by the insertion of the screening operator $\mathscr{D}$ (see the discussion around \eqref{eq:gamma-simple-pole}).

\subsection{Evaluating the path integral}\label{sec:evaluation}

Now that we have established the localization of the worldsheet path integral near the conformal boundary of $\text{AdS}_3$, we are finally in a position to calculate correlation functions in this regime. For the sake of simplicity we will restrict ourselves to worldsheets of genus zero, leaving the study of higher-genus worldsheets to future work. This simplifies our task dramatically not only because CFT correlators on the sphere are simpler to work with than their higher-genus cousins, but also because the moduli space $\mathfrak{M}_{0,n}(\mathbb{CP}^1,N)$ is actually a product:\footnote{We are being purposefully cavalier about the boundary of the moduli space, which will not play a particularly important role in what follows.}
\begin{equation}
\mathfrak{M}_{0,n}(\mathbb{CP}^1,N)\cong\mathfrak{M}_{0,n}\times\mathfrak{F}_N\,.
\end{equation}
This is due to the Riemann sphere having no moduli other than the locations of the marked points, and the definition of $\mathfrak{F}_N$ does not depend on this data. The practical upshot is that for $g=0$ the path integral measure will be a simple product of the measure on $\mathfrak{M}_{0,n}$ (defined by the superconformal ghost system) and the measure on $\mathfrak{F}_N$ (which we will construct shortly). With this in mind, the computation of correlation functions will be broken up into four simple steps:
\begin{enumerate}

    \item The computation of the correlator $\braket{\cdots}_{\Phi}$.

    \item The construction of a natural measure on $\mathfrak{F}_N$.

    \item Integrating against the delta functions $\delta_{w_i|w_i}(\Gamma-x_i)$.

    \item Finding an appropriate set of coordinates for the remaining $m|m$-dimensional integral and computing the associated Jacobian.

\end{enumerate}
The end result will be an explicit expression for the worldsheet correlation functions in the $\Phi\to\infty$ limit.

\paragraph{\boldmath The $\Phi$ correlator:} The easiest piece to compute is the correlator of the scalar $\Phi$, which is readily done using the superspace analogue of Wick's theorem. The result is
\begin{equation}
\begin{split}
\Braket{\cdots}_{\Phi}&=\bigg|\prod_{i<j}I_b(\boldsymbol{z}_i,\boldsymbol{z}_j)^{-(w_i/Q-Qj_i)(w_j/Q-Qj_j)}\prod_{i,a}I_b(\boldsymbol{z}_i,\boldsymbol{\lambda}_a)^{2w_i/Q^2-2j_i}\\
&\hspace{2cm}\times\prod_{a<b}I_b(\boldsymbol{\lambda}_a,\boldsymbol{\lambda}_b)^{-4/Q^2}\bigg|^2\delta\left(\frac{2N}{Q}+\sum_{i=1}^{n}\left(Qj_i-\frac{w_i}{Q}\right)-Q\right)\,,
\end{split}
\end{equation}
where the bosonic interval $I_b$ is defined in equation \eqref{eq:intervals}.

\paragraph{\boldmath The measure on $\mathfrak{F}_N$:} The measure on $\mathfrak{F}_N$ is found by first constructing a measure on $\mathfrak{F}_N(\boldsymbol{\lambda}_a)$ and then simply integrating over the locations of the poles $\boldsymbol{\lambda}_a$. As mentioned in the discussion around equation \eqref{eq:gamma-g0-expansion}, the space $\mathfrak{F}_N(\boldsymbol{\lambda}_a)$ has $N+1$ natural commuting coordinates and $N$ natural anticommuting coordinates given by the coefficients in the general expansion of a chiral superfield with $N$ poles. The most simple measure that one can write down is
\begin{equation}\label{eq:fn-measure}
\mathrm{d}^{(2N+2|2N)}\Gamma=\mathrm{d}^2\mathfrak{a}\prod_{a=1}^{N}\mathrm{d}^2\mathfrak{c}_a\mathrm{d}^2\mathfrak{d}_a\,.
\end{equation}
While we do not have a first principles derivation of this fact, it was shown in \cite{Knighton:2024qxd} that the analogous construction in bosonic string theory leads to a path integral measure that correctly reproduces dual CFT correlation functions. Following this line of reasoning we will take \eqref{eq:fn-measure} as the measure with which we integrate over $\mathfrak{F}_N(\boldsymbol{\lambda}_a)$. With this in mind, we can write the $\Omega,\Gamma$ path integral for a fixed value of $N$ as
\begin{equation}
\int\mathrm{d}^{(2N+2|2N)}\Gamma\left|\prod_{a=1}^{N}(\mathfrak{c}_a)^{-k}\prod_{i=1}^{n}\left(\frac{D^{2w_i}\Gamma(\boldsymbol{z}_i)}{w_i!}\right)^{-m_i-j_i}\right|^2\prod_{i=1}^{n}\delta^{(2)}_{w_i|w_i}(\Gamma-x_i)\,.
\end{equation}

\paragraph{The Jacobian:} Our final task is to put them together and integrate over the locations of the poles $\boldsymbol{\lambda}_a$ and the moduli space $\mathfrak{M}_{0,n}$. The expression we need to evaluate is (at fixed $N$)
\begin{equation}
\begin{split}
\int_{\mathfrak{M}_{0,n}}\mathrm{d\mu}\int\mathrm{d}^{2N|2N}\boldsymbol{\lambda}\int\mathrm{d}^{(2N+2|2N)}\Gamma\left|\prod_{a=1}^{N}(\mathfrak{c}_a^{\Gamma})^{-k}\prod_{i=1}^{n}(\mathfrak{a}_i^{\Gamma})^{-m_i-j_i}\right|^2\prod_{i=1}^{n}\delta^{(2)}_{w_i|w_i}(\Gamma-x_i)\Braket{\cdots}_{\Phi}&\,,
\end{split}
\end{equation}
where $\mathfrak{a}_i$ is defined in equation \eqref{eq:ai-definition}. As we discussed in Section \ref{sec:localization-argument}, we can use the delta functions $\delta^{(2)}_{w_i|w_i}(\Gamma-x_i)$ to reduce the total number of integrals to perform. Of course, this process induces a Jacobian, which we will describe now.

In the current basis, the total path integral measure is given by
\begin{equation}
|I_b(\boldsymbol{z}_1,\boldsymbol{z}_2)I_b(\boldsymbol{z}_2,\boldsymbol{z}_3)|^2\mathrm{d}^2\theta_3\prod_{i=4}^{n}\mathrm{d}^{2|2}\boldsymbol{z}_i\,\mathrm{d}^2\mathfrak{a}\prod_{a=1}^{N}\mathrm{d}^{2|2}\boldsymbol{\lambda}_a\mathrm{d}^{2}\mathfrak{c}_a\mathrm{d}^2\mathfrak{d}_a\,.
\end{equation}
However, the natural variables of the delta functions are the Taylor series coefficients
\begin{equation}
\{D^{r}\Gamma(\boldsymbol{z}_i):r=0,\ldots,2w_i-1\,,\,\,\,i=1,\ldots,n\}\,.
\end{equation}
After integrating over the delta functions (which fixes these coefficients to vanish), we are left with an integral over $m|m$ parameters, for which we also need to specify an appropriate set of coordinates. We argued in Section \ref{sec:localization-argument} that there is in fact a natural set of such parameters, namely the points on the worldsheet which solve the equations
\begin{equation}
D^2\Gamma(\boldsymbol{\zeta}_\ell)=D^3\Gamma(\boldsymbol{\zeta}_\ell)=0\,.
\end{equation}
and which are distinct from the points $\boldsymbol{z}_i$. Once these points are identified, we can use $\Gamma(\boldsymbol{\zeta}_\ell)$ and $D\Gamma(\boldsymbol{\zeta}_{\ell})$ as coordinates for the remaining $m|m$-dimensional integrals.

In practical terms, we therefore have to compute the Jacobian of the change of variables
\begin{equation}\label{eq:change-of-variables}
\left(z_4,\ldots,z_n,\mathfrak{a},\lambda_a,\mathfrak{c}_a|\theta_3,\ldots,\theta_n,\eta_a,\mathfrak{d}_a\right)\longrightarrow\left(D^{\ell}\Gamma(\boldsymbol{z}_i),\Gamma(\boldsymbol{\zeta}_\ell),D\Gamma(\boldsymbol{\zeta}_\ell)\right)\,,
\end{equation}
which will allow us to write the path integral measure as
\begin{equation}
\mathrm{d}\mu\prod_{a=1}^{N}\mathrm{d}^{(2|2)}\boldsymbol{\lambda}_a\,\mathrm{d}^{(2N+2|2N)}\Gamma=\mathcal{J}\prod_{i=1}^{n}\prod_{r=0}^{2w_i-1}\mathrm{d}^2(D^{r}\Gamma(\boldsymbol{z}_i))\prod_{\ell=1}^{m}\mathrm{d}^2\Gamma(\boldsymbol{\zeta}_\ell)\mathrm{d}^2(D\Gamma(\boldsymbol{\zeta}_\ell))\,,
\end{equation}
which can be easily integrated against the delta function. At this point, we do not have a first-principles derivation of this Jacobian, but have computed it for a few examples, which always returns the tantalizingly simple result
\begin{equation}\label{eq:jacobian-conjecture}
\mathcal{J}=|\mathfrak{C}|^2\,,
\end{equation}
where $\mathfrak{C}$ is the coefficient appearing in the decomposition of $\partial\Gamma$ in equation \eqref{eq:gamma-rational-localized}. While we suspect that it should be possible to compute $\mathcal{J}$ directly for generic values of $N$ and $w_i$, we will for the moment state \eqref{eq:jacobian-conjecture} as a conjecture.\footnote{One could in principle compute $\mathcal{J}$ by inductive row-reduction to compute the Berezinian of the matrix defining the change of variables \eqref{eq:change-of-variables}. This has been done in the bosonic case, see Appendix A of \cite{Knighton:2024qxd}.}

\paragraph{The full correlator:} With the $\Phi$ correlator and Jacobian in hand, we can now write down the final answer for the correlation function. First, however, it will be convenient to write the $\Phi$ correlator in terms of data of the field $\Gamma$. Using the parametrization \eqref{eq:gamma-rational-localized} of $\partial\Gamma$, we can write
\begin{equation}
\begin{split}
w_i\mathfrak{a}_i&=\mathfrak{C}\frac{\prod_{j\neq i}I_b(\boldsymbol{z}_i,\boldsymbol{z}_j)^{w_j-1}\prod_{\ell=1}^{m}I_b(\boldsymbol{z}_i,\boldsymbol{\zeta}_\ell)}{\prod_{a=1}^{N}I_b(\boldsymbol{z}_i,\boldsymbol{\lambda_a})^2}\,,\\
2\mathfrak{b}_{\ell}&=\mathfrak{C}\frac{\prod_{i=1}^{n}I_b(\boldsymbol{\zeta}_{\ell},\boldsymbol{z}_i)^{w_i-1}\prod_{m\neq\ell}I_b(\boldsymbol{\zeta}_\ell,\boldsymbol{\zeta}_m)}{\prod_{a=1}^{N}I_b(\boldsymbol{\zeta}_\ell,\boldsymbol{\lambda}_a)^2}\,,\\
-\mathfrak{c}_a&=\mathfrak{C}\frac{\prod_{i=1}^{n}I_b(\boldsymbol{\lambda}_a,\boldsymbol{z}_i)^{w_i-1}\prod_{\ell=1}^{m}I_b(\boldsymbol{\lambda}_a,\boldsymbol{\zeta}_\ell)}{\prod_{b\neq a}I_b(\boldsymbol{\lambda}_a,\boldsymbol{\lambda}_b)^2}\,.
\end{split}
\end{equation}
Using these relations, we can write the $\Phi$ correlator in terms of the analytic data of the map $\Gamma$. The manipulation is essentially the same as the one done for bosonic strings in \cite{Knighton:2024qxd}, and we will simply quote the result:
\begin{equation}\label{eq:phi-correlator-rewritten}
\begin{split}
\braket{\cdots}_{\Phi}=\bigg|&\mathfrak{C}^{\frac{k-2}{2}}\prod_{a=1}^{N}(\mathfrak{c}_a)^{\frac{k}{2}}\prod_{i=1}^{n}(w_i\mathfrak{a}_i)^{-\frac{k(w_i+1)}{4}+j_i}\prod_{\ell=1}^{m}(2\mathfrak{b}_{\ell})^{\frac{k}{4}}\prod_{i<j}I_b(\boldsymbol{z}_i,\boldsymbol{z}_j)^{-q_iq_j}\\
&\times\prod_{i,\ell}I_b(\boldsymbol{z}_i,\boldsymbol{\zeta}_\ell)^{-q_i/Q}\prod_{\ell<m}I_b(\boldsymbol{\zeta}_\ell,\boldsymbol{\zeta_{m}})^{-\frac{1}{Q^2}}\bigg|^2\delta\left(\sum_{i=1}^{n}q_i+\frac{m}{Q}-\mathcal{Q}\right)\,,
\end{split}
\end{equation}
where the momenta $q_i$ and background charge $\mathcal{Q}$ are defined by
\begin{equation}
q_i=Qj_i-\frac{1}{Q}\,,\quad\mathcal{Q}=Q-\frac{2}{Q}=-\sqrt{\frac{2(k-1)^2}{k}}\,.
\end{equation}
The key now is to realize the Wick contractions appearing in \eqref{eq:phi-correlator-rewritten}, as well as the momentum-conserving delta function, are the outcome of computing the correlator
\begin{equation}
\Braket{\prod_{i=1}^{n}e^{-q_i\Phi}(\boldsymbol{z}_i)\prod_{\ell=1}^{m}e^{-\Phi/Q}(\boldsymbol{\zeta}_{\ell})}
\end{equation}
for a scalar field with background charge $\mathcal{Q}$ placed at the north pole.

Putting everything together, we can write the full worldsheet correlation function in the compact form
\begin{tcolorbox}
\begin{equation}
\begin{split}
\sum_{m=0}^{\infty}\frac{p^{N}}{m!}\int\mathrm{d}^{(2m|2m)}\boldsymbol{\xi}\,\Bigg|&(\mathfrak{C})^{\frac{k}{2}}\prod_{a=1}^{N}(\mathfrak{c}_a)^{-\frac{k}{2}}\prod_{i=1}^{n}\left(w_i^{-\frac{k(w_i+1)}{4}+j_i}\mathfrak{a}_i^{-h_i+\frac{k(w_i-1)}{4}}\right)\\
&\times\prod_{\ell=1}^{m}(2\mathfrak{b}_{\ell})^{\frac{k}{4}}\Bigg|^2
\Braket{\prod_{i=1}^{n}e^{-q_i\Phi}(\boldsymbol{z}_i)\prod_{\ell=1}^{m}e^{-\Phi/Q}(\boldsymbol{\zeta}_{\ell})}\,.
\label{eq:final worldsheet path integral result}
\end{split}
\end{equation}
\end{tcolorbox}
\noindent Here, we have introduced the symbol $\boldsymbol{\xi}_{\ell}$ for the pair $(\Gamma(\boldsymbol{\zeta}_\ell),D\Gamma(\boldsymbol{\zeta}_\ell))$, and the integration measure is
\begin{equation}
\int\mathrm{d}^{(2m|2m)}\boldsymbol{\xi}:=\int\prod_{\ell=1}^{m}\mathrm{d}^2\Gamma(\boldsymbol{\zeta}_\ell)\,\mathrm{d}^2D\Gamma(\boldsymbol{\zeta}_{\ell})\sum_{\Gamma}\,,
\end{equation}
where the sum is over all inequivalent solutions $\Gamma$ to the constraint equations with a fixed value of $\boldsymbol{\xi}_{\ell}$. The sum over $m$ replaces the sum over $N$ in equation \eqref{eq:general-amplitude}. A factor of $\frac{1}{N!}$ was included in \eqref{eq:general-amplitude} because the locations of the poles of $\Gamma$ were unordered. Similarly, in the new coordinates of \eqref{eq:change-of-variables}, the extra branch points are unordered and hence a factor of $1/m!$ is needed to avoid overcounting in the integration over $\mathfrak{M}_{0,n}(\mathbb{CP}^1,N)$. Equation \eqref{eq:final worldsheet path integral result} is the main computational result of this paper.

\subsection{Integrating out the odd coordinates}

Equation \eqref{eq:final worldsheet path integral result} gives a compact formula for the contribution of string correlators from worldsheets which live arbitrarily close to the conformal boundary of $\text{AdS}_3$ and which are unexcited in the compact directions. It involves an integration over $m$ commuting coordinates $\Gamma(\boldsymbol{\zeta}_\ell)$ and $m$ anticommuting coordinates $D\Gamma(\boldsymbol{\zeta}_\ell)$. In a forthcoming work, we will show how \eqref{eq:final worldsheet path integral result} reproduces correlation functions in the dual CFT. To this end, it will be useful to understand how exactly the odd integrals arise and how to perform the integration. Since it will also illucidate the meaning of the various terms arising in \eqref{eq:final worldsheet path integral result}, we will briefly review the key points here, leaving a more complete discussion to future work \cite{Knighton:2026xxx}.

To see how such an integration might go, let us first write down the most general solution for the super-holomorphic map $\Gamma$. Written in components $\Gamma=\gamma+\theta\psi$, the condition $\overline{D}\Gamma=0$ ensures $\gamma$ and $\psi$ are holomorphic. Moreover, $\gamma$ and $\psi$ respectively have simple and double poles at $z=\lambda_a$. Finally, the behavior of $\Gamma$ near $\boldsymbol{z}_i$ implies the tower of constraints
\begin{equation}
\begin{split}
\gamma(z_i)+\theta_i\psi(z_i)=x_i\,,&\quad\psi(z_i)+\theta_i\partial\gamma(z_i)=0\,,\\
\partial\gamma(z_i)+\theta_i\partial\psi(z_i)=0\,,&\quad\partial\psi(z_i)+\theta_i\partial^2\gamma(z_i)=0\,,\\
&\vdots\\
\partial^{w_i-1}\gamma(z_i)+\theta_i\partial^{w_i-1}\psi(z_i)=0\,,&\quad\partial^{w_i-1}\psi(z_i)+\theta_i\partial^{w_i}\gamma(z_i)=0\,.
\end{split}
\end{equation}
These constraints can be manipulated into constraints for $\gamma$ and $\psi$ separately, namely
\begin{equation}\label{eq:split-Gamma-constraints}
\begin{gathered}
\gamma(z_i)=x_i\,,\quad\partial\gamma(z_i)=0\,,\quad\cdots\quad\partial^{w_i-1}\gamma(z_i)=0\,,\\
\psi(z_i)=0\,,\quad\partial\psi(z_i)=0\,,\quad\cdots\,\quad\partial^{w_i-2}\psi(z_i)=0\,,
\end{gathered}
\end{equation}
along with one final constraint which determines $\theta_i$ in terms of $\gamma,\psi$:
\begin{equation}\label{eq:determine-thetai}
\theta_i=-\frac{\partial^{w_i-1}\psi(z_i)}{\partial^{w_i}\gamma(z_i)}\,.
\end{equation}
These constraints are suggestive: the first row of \eqref{eq:split-Gamma-constraints} implies that $\gamma$ is an ordinary (bosonic) covering map $\gamma:\Sigma_{\text{red}}\to\mathbb{CP}^1$ of degree $N$, while the second row says that $\psi$ is a spinor with double poles at $z=\lambda_a$ and zeroes of order $w_i-1$ at $z=z_i$. Finally, \eqref{eq:determine-thetai} implies that the first nontrivial coefficients of the Taylor series of $\psi$ and $\partial\gamma$ are not independent, but are related by the coordinate $\theta_i$.

The extra branch points $\boldsymbol{\zeta}_\ell=(\zeta_\ell,\rho_\ell)$ can be interpreted in a similar fashion. At these special points, $\gamma$ and $\psi$ satisfy
\begin{equation}
\partial\gamma(\zeta_\ell)+\rho_{\ell}\partial\psi(\zeta_\ell)=0\,,\quad\partial\psi(\zeta_\ell)+\rho_\ell\partial^2\gamma(\zeta_\ell)=0\,,
\end{equation}
which again can be rewritten into the pair of conditions
\begin{equation}\label{eq:split-extra-branch-points}
\partial\gamma(\zeta_\ell)=0\,,\quad\rho_\ell=-\frac{\partial\psi(\zeta_\ell)}{\partial^2\gamma(\zeta_\ell)}\,.
\end{equation}
This means that the even coordinate $\zeta_\ell$ is a simple branch point of $\gamma$, while the odd coordiante $\rho_\ell$ is determined by the behavior of $\psi$ near that point. The coordinates we integrate over in \eqref{eq:final worldsheet path integral result} are then
\begin{equation}
\begin{gathered}
\Gamma(\boldsymbol\zeta_\ell)=\gamma(\zeta_\ell)+\rho_\ell\psi(\zeta_\ell)=\gamma(\zeta_\ell)+\frac{\psi\partial\psi}{\partial^2\gamma}(\zeta_\ell)\,,\\
D\Gamma(\boldsymbol{\zeta}_\ell)=\psi(\zeta_\ell)+\rho_\ell\partial\gamma(\zeta_\ell)=\psi(\zeta_\ell)\,.
\end{gathered}
\end{equation}
We also note that the Taylor series coefficients $\mathfrak{a}_i,\mathfrak{b}_\ell$ and the residues $\mathfrak{c}_a$ appearing in \eqref{eq:final worldsheet path integral result} are completely encoded in $\gamma$:
\begin{equation}
\begin{split}
\gamma(z)\sim x_i+\mathfrak{a}_i(z-z_i)^{w_i}+\cdots\,,&\quad z\to z_i\,,\\
\gamma(z)\sim\gamma(\zeta_\ell)+\mathfrak{b}_\ell(z-\zeta_\ell)^2+\cdots\,,&\quad z\to\zeta_\ell\,,\\
\gamma(z)\sim\frac{\mathfrak{c}_a}{z-\lambda_a}+\cdots\,,&\quad z\to\lambda_a\,,
\end{split}
\end{equation}
while the constant $\mathfrak{C}$ is read off from the relation
\begin{equation}
\partial\gamma(z)=\mathfrak{C}\,\frac{\prod_{i=1}^{n}(z-z_i)^{w_i-1}\prod_{\ell=1}^{m}(z-\zeta_\ell)}{\prod_{a=1}^{N}(z-\lambda_a)^2}\,.
\end{equation}

Now let us use global superconformal symmetry to fix $z_1,z_2,z_3,\theta_1,\theta_2$, and assume that for a given set of points $x_i$ and $\xi_\ell=\gamma(\zeta_\ell)$, we have constructed a bosonic covering map $\gamma$ of degree $N$ satisfying equations \eqref{eq:split-Gamma-constraints} and \eqref{eq:split-extra-branch-points}. The existence of such a map fixes $z_4,\ldots,z_n$, the locations of the poles $\lambda_a$, and the locations of the extra branch points $\zeta_\ell$, and there are only a finite number of such maps. Given this covering map, we can write $\psi$ as
\begin{equation}
\psi(z)=\frac{\prod_{i=1}^{n}(z-z_i)^{w_i-1}}{\prod_{a=1}^{N}(z-\lambda_a)^2}P(z)\,,
\end{equation}
where $P(z)$ is a polynomial with anticommuting coefficients. The degree of $P$ is readily worked out by demanding that $\psi$ is constant at $z=\infty$, making sure to take into account the conformal transformation of $\psi$ under the coordinate change $z=1/u$, or equivalently by noting that a spinor on a genus $g$ surface has $g-1$ more zeroes than poles. In either case, at $g=0$ the degree of $P$ is read off from the constraint
\begin{equation}
\text{deg}\,P+\sum_{i=1}^{n}(w_i-1)-2N=-1\,,
\end{equation}
or equivalently $\text{deg}\,P=m+1$. Thus, $\psi$ has $m+2$ undetermined anticommuting coefficients, two of which are eliminated by imposing
\begin{equation}\label{eq:theta1-theta2-constraint}
\theta_i=-\frac{\partial^{w_i-1}\psi(z_i)}{\partial^{w_i}\gamma(z_i)}=-\frac{1}{\mathfrak{C}}P(z_i)\prod_{\ell=1}^{m}(z_i-\zeta_\ell)^{-1}
\end{equation}
for $i=1,2$. We are thus left with $m$ anticommuting degrees of freedom, precisely the number of odd integrals in \eqref{eq:final worldsheet path integral result}. Note that once the polynomial $P$ is fixed, the worldsheet moduli $\theta_3,\ldots,\theta_n$ and $\rho_\ell$ are completely determined by equations \eqref{eq:determine-thetai} and \eqref{eq:split-extra-branch-points}. The odd coordinates integrated over in \eqref{eq:final worldsheet path integral result} are then given by
\begin{equation}
D\Gamma(\boldsymbol{\zeta}_\ell)=\frac{\prod_{i=1}^{n}(\zeta_\ell-z_i)^{w_i-1}}{\prod_{a=1}^{N}(\zeta_\ell-\lambda_a)^2}P(\zeta_\ell)=\frac{2\mathfrak{b}_\ell}{\mathfrak{C}}\prod_{m\neq\ell}(\zeta_{\ell}-\zeta_m)^{-1}P(\zeta_\ell)\,.
\end{equation}
We can thus trade the integration over the odd coordinates $D\Gamma(\boldsymbol{\zeta}_\ell)$ for an integration over the $m$ undetermined coefficients of the anticommuting polynomial $P(z)$ at the cost of a Jacobian.

The task of integrating out the odd coordinates in \eqref{eq:final worldsheet path integral result} now takes the following form. We consider an arbitrary polynomial $P(z)$ that solves \eqref{eq:theta1-theta2-constraint} for $i=1,2$. We then determine $\theta_3,\ldots,\theta_n$ and $\rho_\ell$ from \eqref{eq:theta1-theta2-constraint} and
\begin{equation}
\rho_\ell=-\frac{\partial\psi(\zeta_\ell)}{\partial^2\gamma(\zeta_\ell)}\,.
\end{equation}
This will determine all anticommuting objects in \eqref{eq:final worldsheet path integral result} in terms of the coefficients of $P(z)$, which can then be integrated over. While we have not yet worked out the end result of this process, it is an algorithmically simple procedure that can be implemented via computer algebra in specific cases.

\paragraph{\boldmath The simple case of \texorpdfstring{$m=0$}{m=0}:}

The simplest case in which we can compute the correlator \eqref{eq:final worldsheet path integral result} by hand is when the $\text{SL}(2,\mathbb{R})$ spins $j_i$ are chosen to satisfy \eqref{eq:m=0-j-constraint}, so that $m=0$. In this case, there are no integrals to be performed, and the worldsheet correlator reduces to a finite sum over maps $\Gamma$. Since there is in particular no integration over odd moduli, we can write the full correlator in terms of purely bosonic data.

In this particular case, the polynomial $P(z)$ is linear and thus has two odd coefficients. These can be determined by \eqref{eq:theta1-theta2-constraint}, which in this case takes the particularly simple form
\begin{equation}
P(z_1)=-\mathfrak{C}\theta_1\,,\quad P(z_2)=-\mathfrak{C}\theta_2\,.
\end{equation}
This uniquely determines the polynomial to be
\begin{equation}
P(z)=\frac{\mathfrak{C}}{z_{12}}\left(\theta_1(z_2-z)-\theta_2(z_1-z)\right)\,,
\end{equation}
in terms of which the remaining odd coordinates are completely fixed:
\begin{equation}
\theta_i=-\frac{\theta_1z_{2i}-\theta_2z_{1i}}{z_{12}}\,.
\end{equation}
Particularly simple is when we use global conformal symmetry to fix $\theta_1=\theta_2=0$, which immediately implies
\begin{equation}\label{eq:thetas-vanish}
P(z)=0\,,\quad\theta_3=\cdots=\theta_n=0\,.
\end{equation}
That is, all anticommuting quantities vanish, and we are left with the simple result
\begin{tcolorbox}
\begin{equation}\label{eq:m=0-correlator}
\mathcal{A}_{0,n}=\sum_{\gamma}\left|\mathfrak{C}^{\frac{k}{2}}\prod_{a=1}^{N}\mathfrak{c}_a^{-\frac{k}{2}}\prod_{i=1}^{n}w_i^{-\frac{k(w_i+1)}{4}+j_i}\mathfrak{a}_i^{-h_i+\frac{k(w_i-1)}{4}}\right|^2\Braket{\prod_{i=1}^{n}e^{-q_i\phi}(z_i)}\,.
\end{equation}
\end{tcolorbox}
\noindent where the correlator is that of an ordinary scalar linear dilaton with background charge $\mathcal{Q}$. In terms of the momenta $q_i=Qj_i-1/Q$, this solution is valid if
\begin{equation}
\sum_{i=1}^{n}q_i+\mathcal{Q}=0\,.
\end{equation}
It is worth taking a moment to appreciate how simple the correlation function \eqref{eq:m=0-correlator} is. We began with a supersymmetric worldsheet CFT, computed an $n$-point correlator of spectrally-flowed vertex operators, and integrated over the moduli space $\mathfrak{M}_{0,n}$. In a conventional approach to superstring perturbation theory, one would first integrate out the odd moduli, which introduces complications in the form of $(n-2)$ picture-changing operators. However, in the $\text{AdS}_3$ worldsheet correlator we have just computed, picture-changing never becomes an issue, since the integral over $\mathfrak{M}_{0,n}$ \textit{localizes} onto a locus for which all odd coordinates vanish -- essentially, the worldsheet CFT correlation function contains a non-obvious delta function $\delta(\theta_i)$ for each odd coordinate. If we were to instead integrate out the odd moduli first, we would still have obtained the answer \eqref{eq:m=0-correlator}, however, the presence of picture-changing operators would have made the computation significantly more tedious (see for example \cite{Sriprachyakul:2024gyl,Yu:2024kxr}).

\section{Heterotic strings}\label{sec:heterotic}

So far, our discussion has been concerned with type II superstrings in $\text{AdS}_3$. The worldsheet theory for heterotic strings on $\text{AdS}_3$ is constructed in much the same way as for type II strings, with the exception that the appropriate superspace has only one (left-moving) supercoordinate. In this section we will show how the analysis of the previous sections is naturally extended to the heterotic case by writing down the sigma model, currents, vertex operators, and screening operators explicitly and using them to calculate worldsheet correlation functions. The discussion is essentially parallel to the type II case, with some important differences that we will highlight along the way.

\subsection{The heterotic sigma model}

Let us take the heterotic superspace coordinates to be $(z,\bar{z},\theta)$, and define as usual $D=\partial_{\theta}+\theta\partial_z$ and $\overline{\partial}=\partial_{\bar{z}}$. For a target spacetime with target metric $G_{\mu\nu}$ and Kalb-Ramond field $B_{\mu\nu}$, the heterotic Polyakov action is
\begin{equation}
S_{\text{Het}}=\frac{1}{2\pi\alpha'}\int\mathrm{d}^{2|1}\boldsymbol{z}\,(G_{\mu\nu}(X)+B_{\mu\nu}(X))DX^{\mu}\overline{\partial}X{^\nu}\,.
\end{equation}
In the case of $\text{AdS}_3$ in Poincar\'e coordinates, we find
\begin{equation}
S_{\text{Het}}=\frac{k}{2\pi}\int\mathrm{d}^{2|1}\boldsymbol{z}\,(D\Phi\overline{\partial}\Phi+e^{2\Phi}D\bar{\Gamma}\overline{\partial}\Gamma)\,,
\end{equation}
where we focus only on the $\text{AdS}_3$ component. As in the type II case, we can pass to a first-order formulation by introducing an anticommuting chiral superfield $\Omega$ of weight $(1/2,0)$ and a commuting field $\overline{\beta}$ of weight $(0,1)$ as Lagrange multipliers, in terms of which we write the action as
\begin{equation}
S_{\text{Het}}=\frac{1}{2\pi}\int\mathrm{d}^{2|1}\boldsymbol{z}\left(kD\Phi\overline{\partial}\Phi+\Omega\overline{\partial}\Gamma+\bar\beta D\bar\Gamma-\mu\Omega\bar\beta\,e^{-2\Phi}\right)\,.
\end{equation}
The fields entering the action have the superspace decompositions
\begin{equation}
\Phi=\phi+\theta\lambda\,,\quad \Gamma=\gamma+\theta\psi\,,\quad \bar\Gamma=\bar\gamma\,,\quad \Omega=\omega+\theta\beta\,.
\end{equation}

\paragraph{Decoupling the fields:} Similarly to the case of type II, we now want to decouple the path integral measure, which schematically takes the form
\begin{equation}
\mathcal{D}(e^{2\Phi}\Gamma)\mathcal{D}(e^{2\Phi}\bar\Gamma)\mathcal{D}\Phi\,.
\end{equation}
We can exchange this interacting measure for a factorized measure at the expense of introducing the Jacobian
\begin{equation}
\mathcal{D}(e^{2\Phi}\Gamma)\mathcal{D}(e^{2\Phi}\bar\Gamma)\mathcal{D}\Phi=\mathcal{D}\Gamma\,\mathcal{D}\bar\Gamma\,\mathcal{D}\Phi\,\exp\left(\frac{1}{2\pi}\int\mathrm{d}^{2|1}\boldsymbol{z}\left(D\Phi\overline{\partial}\Phi+\mathcal{R}\Phi\right)\right)\,,
\end{equation}
where $\mathcal{R}=\theta R$ is the superspace curvature.\footnote{Just as in the type II case, $\mathcal{R}$ generally has a bottom component involving the worldsheet gravitino, which we have gauged away.} We derive this Jacobian in Appendix \ref{app:measure}. Including this Jacobian into the path integral and appropriately rescaling $\Phi$, we wind up with the first-order action
\begin{equation}
S_{\text{Het}}=\frac{1}{2\pi}\int\mathrm{d}^{2|1}\boldsymbol{z}\left(\frac{1}{2}D\Phi\overline{\partial}\Phi-\frac{Q}{4}\mathcal{R}\Phi+\Omega\overline{\partial}\Gamma+\bar\beta D\bar\gamma-\mu\Omega\bar\beta\,e^{-Q\Phi}\right)\,,
\end{equation}
where the background charge is now
\begin{equation}
Q=\sqrt{\frac{2}{k-1}}\,.
\end{equation}

Just as in bosonic and type II strings, the worldsheet fields are free in the regime $\Phi\to\infty$. The OPEs between these free fields are easily derived, and take the same form as in equations \eqref{eq:ii-superspace-opes} and \eqref{eq:free-field-opes}. The left-moving stress tensor is the same as the type II case (see equations \eqref{eq:free-field-stress-tensor} and \eqref{eq:ii-superspace-stress-tensor}), while the right-moving stress tensor is
\begin{equation}
\overline{T}=-\frac{1}{2}(\overline{\partial}\phi)^2-\frac{Q}{2}\overline{\partial} {^2\phi}-\bar\beta\bar\partial\bar\gamma\,.
\end{equation}
Given the left- and right-moving superconformal generators, we can read off the left- and right-moving central charges of the worldsheet theory:
\begin{equation}\label{eq:heterotic-ads3-central-charge}
c=\frac{9}{2}+\frac{6}{k-1}\,,\quad\bar{c}=3+\frac{6}{k-1}\,.
\end{equation}

\paragraph{The reduced action:} For completeness, we can integrate over $\theta$ and write the action in terms of the reduced fields
\begin{equation}
\begin{split}
S_{\text{Het}}=\frac{1}{2\pi}\int\mathrm{d}^2z\bigg(&\frac{1}{2}\partial\phi\overline{\partial}\phi-\frac{1}{2}\lambda\overline{\partial}\lambda-\frac{Q}{4}R\phi\\
&+\beta\overline{\partial}\gamma+\omega\overline{\partial}\psi+\bar\beta\partial\bar\gamma-\mu(\beta+Q\omega\lambda)\bar\beta\,e^{-Q\phi}\bigg)\,.
\end{split}
\end{equation}
In the $\phi\to\infty$ limit, we are left with a free field system, with central charge
\begin{equation}
c=\frac{9}{2}+3Q^2\,,\quad \bar{c}=3+3Q^2
\end{equation}
in agreement with \eqref{eq:heterotic-ads3-central-charge}.

\paragraph{The current algebra:} The heterotic worldsheet theory possesses an $\mathfrak{sl}(2,\mathbb{R})^{(1)}_L\times\mathfrak{sl}(2,\mathbb{R})_R$ symmetry with currents
\begin{equation}
J^+=\Omega\,,\quad J^3=(\Omega\Gamma)-\frac{1}{Q}D\Phi\,,\quad J^-=(\Omega(\Gamma\Gamma))-\frac{2}{Q}\Gamma D\Phi-(k-1)D\Gamma
\end{equation}
in the left-moving sector and
\begin{equation}
\bar{\jmath}{\,^+}=\bar\beta\,,\quad\bar{\jmath}{\,^3}=(\bar\beta\bar\gamma)-\frac{1}{Q}\overline{\partial}\phi\,,\quad\bar{\jmath}{\,^-}=(\bar\beta(\bar\gamma\bar\gamma))-\frac{2}{Q}\bar\gamma\overline{\partial}\phi-(k+1)\overline{\partial}\bar\gamma
\end{equation}
in the right-moving sector. A quick check of the OPEs satisfied by the currents shows that left-moving supercurrents $J^a$ generate $\mathfrak{sl}(2,\mathbb{R})_{k-1}^{(1)}$ while the right-moving currents $\bar\jmath\,^a$ generate $\mathfrak{sl}(2,\mathbb{R})_{k+1}$. Indeed, the central charges of these algebras obtained from the Sugawara construction agree with equation \eqref{eq:heterotic-ads3-central-charge}. We are lead to the conclusion:
\begin{tcolorbox}
\textit{Heterotic strings on $\text{AdS}_3$ are described by the WZW models $\mathfrak{sl}(2,\mathbb{R})^{(1)}_{k-1}$ in the left-moving sector and $\mathfrak{sl}(2,\mathbb{R})_{k+1}$ in the right-moving sector, where $k=L^2/\alpha'$.}
\end{tcolorbox}
\noindent Note that this is slightly different than what has been claimed in the literature \cite{Kutasov:1998zh}, albeit only by a shift $k\to k+1$ which is invisible in the semiclassical regime $k\to\infty$.

\paragraph{Vertex operators:} Vertex operators in the heterotic worldsheet theory are built in the obvious way by taking the left-moving vertex operator in the supersymmetric sigma model and the right-moving operator in the bosonic sigma model. We can specifically write them as
\begin{equation}
\mathscr{V}_{m,j}^{w}(z|\theta;x_i)\bar{V}_{\bar{m},j}^{w}(\bar{z};\bar{x}_i)\,.
\end{equation}
The left-moving piece is simply the vertex operator constructed in Section \ref{sec:localization} (with a shift $k\to k+1$), while the right-moving piece was constructed in \cite{Knighton:2023mhq}. These operators have left- and right-moving conformal weights
\begin{equation}
\Delta=\frac{j(1-j)}{k-1}-hw+\frac{(k-1)w^2}{4}\,,\quad\bar{\Delta}=\frac{j(1-j)}{k-1}-\bar{h}w+\frac{(k+1)w^2}{4}\,,
\end{equation}
where $h=m+(k-1)w/2$ and $\bar{h}=\bar{m}+(k+1)w/2$.

\paragraph{Screening operator:} We can also construct a screening operator entirely analogously to the type II case, by taking the left-moving piece to be the operator $\mathscr{D}$ constructed above (but with a shift $k\to k-1$), while we take the right-moving piece $\bar{D}$ to be the operator constructed in \cite{Knighton:2023mhq} (but with a shift $k\to k+1$). Explicitly, we define
\begin{equation}
\begin{split}
\mathscr{D}(\boldsymbol{z})&=e^{-2\Phi/Q}\left(\text{Res}\,\Gamma\right)^{-(k-1)}\delta_{1|1}(\Omega)\\
\bar D(\bar z)&=e^{-2\Phi/Q}\left(\text{Res}\,\bar\gamma\right)^{-k}\delta(\bar\beta)
\end{split}
\end{equation}
The full screening operator is the integral
\begin{equation}
\int\mathrm{d}^{2|1}\boldsymbol{z}\,\mathscr{D}(z|\theta)\bar{D}(\bar z)\,.
\end{equation}
The integration makes sense since the integrand has conformal dimension $(\frac{1}{2},1)$.

\paragraph{Compactification:}

Now that we have constructed the heterotic sigma model on $\text{AdS}_3$, we need to compactify on some manifold $\mathcal{N}$. This will produce a string theory described by the supersymmetric sigma model
\begin{equation}
\mathfrak{sl}(2,\mathbb{R})^{(1)}_{k-1}\times\mathcal{N}_L^{(1)}
\end{equation}
in the left-moving sector and the bosonic sigma model
\begin{equation}
\mathfrak{sl}(2,\mathbb{R})_{k+1}\times\mathcal{N}_R\times\mathbb{R}^{16}/\Lambda
\end{equation}
in the right-moving sector. The self-dual lattice $\Lambda$ will either by the $\text{E}_8\times\text{E}_8$ or $\text{Spin}(32)/\mathbb{Z}_2$ lattice, depending on which heterotic model we are considering. Here, $\mathcal{N}_L^{(1)}$ and $\mathcal{N}_R$ are a supersymmetric left-moving and bosonic right-moving sigma model on $\mathcal{N}$, respectively.

Criticality of the model requires that the central charges of the worldsheet CFT are $c_L=15$ and $c_R=26$. This in particular implies
\begin{equation}
c(\mathcal{N}_L^{(1)})=\frac{21}{2}-\frac{6}{k-1}\,,\quad c(\mathcal{N}_R)=7-\frac{6}{k-1}\,.
\end{equation}
For example, we can compactify on $\text{S}^3\times\mathcal{M}_4$ with $\mathcal{M}_4=\mathbb{T}^4$ or $\text{K}3$, for which
\begin{equation}
\mathcal{N}_L^{(1)}=\mathfrak{su}(2)^{(1)}_{k-1}\times(\mathcal{M}_4)^{(1)}\,,\quad\mathcal{N}_R=\mathfrak{su}(2)_{k-3}\times\mathcal{M}_4\,.
\end{equation}
We could also compactify on $\text{S}^3\times\text{S}^3\times\text{S}^1$, for which
\begin{equation}
\begin{gathered}
\mathcal{N}_L^{(1)}=\mathfrak{su}(2)_{k_1-1}^{(1)}\times\mathfrak{su}(2)_{k_2-1}^{(1)}\times\mathfrak{u}(1)^{(1)}\,,\\
\mathcal{N}_R=\mathfrak{su}(2)_{k_1-3}\times\mathfrak{su}(2)_{k_2-3}\times\mathfrak{u}(1)\,,
\end{gathered}
\end{equation}
subject to the criticality condition
\begin{equation}
\frac{1}{k-1}=\frac{1}{k_1-1}+\frac{1}{k_2-1}\,.
\end{equation}
In all of these examples, there is an isomorphism
\begin{equation}\label{eq:nl-nr-isomorphism}
\mathcal{N}_L^{(1)}\cong\mathcal{N}_R\times(\text{7 free fermions})\,,
\end{equation}
which naturally explains the relation $c(\mathcal{N}_L^{(1)})-c(\mathcal{N}_R)=\frac{7}{2}$ between the left- and right-moving central charges. In what follows, we will keep the compactification $\mathcal{N}$ arbitrary, but we will assume that \eqref{eq:nl-nr-isomorphism} always holds.

\subsection{Heterotic correlators}

Just as in the type IIB case, we can calculate correlation functions of the heterotic string in the limit $\Phi\to \infty$. The derivation will largely mirror that of Section \ref{sec:localization}, and so we will be brief.

At genus $g$, the correlation function we wish to calculate is
\begin{equation}\label{eq:heterotic-correlator}
\sum_{N=0}^{\infty}\frac{p^N}{N!}\int_{\mathfrak{M}_{g,n}^{\text{het}}}\mathrm{d}\mu\Braket{\left(\int\mathscr{D}\bar{D}\right)^N\prod_{i=1}^{n}\mathscr{V}_{m_i,j_i}^{w_i}(\boldsymbol{z}_i;x_i)\bar{V}^{w_i}_{\bar{m}_i,j_i}(\bar{z}_i;\bar{x}_i)}\,.
\end{equation}
The heterotic moduli space $\mathfrak{M}_{g,n}^{\text{het}}$ merits some explanation. From a holomorphic point of view, we can construct the heterotic string from two separate worldsheets $\Sigma_L\times\Sigma_R$, where $\Sigma_L$ is a super Riemann surface and $\Sigma_R$ is an ordinary Riemann surface. The heterotic worldsheet $\Sigma$ is then a subset
\begin{equation}
\Sigma\subset\Sigma_L\times\Sigma_R
\end{equation}
of real dimension $2|1$ such that $\Sigma_{\text{red}}$ lies `sufficiently closely' to the diagonal of $(\Sigma_L)_{\text{red}}\times\Sigma_R$ \cite{Witten:2012ga}. In practice, this means that we treat the left-moving coordinates $(z|\theta)$ and right-moving coordinate $\bar{z}$ independently until the end of any calculation, upon which we impose that $z$ and $\bar{z}$, as well as the left- and right-moving complex structure moduli, are complex conjugates. The moduli space $\mathfrak{M}_{g,n}^{\text{het}}$ is constructed in a similar fashion, namely as a submanifold
\begin{equation}
\mathfrak{M}_{g,n}^{\text{het}}\subset\mathfrak{M}_{g,n}^{(L)}\times\mathcal{M}_{g,n}^{(R)}
\end{equation}
of real dimension $2(n+3g-3)|n+2g-2$.

For $g=0$, the worldsheet has a global symmetry which is a real form of $\text{PSL}(2|1,\mathbb{C})_L\times\text{PSL}(2,\mathbb{C})_R$ and which can be used to fix $(z_1,z_2,z_3)$ and $(\theta_1,\theta_2)$ on the left and $(\bar{z}_1,\bar{z}_2,\bar{z}_3)$ on the right. After fixing the global superconformal symmetry, the measure on $\mathfrak{M}_{0,n}^{\text{het}}$ is determined by the left- and right-moving conformal ghost systems:
\begin{equation}
\begin{split}
\mathrm{d}\mu
&=I_b(\boldsymbol{z}_1,\boldsymbol{z}_3)I_b(\boldsymbol{z}_2,\boldsymbol{z}_3)\bar{z}_{12}\bar{z}_{23}\bar{z}_{13}\,\mathrm{d}\theta_3\prod_{i=4}^{n}\mathrm{d}\boldsymbol{z}_i\,\mathrm{d}\bar{z}_i\,.
\end{split}
\end{equation}
The calculation of the path integral now proceeds as before. Integrating out $\Omega$ localizes the left-moving sector to fields $\Gamma$ which are holomorphic, have poles at the locations of the screening operators, and which satisfy $D^{\ell}\Gamma(\boldsymbol{z}_i)=0$ for $\ell=1,\ldots,2w_i-1$ and $\Gamma(\boldsymbol{z}_i) = x_i$. Meanwhile, the right-moving sector localizes to antiholomorphic maps $\bar{\gamma}(\bar{z})$ which have simple poles at the locations of the screening operators and which satisfy
\begin{equation}
\bar{\gamma}=\bar{x}_i+\bar{a}_i(\bar{z}-\bar{z}_i)^{w_i}+\cdots\,.
\end{equation}
Since the number $N$ of screening operators inserted in the correlator at any order in perturbation theory is the same in the left- and right-moving sectors, the maps $\Gamma$ and $\bar\gamma$ have the same degree (in the sense that they have the same number of poles), and they have the same branching behavior near the locations of the vertex operators. Thus, combining the localization argument above from the type II case with the bosonic localization argued for in \cite{Knighton:2023mhq,Knighton:2024qxd}, we find that the worldsheet path integral localizes to a moduli space of real dimension $2m|m$, where $m$ is given by \eqref{eq:number-of-extra-branch-points}. For a given set of $\text{SL}(2,\mathbb{R})$ spins $j_i$, the number of branch points $m$ is constrained by momentum conservation of the $\Phi$ field. Repeating the logic that lead to equation \eqref{eq:general-j-constraint} for the left-moving sector, we find
\begin{equation}\label{eq:heterotic-j-constraint}
m=-Q^2\left(\sum_{i=1}^{n}j_i-\frac{1}{Q^2}(n+2g-2)+(g-1)\right)\,,
\end{equation}
where we recall that $Q=\sqrt{2/(k-1)}$ for heterotic strings. This is easily checked to be consistent with the bosonic constraint in the $\mathfrak{sl}(2,\mathbb{R})_{k+1}$ WZW model (see \cite{Knighton:2024qxd} for details). 

Computing the correlation function \eqref{eq:heterotic-correlator} in the $\Phi\to\infty$ limit can now be done by following the same set of manipulations as for the type II case in Section \ref{sec:localization}. In the left-moving sector, we have to change the integration variables as in equation \eqref{eq:change-of-variables}. In the right-moving sector, we can write the map $\bar\gamma$ as
\begin{equation}
\bar\gamma(\bar{z})=\bar{a}+\sum_{a=1}^{N}\frac{\bar{c}_a}{\bar{z}-\bar{\lambda}_a}\,,
\end{equation}
and perform the bosonic change of variables
\begin{equation}
(\bar{z}_4,\ldots,\bar{z}_n,\bar{a},\bar{\lambda}_a,\bar{c}_a)\to(\overline{\partial}\,^{\ell}\bar{\gamma}(\bar{z}_i),\bar{\gamma}(\bar{\zeta_\ell}))\,,
\end{equation}
where $\bar{\zeta}_{\ell}$ are the extra branch points of $\bar\gamma$. As conjectured in Section \ref{sec:localization}, the left-moving sector gives a Jacobian of
\begin{equation}
\mathcal{J}_{\text{left}}=\mathfrak{C}\,.
\end{equation}
The Jacobian for the right-moving sector was calculated explicitly in \cite{Knighton:2024qxd} and takes the more complicated form
\begin{equation}
\mathcal{J}_{\text{right}}=\bar{C}^2\prod_{i=1}^{n}(w_i\bar{a}_i)^{-\frac{w_i+1}{2}}\prod_{\ell=1}^{m}(2\bar{b}_\ell)^{-\frac{1}{2}}\,,
\end{equation}
where $\bar{C},\bar{b}_{\ell}$ are the analogues of $\mathfrak{C}$ and $\mathfrak{b}_{\ell}$ for $\bar{\gamma}$, i.e. the coefficients appearing in the expansions
\begin{equation}
\bar{\gamma}(\bar{z})\sim\bar{\xi}_{\ell}+\bar{b}_{\ell}(\bar{z}-\bar{\zeta}_{\ell})^2+\cdots\,,\quad\overline{\partial}\bar{\gamma}(\bar{z})=\bar{C}\frac{\prod_{i=1}^{n}(\bar{z}-\bar{z}_{i})^{w_i-1}\prod_{\ell=1}^{m}(\bar{z}-\bar{\zeta}_{\ell})}{\prod_{a=1}^{N}(\bar{z}-\bar{\lambda}_a)^2}\,.
\end{equation}
Taking the Jacobians into account, as well as the correlator of the scalar $\Phi$ (see Section \ref{sec:localization}), the worldsheet correlator \eqref{eq:heterotic-correlator} can be massaged into the form
\begin{tcolorbox}
\begin{equation}\label{eq:heterotic-correlator-answer}
\begin{split}
\sum_{m=0}^{\infty}\frac{p^N}{m!}\int\mathrm{d}^{2m|m}\boldsymbol{\xi}\,&\mathfrak{C}^{\frac{k-1}{2}}\prod_{a=1}^{N}\mathfrak{c}_a^{-\frac{k-1}{2}}\prod_{i=1}^{n}w_i^{-\frac{(k-1)(w_i+1)}{4}+j_i}\mathfrak{a}_i^{-h_i+\frac{(k-1)(w_i-1)}{4}}\prod_{\ell=1}^{m}(2\mathfrak{b}_\ell)^{\frac{k-1}{4}}\\
&\times \bar{C}^{\frac{k+1}{2}}\prod_{a=1}^{N}\bar{c}_a^{-\frac{k+1}{2}}\prod_{i=1}^{n}w_i^{-\frac{(k+1)(w_i+1)}{4}+j_i}\bar{a}_i^{-\bar{h}_i+\frac{(k+1)(w_i-1)}{4}}\prod_{\ell=1}^{m}(2\bar{b}_{\ell})^{\frac{k-3}{4}}\\
&\times\left\langle\prod_{i=1}^{n}e^{-q_i\Phi}(z_i,\bar{z}_i|\theta_i)\prod_{\ell=1}^{m}e^{-\Phi/Q}(\zeta_\ell,\bar\zeta_\ell|\rho_\ell)\right\rangle\,.
\end{split}
\end{equation}
\end{tcolorbox}

\paragraph{\boldmath The $m=0$ contribution:} Just as in the type II case, the heterotic worldsheet correlator simplifies drastically when we take $m=0$. In this case, there are no extra branch points either in $\Gamma$ or $\bar{\gamma}$. Moreover, we can gauge-fix the left-moving fermionic coordinates to satisfy $\theta_1=\theta_2=0$, in terms of which the remaining fermionic coordinates $\theta_3,\ldots,\theta_n$ as well as the spinor field $\psi(z)$ are also forced to vanish identically (see the discussion around equation \eqref{eq:thetas-vanish}). For this particular gauge choice, $\Gamma(z|\theta)=\gamma(z)$ is just a bosonic map, and $\gamma$ and $\bar{\gamma}$ are simply complex conjugates (assuming $z_{i}$ are taken to be the complex conjugates of $\bar{z}_i$). The end result is that the correlator with $m=0$ can be written in terms of the data of a single map $\gamma(z)$, and takes the form
\begin{tcolorbox}
\begin{equation}\label{eq:heterotic-m=0}
\begin{split}
\mathcal{A}_{0,n}=\sum_{\gamma}&\mathfrak{C}^{\frac{k-1}{2}}\prod_{a=1}^{N}\mathfrak{c}_a^{-\frac{k-1}{2}}\prod_{i=1}^{n}w_i^{-\frac{(k-1)(w_i+1)}{4}+j_i}\mathfrak{a}_i^{-h_i+\frac{(k-1)(w_i-1)}{4}}\\
&\times\bar{\mathfrak{C}}^{\frac{k+1}{2}}\prod_{a=1}^{N}\bar{\mathfrak{c}}_a^{-\frac{k+1}{2}}\prod_{i=1}^{n}w_i^{-\frac{(k+1)(w_i+1)}{4}+j_i}\bar{\mathfrak{a}}_i^{-\bar{h}_i+\frac{(k+1)(w_i-1)}{4}}\Braket{\prod_{i=1}^{n}e^{-q_i\phi}(z_i,\bar{z}_i)}\,.
\end{split}
\end{equation}
\end{tcolorbox}

\section{Discussion}\label{sec:discussion}

In this paper, we reconsidered the path integral formulation of the worldsheet superconformal field theory describing strings on $\text{AdS}_3$, working in superspace from the start. After writing down explicit forms of vertex operators and screening operators in the free field realization of $\mathfrak{sl}(2,\mathbb{R})_k^{(1)}$, we were able to arrive at remarkably simple formulae for the near-boundary limit of correlation functions of NS-sector long strings at tree level. Most interestingly of all, these correlators are captured by the geometric data of `super-holomorphic' maps from the worldsheet to the boundary of $\text{AdS}_3$ -- a natural supersymmetric generalization of the holomorphic covering construction that appears in bosonic string theory.

\subsection{Identifying the dual CFT}

The most natural application of the results of this paper is in the identification of the superconformal field theory dual to superstrings on $\text{AdS}_3\times\mathcal{N}$. Indeed, in bosonic $\text{AdS}_3$ string theory, the analogues of the main formulae \eqref{eq:final worldsheet path integral result} and \eqref{eq:heterotic-correlator-answer} for long-string path integrals perfectly reproduced the perturbative expansion of the dual CFT proposed in \cite{Balthazar:2021xeh,Eberhardt:2021vsx} (see also \cite{Eberhardt:2019qcl} for an earlier proposal), with the sum over $m$ controlling the degree in conformal perturbation theory. Can we expect the same to happen in the supersymmetric case?

Let's start with the type II case. On general grounds, the dual CFT to (perturbative) superstring theory on $\text{AdS}_3\times\mathcal{N}$ with pure NS-NS flux is expected to be a marginal deformation of the symmetric product orbifold \cite{Balthazar:2021xeh,Eberhardt:2021vsx}\footnote{We should mention that it is not entirely clear whether this CFT captures the full (perturbative) dynamics of strings on $\text{AdS}_3$, or whether it just captures the long-string/near-boundary subsector of the bulk theory, or whether the theory is even well-defined for $k>1$. See \cite{Balthazar:2021xeh,Eberhardt:2021vsx,Aharony:2024fid,Chakraborty:2025nlb} for various discussions on this point.}
\begin{equation}\label{eq:cft-dual-conjecture}
\text{Sym}^N(\mathbb{R}_{\mathcal{Q}}^{(1)}\times\mathcal{N})+\int\mathrm{d}^2\xi\,\mathcal{O}_{2,\alpha}(\xi,\bar\xi)\,.
\end{equation}
The seed theory of this orbifold is the effective theory of a single worldsheet instanton (long string) near the boundary of $\text{AdS}_3$ \cite{Seiberg:1999xz}, and the symmetric product approximately describes a gas of long strings. The $\mathscr{N}=(1,1)$ linear dilaton theory is composed of a scalar $\phi$ and two chiral fermions $\lambda,\bar\lambda$ satisfying the action
\begin{equation}
S[\phi,\lambda,\bar\lambda]=\frac{1}{2\pi}\int\mathrm{d}^2x\left(\frac{1}{2}\partial\phi\,\overline{\partial}\phi-\frac{1}{2}\lambda\overline{\partial}\lambda-\frac{1}{2}\bar\lambda\partial\bar\lambda-\frac{\mathcal{Q}}{4}R\phi\right)\,,
\end{equation}
where the background charge is related to the $\text{SL}(2,\mathbb{R})$ level by $\mathcal{Q}=-\sqrt{2(k-1)^2/k}$. Finally, the marginal operator $\mathcal{O}_{2,\alpha}$ is a twist field of length 2 carrying momentm $\alpha$ in the $\phi$ direction. The holographic dictionary relates spectrally-flowed long-string vertex operators in the NS sector on the worldsheet with twisted-sector states in the dual CFT:
\begin{equation}
\mathscr{V}_{m,j}^{w}(z,x)\longleftrightarrow e^{-q\phi}\sigma_{w}(x)\,,
\end{equation}
where $\sigma_w$ is the twisted-sector ground state of the boundary CFT and the momentum $q$ is related linearly to the $\text{SL}(2,\mathbb{R})$ spin:
\begin{equation}\label{eq:q-j-relation}
q=\sqrt{\frac{2}{k}}\left(j-\frac{k}{2}\right)\,.
\end{equation}

Order-by-order in perturbation theory in the marginal deformation $\mathcal{O}_{2,\alpha}$, we would expect that the worldsheet correlation functions reproduce the correlators of the dual CFT. A natural ansatz, and what is indeed the case in the bosonic duality, is that the terms in the sum over $m$ in \eqref{eq:final worldsheet path integral result} reproduce the terms
\begin{equation}\label{eq:dual-cft-m-correlator}
\frac{1}{m!}\int\mathrm{d}^2\xi_1\cdots\mathrm{d}^2\xi_m\Braket{\prod_{i=1}^{n}e^{-q_i\phi}\sigma_{w_i}(x_i,\bar{x}_i)\prod_{\ell=1}^{m}\mathcal{O}_{2,\alpha}(\xi_{\ell},\bar\xi_{\ell})}
\end{equation}
in the perturbative expansion of the dual CFT. The simplest term is $m=0$, for which there is no integral. In this case, the undeformed CFT correlation function is readily computed using the covering space method of Lunin and Mathur \cite{Lunin:2000yv}, and one indeed recovers \eqref{eq:m=0-correlator} (see for example equation (8.1) of \cite{Dei:2019iym}).

Matching the terms with $m>0$ on the worldsheet and CFT sides is somewhat trickier. However, there are hints that it can be done. For example, a strong consistency condition in the dual CFT is that the $j$-constraint \eqref{eq:general-j-constraint} must reproduce the momentum conservation of the linear dilaton $\phi$ in the dual CFT. The latter takes the form
\begin{equation}
\sum_{i=1}^{n}q_i+m\alpha=\mathcal{Q}\,.
\end{equation}
Using \eqref{eq:q-j-relation} and comparing to \eqref{eq:general-j-constraint}, we find that the worldsheet and CFT answers are only compatible if
\begin{equation}
\alpha=\sqrt{\frac{k}{2}}\,,
\end{equation}
exactly the value proposed in \cite{Balthazar:2021xeh,Eberhardt:2021vsx}.

Beyond this simple matching of momentum conservation, there are still conceptual hurdles for matching the correlators \eqref{eq:final worldsheet path integral result} and \eqref{eq:dual-cft-m-correlator}, especially if we insist on identifying the worldsheet with the covering space. For example, the marginal operator $\mathcal{O}_{2,\alpha}$ lifts to a Ramond-sector operator on the cover \cite{Lunin:2001pw}. This feature raises two points of difficulty. Firstly, while on the covering space the existence of a Ramond operator would signal a degeneration of the superconformal structure, we do note see such a degeneration on the string theory side, making a direct identification between the covering space and the worldsheet difficult. Secondly, since $\mathcal{O}_{2,\alpha}$ lives in the Ramond sector on the covering space, there are in general a large number of such operators to choose from. While, in cases with extended spacetime supersymmetry, there is typically a privileged choice,\footnote{See for example equation (6.4) of \cite{Eberhardt:2021vsx} for a natural construction in the $\mathscr{N}=(4,4)$ case.} there is no currently known natural choice for arbitrary $\mathcal{N}$.

Another source of difficulty is that the natural generalizations of `covering maps' are different on the CFT and worldsheet. In the spacetime CFT, it is natural to work in spacetime superspace, for which the generalization of a covering space is a \textit{super-covering map} \cite{Nairz:2025kou}
\begin{equation}
\tilde{\Gamma}:\mathbb{CP}^{1|1}_{\text{covering}}\to\mathbb{CP}^{1|1}_{\text{spacetime}}\,.
\end{equation}
Meanwhile, as we discussed in this paper, the natural geometric object appearing on the worldsheet side is a \textit{super-holomorphic curve} (see Section \ref{sec:super-holomprphic-curves})
\begin{equation}
\Gamma:\mathbb{CP}^{1|1}_{\text{worldsheet}}\to\mathbb{CP}^1_{\text{spacetime}}\,.
\end{equation}
The difference is that the map $\tilde{\Gamma}$ has a super Riemann surface as its target, while $\Gamma$ maps to an ordinary Riemann surface. The process of matching the spacetime and worldsheet correlators, then, will involve relating the geometric data of these two types of maps. While this is in principle possible for very simple cases, it is still conceptually complicated to understand how these objects relate to each other in full generality. (This is, of course, a reflection of a universal and unfortunate fact: there is no covariant quantization of string theory which has both manifest worldsheet and spacetime supersymmetry.) Nevertheless, there are hints that one can match the worldsheet and spacetime correlation functions, and we refer this discussion to future work \cite{Knighton:2026xxx}.

\subsection{The CFT dual of heterotic strings}

Returning to the case of heterotic strings, we can use our expressions for correlation functions to deduce a novel proposal for the CFT dual of heterotic string theory on $\text{AdS}_3\times\mathcal{N}$.\footnote{In the case of heterotic strings on $\text{AdS}_3\times\text{S}^3\times\mathbb{T}^4$ at minimal tension ($k=2$ in our conventions), a CFT dual was proposed in \cite{Eberhardt:2025sbi}.} Our proposal again takes the form of a deformed symmetric product:
\begin{equation}
\text{Sym}^N(\mathcal{S}_L\times\mathcal{S}_R)+\int\mathcal{O}_{2,\alpha}\,.
\end{equation}
The seed theories in the left- and right-moving sectors are themselves different, with the left-moving sector being described by the $\mathscr{N}=1$ chiral sigma model
\begin{equation}
\mathcal{S}_L=\mathbb{R}_{\mathcal{Q}}^{(1)}\times\mathcal{N}^{(1)}\,,
\end{equation}
while the right-moving sector is the bosonic model
\begin{equation}
\mathcal{S}_R=\mathbb{R}_{\mathcal{Q}}\times\mathcal{N}\times\mathbb{R}^{16}/\Lambda\,.
\end{equation}
Here, $\mathbb{R}^{16}/\Lambda$ is the 16-dimensional torus that determines the precise heterotic string theory in question ($\Lambda$ is the $E_8\times E_8$ or the $\text{Spin}(32)/\mathbb{Z}_2$ lattice). The background charge is given by
\begin{equation}
\mathcal{Q}=-\sqrt{\frac{2(k-2)^2}{k-1}}\,.
\end{equation}
Interestingly, the central charges of the left- and right-moving seed theories are different:
\begin{equation}
c(\mathcal{S}_L)=6(k-1)\,,\quad c(\mathcal{S}_R)=6(k+1)\,,
\end{equation}
leading to a gravitational anomaly in the full CFT
\begin{equation}
c_L-c_R=-12N\,.
\end{equation}
This was already predicted in \cite{Kutasov:1998zh}, but it is satisfying to see it arise from a direct CFT calculation.

The holographic dictionary is similar to that of the type II case. We can combine the left- and right-moving chiral bosons into a single boson $\phi$. The holographic dictionary in this case is identical to that in the type II case, except the momentum is now related to the $\text{SL}(2,\mathbb{R})$ spin by
\begin{equation}
q=\sqrt{\frac{2}{k-1}}\left(j-\frac{k-1}{2}\right)\,.
\end{equation}
The momentum $\alpha$ of the marginal operator can also be read off analogously by consistency of $\phi$ momentum conservation with the $j$ constraint \eqref{eq:heterotic-j-constraint}, from which we find
\begin{equation}
\alpha=\sqrt{\frac{k-1}{2}}\,.
\end{equation}
In fact, there is a natural candidate for the deformation operator, which can be seen by computing the conformal weight of the twisted ground state $e^{-\alpha\phi}\sigma_2$. Specifically for the value of $\alpha$ found above, we have
\begin{equation}
h(e^{-\alpha\phi}\sigma_2)=\frac{1}{2}\,,\quad\bar{h}(e^{-\alpha\phi}\sigma_2)=1\,.
\end{equation}
Thus, $\mathcal{O}_{2,\alpha}=G_{-1/2}e^{-\alpha\phi}\sigma_2$ has conformal dimension $(1,1)$ and is exactly marginal by virtue of being a supersymmetric descendant. (Of course, there is still the ambiguity of \textit{which} twist-2 vacuum to take for the left-moving degrees of freedom.) Finally, we note that the $m=0$ heterotic correlator \eqref{eq:heterotic-m=0} has precisely the form one might expect from the correlation function of twisted-sector ground states in a heterotic symmetric product CFT.

\subsection{Open problems and future directions}

\noindent Aside from the identification of the dual CFTs to these backgrounds, there are several interesting directions to be pursued in future work.

\paragraph{Higher-genus correlators and partition functions:} An obvious generalization of the present work is the computation of partition functions and correlation functions on worldsheets of higher genus. In particular, while the localization argument presented in Section \ref{sec:localization} was not particular to sphere correlators, the calculation of the resulting integral was. Two features of sphere correlators that made this computation possible were 1) the fact that the path integral measure $\mathrm{d}\mu$ on the moduli space $\mathfrak{M}_{0,n}$ is readily calculable and 2) that there exists a nice set of coordinates on $\mathfrak{M}_{0,n}$ in terms of the locations $(z_i|\theta_i)$ of the vertex operators. At `low' genus ($g=1$ in particular), facets of these features will persist, but at $g>1$ things very quickly become much more complicated. It is likely that the generalization of our current work to higher-genus correlators will involve new techniques or a different choice of gauge on the worldsheet. Indeed, in bosonic string theory, computing higher-genus partition functions in the near-boundary limit was recently achieved in \cite{Knighton:2024pqh} by fixing the worldsheet metric to be the pullback of the boundary metric under the covering map, as opposed to using conformal gauge. It would be interesting to explore whether there is a gauge choice for the worldsheet metric and graviton which makes such higher-genus computations possible in the supersymmetric case.

\paragraph{Full 3- and 4-point functions:} A key conceptual step in the present work was to treat the $\mathfrak{sl}(2,\mathbb{R})_k^{(1)}$ WZW model in a manifestly supersymmetric fashion. The affine currents of this model are spin-$1/2$ chiral superfields $J^a(z|\theta)$ which obey the superspace current algebra $\mathfrak{sl}(2,\mathbb{R})_k^{(1)}$. Since the vertex operators we are interested in naturally fall into (spectrally-flowed) highest-weight representations, the currents $J^a(z|\theta)$ can be used to derive supersymmetric versions of local Ward identities, and the super-Sugawara construction \eqref{eq:super-sugawara} can also be used to write down supersymmetric versions of the Knizhnik–Zamolodchikov (KZ) equations. 

In the bosonic $\text{SL}(2,\mathbb{R})$ WZW model, the local Ward identities, KZ equations, and crossing symmetry have been shown to be so constraining that they can be used to bootstrap the full form of 2-, 3-, and 4-point functions of spectrally-flowed primaries \cite{Eberhardt:2019ywk,Dei:2021xgh,Dei:2021yom,Dei:2022pkr}, including both the contribution of near-boundary worldsheets considered in the present work and of strings which propagate in the bulk of $\text{AdS}_3$. It does not seem unlikely that a supersymmetric version of this analysis is possible, and that one could write down full closed-form expressions for the 2-, 3-, and 4-point functions for superstrings propagating in $\text{AdS}_3$. (While certain expressions of this type exist in the literature, see for example \cite{Iguri:2022pbp,Yu:2024kxr,Yu:2025qnw,Barone:2025vww}, they are computed via the decomposition \eqref{eq:intro-sl2-decomposition}, and thus are not well-suited to the computation of superstring correlators without the use of picture-changing).

\paragraph{Minimal superstring theories:} A remarkable feature of the present work is that, as far as the authors are aware, it constitutes the first calculation of superstring $n$-point functions on a curved background performed entirely in a manifestly (worldsheet) supersymmetric manner, i.e. without resorting to the use picture-changing operators. Moreover, the method of computing these correlation functions was almost entirely analogous to the corresponding calculations that had recently been performed in bosonic $\text{AdS}_3$ string theory.

This is an extremely promising state of affairs, and suggests that there may be other examples of exactly-solvable bosonic string theories whose supersymmetric counterparts may similarly be solvable by insisting on working directly in superspace.\footnote{A notable example is the computation of the supersymmetric JT gravity path integral using an extension of topological recursion to the moduli space of super Riemann surfaces \cite{Stanford:2019vob}.} For example, the supersymmetric extensions of $(p,q)$ minimal string theories \cite{Seiberg:2003nm}, the Virasoro Minimal String (VMS) \cite{Collier:2023cyw,Johnson:2024fkm,Rangamani:2025wfa,Muhlmann:2025ngz}, and the Complex Liouville String ($\mathbb{C}$LS) \cite{Collier:2024kmo,Collier:2024kwt,Du:2025lya} may be directly solvable using a superspace approach. Particularly interesting are the VMS and $\mathbb{C}$LS, whose bosonic correlation functions can be cleanly expressed in terms of the intersection theory on the Deligne-Mumford compactification $\overline{\mathcal{M}}_{g,n}$ of the moduli space of curves. Perhaps a similar statement is true for their supersymmetric versions.

\paragraph{Correlation functions with Ramond insertions:} In this work, we exclusively considered correlation functions of spectrally flowed NS-sector ground states. There appear to be several technical obstacles to generalizing our work to the computation of correlation functions with R-sector insertions. As we discuss in Appendix \ref{app:srs}, the moduli space of super Riemann surfaces in the presence of NS and (picture $P=-1/2$) R punctures has dimension
\begin{equation}
    \dim \mathfrak{M}_{g,n_{NS},n_{R}} = (3g-3 + n_{NS} + n_{R})|(2g-2 + n_{NS} +\tfrac{1}{2}n_R)\,.
\end{equation}
While there exists a natural forgetful map $\mathfrak{M}_{g,n_{NS},n_R} \to \mathfrak{M}_{g,n_{NS}-1,n_R}$ by integrating over the bosonic and fermionic coordinate of an NS puncture, no such forgetful map exists for the R punctures. Instead, the odd moduli for Ramond punctures are global in nature, meaning they are not simply parameterized by the insertion points of vertex operators. In this way, correlation functions with Ramond punctures are similar to higher-genus correlators. This means we cannot generically remove the need for picture-changing operators in correlation functions with Ramond insertions using the methods developed in the present work.\footnote{One exception to this is $g=0$ with $n_R=2$ and $n_{NS}\geq1$, so our methods may have a simple generalization in this case.}

\acknowledgments
We thank Alexandre Belin, Alejandra Castro, Andrea Dei, Lorenz Eberhardt, Matthias Gaberdiel, Shota Komatsu, Ji Hoon Lee, Kiarash Naderi, Beat Nairz, Ron Reid-Edwards, and Alessandro Sfondrini for useful discussions. We also thank Ofer Aharony and Rajesh Gopakumar for helpful feedback on an early version of the manuscript. The work of BK was supported by STFC consolidated grants ST/T000694/1 and ST/X000664/1. The work of NM was supported by STFC consolidated grants ST/T000694/1 and ST/X000664/1, an EPSRC studentship and the Department of Atomic Energy, Government of India, under project no. RTI4001. The work of VS is supported by a grant from the Swiss National Science Foundation, by the Simons Foundation grant 994306  (Simons Collaboration on Confinement and QCD Strings), by the NCCR SwissMAP that is also funded by the Swiss National Science Foundation, by ISF grant no. 2159/22, by Simons Foundation grant 994296 (Simons Collaboration on Confinement and QCD Strings), by the Minerva foundation with funding from the Federal German Ministry for Education and Research, by the German Research Foundation through a German-Israeli Project Cooperation (DIP) grant ``Holography and the Swampland'', and by the Koshland foundation. 

\appendix

\section{Some super Riemann surface theory}\label{app:srs}

The goal of this appendix is to gather definitions and results in the theory of super Riemann surfaces which are used in the main text. We largely follow the exposition of \cite{Witten:2012ga,Witten:2012bh,Witten:2012bg}.

\subsection{Super Riemann surfaces}\label{app:SRS_subsection}

A \text{complex} supermanifold is a supermanifold $X$ equipped with an even linear map $J:TX\to TX$ such that $J^2=-I$ and $J$ is integrable. Integrability means that the sections of the left- and right-tangent bundles of $X$ (where `left' and `right' is defined with respect to $J$) each form a Lie algebra after reducing modulo odd variables.\footnote{This integrability condition is trivial for the worldsheets of heterotic and type II superstring theory.} The simplest complex supermanifolds are the spaces $\mathbb{C}^{r|s}$, which have global holomorphic coordinates $(z^1,\ldots,z^r|\theta^1,\ldots,\theta^s)$, for which the tangent vectors $\partial_{z^i}$ and $\partial_{\theta^m}$ are eigenvectors of $J$ with eigenvalue $+i$. We say that $X$ has (complex) dimension $r|s$ if it locally looks like $\mathbb{C}^{r|s}$.

A \textit{super Riemann surface} is a complex supermanifold $\Sigma$ of dimension $1|1$ equipped with a superconformal structure -- a subbundle $\mathcal{D}\subset T\Sigma$ which is `completely nonintegrable'. This condition is equivalent to demanding that if $D$ is a nonzero section of $\mathcal{D}$, then $D^2$ is nowhere proportional to $D$. Assuming such a subbundle exists, we can always find local complex coordinates in which a section $D$ takes the form
\begin{equation}\label{eq:D_theta_appendix}
D_{\theta}=\partial_{\theta}+\theta\partial_z\,.
\end{equation}
We include the subscript `$\theta$' here to emphasize the choice of local coordinates. One can readily see that $D_{\theta}^2 = \p_z$, which is linearly independent of $D_{\theta}$.

Superconformal transformations are holomorphic diffeomorphisms of $\Sigma$ which preserve the superconformal structure. 
These are diffeomorphisms
\begin{equation}
    (z|\theta) \mapsto (z'(z|\theta)|\theta'(z|\theta))
\end{equation}
under which $D_{\theta} \mapsto f(z'|\theta')D_{\theta'}$, thus preserving the subbundle $\mathcal{D}$. By the chain rule, $f(z'|\theta') = D_{\theta}\theta'$. Infinitesimal superconformal transformations are generated by vector fields $W$ which satisfy
\begin{equation}
[W,D\}\propto D\,,
\end{equation}
where we recall that the bracket $[\cdot,\cdot\}$ is a commutator if at least one entry is even and an anticommutator if both entries are odd. A convenient basis for the even superconformal transformations is provided by the Virasoro generators
\begin{equation}
    L_n = -z^{n+1}\p_z -\frac{n+1}{2}z^n\theta\p_{\theta}
\end{equation}
with $n\in\mathbb{Z}$, while the odd transformations are generated by the superconformal generators
\begin{equation}
    G_r = z^{r+1/2}(\p_{\theta}-\theta\p_z)
\end{equation}
with $r\in\mathbb{Z}+1/2$. Together, the transformations $L_n$ and $G_r$ form a basis of even and odd superconformal vector fields on $\Sigma$. We denote by $\mathcal{S}$ the sheaf of all such superconformal vector fields.

\vspace{0.2cm}

So far, we have concentrated on local properties of the super Riemann surface, describing local coordinates on an open set $U\subset \Sigma$ for which $U \subset \mathbb{C}^{1|1}$. To construct the full super Riemann surface, $\Sigma$, one must cover $\Sigma$ with such open sets and construct transition functions (using superconformal transformations) between charts, as one would do for bosonic manifolds. Generically, these transition functions between charts $U_{\alpha}$ and $U_{\beta}$ will take the form
\begin{equation}\label{eq:transition_functions}
    \begin{aligned}
        z_{\alpha} &= u_{\alpha\beta}(z_{\beta}) + \theta_{\beta}\zeta_{\alpha\beta}(z_{\beta})\,,\\
        \theta_{\alpha} &= \eta_{\alpha\beta}(z_{\beta}) + \theta_{\beta}v_{\alpha\beta}(z_{\beta})\,,
    \end{aligned}
\end{equation}
where $\zeta_{\alpha\beta}(z_{\beta})$ and $\eta_{\alpha\beta}(z_{\beta})$ are fermionic.\footnote{If one is studying a single SRS, then the $\theta_{\alpha}$ are the only (holomorphic) Grassman odd coordinates. In this case, $\eta_{\alpha\beta}=\zeta_{\alpha\beta}=0$ necessarily. However, in perturbative superstring theory, we are instead studying families of SRSs, where we view the SRSs as fibers over the moduli space of SRSs, $\mathfrak{M}_g$. Then the transition functions are parameterized by the moduli, allowing for non-zero $\eta_{\alpha\beta}$ and $\zeta_{\alpha\beta}$. In this setting, the reduced space is defined with respect to the total space of the SRSs fibered over $\mathfrak{M}_g$, by setting all $\theta_{\alpha}$ and odd moduli to zero. \label{footnote:SRS_gluing}} The generalization to higher-dimensional complex supermanifolds is clear, using coordinate charts with holomorphic coordinates $(z^1_{\alpha},\ldots,z^r_{\alpha}|\theta^1_{\alpha},\ldots,\theta^s_{\alpha})$.

For any complex supermanifold, $X$, one may define the \textit{reduced space}, $X_{\text{red}}$, by setting all of the fermionic coordinates $(\theta^1_{\alpha}, \dots, \theta^s_{\alpha})$ to zero. This is consistent with the gluing rules \eqref{eq:transition_functions} between charts (see footnote \ref{footnote:SRS_gluing}) and hence leads to a bosonic manifold of the same bosonic dimension as $X$. For super Riemann surfaces, we denote the reduced space by $\Sigma_{\text{red}}$. There is a natural embedding map $i: X_{\text{red}} \xhookrightarrow{} X$ that sends the point $(z_1, \dots, z_r)$ in $X_{\text{red}}$ to $(z_1, \dots, z_r|0,\dots,0)$ in $X$ for each chart. One can moreover define a holomorphic projection map $\pi : X \to X_{\text{red}}$, but this is not uniquely specified. We say that the manifold $X$ is \textit{holomorphically projected} if $\pi \circ i = 1$ as a function on $X_{\text{red}}$. Moreover, we say that $X$ is \textit{holomorphically split} if $X$ is holomorphically isomorphic to the total space of a purely fermionic vector bundle $V \to X_{\text{red}}$. On the level of transition functions, $X$ is holomorphically projected if one can construct local coordinate charts such that all bosonic transition functions are independent of the odd variables (i.e. the bosonic transition functions do not mix even and odd variables). It is holomorphically split if, in addition, all fermionic transition functions depend linearly on the odd variables.

Specifically, a super Riemann surface is said to be split if there exist local coordinate charts such that $\zeta_{\alpha\beta}(z_{\beta}) = \eta_{\alpha\beta}(z_{\beta}) =0$ for all transition functions. For a split (or projected) super Riemann surface, one can consistently `forget' the fermionic coordinates and globally project onto $\Sigma_{\text{red}}$. In this setting, $\Sigma$ is given by the total space of the line bundle $\Pi K^{-1/2}$ over $\Sigma_{\text{red}}$ for heterotic superstring theory and the total space of the vector bundle $\Pi K^{-1/2} \oplus \Pi \bar{K}^{-1/2}$ over $\Sigma_{\text{red}}$ for type II superstring theory. That is to say, the local coordinates $\theta$ (and $\bar{\theta}$) describe a fermionic section of $K^{-1/2}$ (and $\bar{K}^{-1/2}$), where $K$ is the canonical bundle of $\Sigma_{\text{red}}$. The choice of square root defines a spin structure on the bosonic submanifold, of which there are $2^{2g}$ inequivalent choices for Riemann surface of genus $g$. Hence, the reduced space $\Sigma_{\text{red}}$ is a bosonic Riemann surface with a choice of spin structure, otherwise known as a `spin curve'. It is often assumed in RNS superstring theory that the worldsheet is a split (or at least projected) super Riemann surface, since this allows one to consistently integrate out the fermionic coordinates and define the theory on the reduced space.

\subsection{Punctures}

We would now like to describe the vertex operators of RNS superstring theory in superspace. These vertex operators are inserted at punctures on the super Riemann surface. For bosonic Riemann surfaces, a puncture is simply a marked point $z=z_i$, however there are two types of puncture on a super Riemann surface corresponding to the NS and R sectors. In this section, we will focus only on the holomorphic sector, such that the discussion is relevant for both type II and heterotic strings.

\paragraph{The NS sector:} Vertex operators in the NS sector take the form 
\begin{equation}\label{eq:NS_vo}
    \mathcal{V}(z_i,\theta_i) = \delta_{1|1}(C)\mathscr{V}(z_i,\theta_i)\,,
\end{equation}
where $C = c + \theta\hat{\gamma}$ are the usual ghosts of RNS superstring theory and $\mathscr{V}$ depends only on the matter sector. This is the canonical choice $P=-1$ for the picture number and the vertex operator is physical provided
\begin{equation}
    \begin{aligned}
        L_n \mathscr{V} &= \frac{1}{2}\delta_{n,0}\mathscr{V}, &\text{for all } n \geq 0\,,\\
        G_r\mathscr{V} &= 0,  &\text{for all } r\geq \frac{1}{2}\,.
    \end{aligned}
\end{equation}
These constraints ensure that the total expression \eqref{eq:NS_vo} is superconformally invariant.

The vertex operator is inserted at an NS puncture, which is simply a marked point $(z_i,\theta_i)$ on the super Riemann surface. This is in close analogy with vertex operators in bosonic string theory. Yet, it is a familiar concept in bosonic string theory that integrated vertex operators also provide a conformally invariant combination. A similar statement holds for NS punctures in superspace: one can define a superconformally invariant vertex operator by integrating over the full super Riemann surface:
\begin{equation}\label{eq:integrated_vo}
    \int_{\Sigma}\mathrm{d}^{2|2}\boldsymbol{z}\, \mathscr{V}(\boldsymbol{z})\,.
\end{equation}

\paragraph{The R sector:} As was discussed in Section \ref{sec:ramond_sector}, vertex operators in the Ramond sector are associated to Ramond punctures, otherwise known as Ramond divisors. These are submanifolds $\mathscr{F}\subset \Sigma$ of dimension $0|1$ on which the superderivative degenerates, $D^2=0$. If the Ramond divisor is at $z=0$, then the superconformal structure $\mathcal{D}$ is locally generated by
\begin{equation}
    D_{\theta} = \p_{\theta} + z\theta\p_z\,,
\end{equation}
such that $D_{\theta}^2 = z\p_z$ which vanishes on the divisor. As discussed in Section \ref{sec:ramond_sector}, one can perform a superconformal transform to bring $D_{\theta}$ back into the canonical form \eqref{eq:D_theta_appendix}, but the coordinates will no longer be single-valued in a neighborhood of the Ramond divisor. In terms of $z$ and $\theta$ a basis of superconformal transformations is
\begin{equation}
\begin{aligned}
    L_n &= -z^{n+1}\p_z -\frac{n}{2}z^n\theta\p_{\theta}\,,\\
    G_r &= z^{r}(\p_{\theta}-z\theta\p_z)\,,
\end{aligned}
\end{equation}
for $n, r\in\mathbb{Z}$. Note that $G_0 = \p_{\theta}$ on $z=0$, such that it generates translations along the divisor.

Given that Ramond punctures are specified by a divisor, there a two natural ways to insert Ramond vertex operators. One either associates it to the whole divisor or one associates to a point on the divisor. The difference amounts toa choice of picture number, with $P=-1/2$ representing the whole divisor and $P=-3/2$ a point on the divisor. Whichever choice one makes will change the moduli space over which one integrates in the path integral --- see Section \ref{app:moduli-space}. In particular, the $P=-1/2$ expression is recovered from $P=-3/2$ by integrating over the divisor. That is to say,
\begin{equation}
    \mathscr{V}(z,\theta_{}) 
        = \mathscr{V}(z) + \theta G_0 \mathscr{V}(z) \,.
\end{equation}
encodes the $P=-1/2$ expression $G_0\mathscr{V}$ as in \eqref{eq:P=-3/2_expansion}. Note that this is consistent with $G_0 = \p_{\theta}$ on the divisor. In picture $P=-3/2$, the physical state conditions are
\begin{equation}
    \begin{aligned}
        L_n \mathscr{V} &= \frac{5}{8}\delta_{n,0}\mathscr{V}, &\text{for all } n \geq 0\,,\\
        G_r\mathscr{V} &= 0,  &\text{for all } r\geq 1\,,
    \end{aligned}
\end{equation}
whilst in $P=-1/2$ we additionally require $G_0\mathscr{V}=0$. Under these conditions, both $ce^{-3\varphi/2}\mathscr{V}(z_i,\theta_i)$ and $ce^{-\varphi/2}\mathscr{V}(z_i)$ are superconformally invariant. However, there does not exist an integrated vertex operator analogous to \eqref{eq:integrated_vo} which removes one even and one odd modulus from the moduli space.

\subsection{The moduli space of super Riemann surfaces}\label{app:moduli-space}

It is a familiar concept in bosonic string theory that one integrates over the moduli space of Riemann surfaces, the space of complex structures modulo diffeomorphisms. Analogously, in RNS superstring theory, one integrates over the moduli space of super Riemann surfaces, the space of supercomplex structures modulo superdiffeomorphisms. A detailed understanding of this moduli space is not required for the current work. What is important for our needs is the dimension of the moduli space and for this reason, we refer the interested reader to the references \cite{Witten:2012bh,Witten:2012ga} for details beyond dimension counting.

Let us first consider the case of a genus $g$ super Riemann surface in the absence of punctures. The associated moduli space, $\mathfrak{M}_g$, can be studied through deformation theory, where one studies infinitesimal deformations of the gluing rules \eqref{eq:transition_functions}. For a given super Riemann surface $\Sigma$, such deformations live in the tangent space
\begin{equation}\label{eq:tangent_of_moduli_space}
    T\mathfrak{M}_g|_{\Sigma} = \text{H}^{1}(\Sigma,\mathcal{S})\,,
\end{equation}
where $\mathcal{S}$ is the sheaf of superconformal vector fields on $\Sigma$. 
The dimension of this space is invariant under changes to the odd moduli of $\Sigma$, and hence can be computed at a convenient point where $\Sigma$ is a split super Riemann surface. The tangent space then decomposes $T\mathfrak{M}_g = T_+\mathfrak{M}_g \oplus T_-\mathfrak{M}_g$ into even and odd subspaces. The dimension of each of these subspaces is readily computed using the Riemann-Roch theorem. Typically, in superstring perturbation theory, we focus on cases where the moduli space has no continuous automorphisms i.e. we fix global superconformal transformations. In the absence of punctures, this is given by $g\geq 2$, for which the Riemann-Roch theorem implies
\begin{equation}
    T_+\mathfrak{M}_g|_{\Sigma} = \text{H}^1(\Sigma_{\text{red}},T\Sigma_{\text{red}}) = T\mathcal{M}_g|_{\Sigma_{\text{red}}}\,
\end{equation}
is of dimension $3g-3$, whilst
\begin{equation}
    T_-\mathfrak{M}_g|_{\Sigma} = \Pi\,\text{H}^1\left(\Sigma_{\text{red}},T\Sigma_{\text{red}}^{1/2}\right)\,,
\end{equation}
is of dimension $2g-2$. In the above, $\mathcal{M}_g$ denotes the moduli space of bosonic Riemann surfaces, whilst $\Pi$ denotes parity reversal. Thus, for $g\geq 2$, we have that
\begin{equation}
    \dim \mathfrak{M}_g = (3g-3)|(2g-2)\,.
\end{equation}

One can also consider the moduli space of punctured super Riemann surfaces, $\mathfrak{M}_{g,n_{\text{NS}},n_{\text{R}}}$, with $n_{\text{NS}}$ NS punctures and $n_{\text{R}}$ R punctures.\footnote{When there are no Ramond punctures, we simply write $\mathfrak{M}_{g,n}$.} The NS punctures are given by points $(z_i,\theta_i)$ on $\Sigma$ and these points can be varied independently of all other moduli. Thus, an NS puncture increases the dimension of the moduli space by $1|1$. This is inherently related to the existence of integrated NS-sector vertex operators \eqref{eq:integrated_vo}. This integral over the location of the NS puncture acts as a forgetful map
\begin{equation}
    \mathfrak{M}_{g,n_{\text{NS}},n_{\text{R}}} \to \mathfrak{M}_{g,n_{\text{NS}}-1,n_{\text{R}}}\,.
\end{equation}
However, such a forgetful map does not exist for Ramond punctures. Whilst it is true that the even modulus is parameterised by the location $z_i$ of the Ramond puncture on the worldsheet, the odd moduli are much more intricate. Indeed, Ramond punctures correspond to degenerations of the superconformal structure such that they are an intrinsic part of the geometry of the super Riemann surface. For instance, in the presence of a Ramond divisor, $\mathscr{F}$, one would have to restrict the sheaf $\mathcal{S}$ in \eqref{eq:tangent_of_moduli_space} to superconformal vector fields on $\Sigma$ that leave $\mathscr{F}$ fixed. This adds an additional $\frac{n_{\text{R}}}{2}$ odd moduli. In particular, any odd modulus associated to a particular Ramond divisor cannot be varied independently of all other odd moduli. A second point of note is that there are two ways to define Ramond vertex operators and thus two possible moduli spaces. If we associate each vertex operator to the full Ramond divisor (the $P=-1/2$ prescription), then each operator contributes one even modulus along with the $\frac{1}{2}$ odd modulus mentioned above. The total moduli space has dimension
\begin{equation}
    \dim \mathfrak{M}_{g,n_{\text{NS}},n_{\text{R}}} = (n_{\text{NS}} + n_{\text{R}} + 3g-3)|(n_{\text{NS}} + \tfrac{1}{2}n_{\text{R}} + 2g-2)\,.
\end{equation}
If we additionally label each vertex operator by a point on the divisor, we add $n_{\text{R}}$ odd moduli. We denote this second moduli space by $\mathfrak{M}_{g,n_{\text{NS}},n_{\text{R}}}'$ and it has dimension
\begin{equation}
    \dim \mathfrak{M}_{g,n_{\text{NS}},n_{\text{R}}}' = (n_{\text{NS}} + n_{\text{R}} + 3g-3)|(n_{\text{NS}} + \tfrac{3}{2}n_{\text{R}} + 2g-2)\,.
\end{equation}
Of course, there exists a natural forgetful map $\mathfrak{M}_{g,n_{\text{NS}},n_{\text{R}}}' \to \mathfrak{M}_{g,n_{\text{NS}},n_{\text{R}}}$ by integrating over the locations on the divisors.
\\

We can now return to an issue raised at the end of \ref{app:SRS_subsection}, which is that it is commonplace in the study of RNS superstring theory to assume that the worldsheet is a split (or at least projected) super Riemann surface. One then typically integrates out all of the odd coordinates and moduli and gauge-fixes the worldsheet theory. This includes a gauge-fixing of the worldsheet gravitino, which inserts picture changing operators at particular points on the worldsheet.

Yet, there are several shortcomings to this approach. Strictly speaking, in RNS superstring theory we are studying a family of super Riemann surfaces parameterized by the moduli space of punctured super Riemann surfaces, $\mathfrak{M}_{g,n_{\text{NS}},n_{\text{R}}}$, which we integrate over. The gauge choices made for the gravitino will not generically be consistent throughout this moduli space implying, for instance, that we cannot fix the locations of PCOs globally across $\mathfrak{M}_{g,n_{\text{NS}},n_{\text{R}}}$. This shortcoming of the picture changing formalism can be remedied through the use of `vertical integration' \cite{Sen:2014pia,Sen:2015hia}, where one takes different gauge choices in different regions of $\mathfrak{M}_{g,n_{\text{NS}},n_{\text{R}}}$. However, there is an even more fundamental problem which is that $\mathfrak{M}_{g,n_{\text{NS}},n_{\text{R}}}$ generically is not a split (nor projected) supermanifold \cite{Donagi:2013dua}. Hence, the transition functions between charts on the super Riemann surface will generically mix the even coordinates on the SRS with odd moduli. In other words, in general, one cannot consistently integrate out all of the odd variables and reduce RNS superstring theory to the study of a worldsheet CFT defined on a spin curve, combined with an integral over the (bosonic) moduli space of (punctured) spin curves.

\subsection{Superstring perturbation theory}

Analogously to bosonic string theory, amplitudes in superstring theory with NS-sector vertex operators are defined via integration over the moduli space $\mathfrak{M}_{g,n}$.\footnote{Strictly speaking, one actually integrates over a cycle $\Gamma\subset\mathfrak{M}_{g,n}\times\overline{\mathfrak{M}}_{g,n}$ which is `close' to the diagonal. This accounts for the fact that the left- and right-moving fermionic moduli may not be complex conjugates.} As such, we need to define a measure in the moduli space to integrate over. As in bosonic string theory, this is achieved with the help of a ghost system that gauges the superconformal symmetry on the worldsheet. Since we only work at tree level, we briefly explain this procedure, referring the reader to \cite{Witten:2012bh} for a much more detailed account of the subject.

\paragraph{The superconformal ghost system:} The ghost system on the worldsheet consists of two superfields $B,C$ with action
\begin{equation}
\int_{\Sigma} \mathrm{d}^{2|2}\boldsymbol{z}\, B\overline{D}C\,.
\end{equation}
Since we are gauging the superconformal symmetry, which is generated by an anticommuting superfield $\mathscr{T}$ of weight $(\frac{3}{2},0)$, the $B$ ghost is a commuting superfield of weight $(\frac{3}{2},0)$. Meanwhile the $C$ ghost is anticommuting and has weight $(-1,0)$, so as to make the action well-defined. Expanding the ghosts in superspace, we have
\begin{equation}
B=\hat{\beta}+\theta b\,,\quad C=c+\theta\hat{\gamma}\,,
\end{equation}
where $b,c$ and $\hat\beta,\hat\gamma$ are the familiar commuting and anticommuting superconformal ghosts.

\paragraph{A measure on the moduli space:} As discussed above, the tangent space to the moduli space $\mathfrak{M}_{g,n}$ near a surface $\Sigma$ is naturally isomorphic to the cohomology group $\text{H}^1(\Sigma,\mathcal{S})$. In terms of differential forms, its parity reversal, $\Pi\text{H}^1(\Sigma,\mathcal{S})$, is spanned by super-Beltrami differentials $M$, which in superspace can be taken to have the form
\begin{equation}
M=\bar{\theta}\mu\indices{^z_{\bar{z}}}+\theta\bar\theta\chi\indices{^\theta_{\bar{z}}}\,,
\end{equation}
where $\mu\indices{^z_{\bar{z}}}$ is an ordinary Beltrami differential and $\chi\indices{^\theta_{\bar{z}}}$ is an infinitesimal deformation in the worldsheet gravitino (see for example \S 3J of \cite{DHoker:1988pdl}). Under an infinitesimal change in the worldsheet moduli, the worldsheet action responds via the stress tensor:
\begin{equation}
S\to S+\frac{1}{2\pi}\int_{\Sigma}\mathrm{d}^{2|2}\boldsymbol{z}\,M\mathscr{T}\,.
\end{equation}
Now, near the surface $\Sigma$, we can choose a local coordinate system $(\vec{\tau}|\vec{\sigma})$ on $\mathfrak{M}_{g,n}$. For small deformations, we can parametrize the Beltrami differentials as
\begin{equation}
M=\sum_{a=1}^{n+3g-3}\delta\tau_aM_a+\sum_{\alpha=1}^{n+2g-2}\delta\sigma_{\alpha}M_{\alpha}
\end{equation}
with $M_a$ anticommuting and $M_{\alpha}$ commuting. We can think of $M_a$ and $M_{\alpha}$ as odd and even basis elements of $\Pi T_{\Sigma}\mathfrak{M}_{g,n}$.

Given a particular set of coordinates near $\Sigma$, we can couple the $B$ ghosts to the basis elements $M_a,M_{\alpha}$ by constructing the non-local operators
\begin{equation}
\Psi_{a}=\frac{1}{2\pi}\int_{\Sigma}\mathrm{d}^{2|2}\boldsymbol{z}\,M_{a}B\,,\quad\Phi_{\alpha}=\frac{1}{2\pi}\int_{\Sigma}\mathrm{d}^{2|2}\boldsymbol{z}\,M_{\alpha}B\,.
\end{equation}
The measure on the moduli space is then given by the delta function\footnote{This is essentially the content of equations (12.5.24) and (12.5.25) of \cite{Polchinski:1998rr}, phrased in the language of Beltrami differentials.}
\begin{equation}
\delta^{(2)}(\Phi_{1},\ldots,\Phi_{n+2g-2}|\Psi_1,\ldots,\Psi_{n+3g-3})\,.
\end{equation}
This delta function defines a top form on the moduli space $\mathfrak{M}_{g,n}$ that can be integrated against.

\paragraph{The sphere measure:} We can use the general framework above to define a measure on $\mathfrak{M}_{0,n}$ to compute tree-level diagrams. Since we can use the global superconformal symmetry to fix the points $z_1,z_2,z_3$ and $\theta_1,\theta_2$, the coordinates on the moduli space are the remaining points $(z_4,\ldots,z_n|\theta_3,\ldots,\theta_n)$. An infinitesimal variation in the bosonic position of an operator is accomplished by inserting the contour integral
\begin{equation}
\delta z_i\oint_{\mathscr{F}_i}\frac{\mathrm{d}z\,\mathrm{d}\theta}{2\pi i}\mathscr{T}(z|\theta)=\frac{\delta z_i}{2\pi}\int_{\Sigma}\mathrm{d}^{2|2}\boldsymbol{z}\,\mathscr{T}(z|\theta)\,\bar\theta\,\overline{\partial}\delta(z-z_i-\theta\theta_i)
\end{equation}
into the path integral, where the contour is taken around the divisor $\mathscr{F}_i$ defined by the equation $z-z_i-\theta\theta_i=0$, and we have used Stokes' theorem to write the result as an integral over $\Sigma$. Since the divisor $\mathscr{F}_i$ is generated by the action of $D_{\theta}$ on the point $(z_i,\theta_i)$, this deformation manifestly preserves the superconformal structure of $\Sigma$. This provides us with a set
\begin{equation}
M_a=\bar\theta\,\overline{\partial}\delta(z-z_a-\theta\theta_a)\,,\quad a=4,\ldots,n\,.
\end{equation}
of $n-3$ super-Beltrami differentials associated to the even moduli. Similarly, shifting the coordinate $\theta_i$ is achieved by the contour integral
\begin{equation}
\begin{split}
&\delta\theta_i\oint_{\mathscr{F}_i}\frac{\mathrm{d}z\,\mathrm{d}\theta}{2\pi i}\,(\theta-\theta_i)\mathscr{T}(z|\theta)\\
&\hspace{2cm}=\frac{\delta\theta_i}{2\pi}\int_{\Sigma}\mathrm{d}^{2|2}\boldsymbol{z}\,\mathscr{T}(z|\theta)\,\bar\theta(\theta-\theta_i)\overline{\partial}\delta(z-z_i-\theta\theta_i)\,,
\end{split}
\end{equation}
which yields a basis
\begin{equation}
M_{\alpha}=\bar\theta(\theta-\theta_\alpha)\overline{\partial}\delta(z-z_\alpha-\theta\theta_\alpha)\,,\quad\alpha=3,\ldots,n\,.
\end{equation}
of $n-2$ super-Beltrami differentials associated to the odd moduli.

With the Beltrami differentials in hand, we can define the differentials by integrating against the $B$-ghosts, which yields the result
\begin{equation}
\Psi_a=\oint_{\mathscr{F}_a}\frac{\mathrm{d}z\,\mathrm{d}\theta}{2\pi i}B(z|\theta)\,,\quad\Phi_{\alpha}=\oint_{\mathscr{F}_\alpha}\frac{\mathrm{d}z\,\mathrm{d}\theta}{2\pi i}(\theta-\theta_\alpha)B(z|\theta)
\end{equation}
Now, a matter vertex operator in the NS sector is an anticommuting primary superfield $\mathscr{V}$ with conformal weight $(1/2,1/2)$. The basic partition function for a fixed set of worldsheet moduli is the `unintegrated' correlator
\begin{equation}
\left\langle\prod_{i=1}^{n}\delta^{(2)}_{1|1}(C)\mathscr{V}_i(z_i,\bar{z}_i|\theta_i,\bar\theta_i)\right\rangle\,.
\end{equation}
To integrate this over the moduli space, we insert the delta function
\begin{equation}
\prod_{i=4}^{n}\delta^{(2)}\left(\oint_{\mathscr{F}_i}\frac{\mathrm{d}z\,\mathrm{d}\theta}{2\pi i}B(z|\theta)\right)\prod_{i=3}^{n}\delta^{(2)}\left(\oint_{\mathscr{F}_i}\frac{\mathrm{d}z\,\mathrm{d}\theta}{2\pi i}(\theta-\theta_i)B(z|\theta)\right)
\end{equation}
into the path integral, then integrate over the moduli $z_4,\ldots,z_n|\theta_3,\ldots,\theta_n$. The contour integrals around the insertion points can be performed with the aid of the OPEs between the $B$ and $C$ ghosts, and we end up with the integrated correlator
\begin{equation}
\int_{\mathfrak{M}_{0,n}}\mathrm{d}\mu\,\left\langle\prod_{i=1}^{n}\mathscr{V}_i(z_i|\theta_i)\right\rangle\,,
\end{equation}
where the measure is given by the ghost correlator
\begin{equation}
\mathrm{d}\mu=\mathrm{d}^2z_4\cdots\mathrm{d}^2z_n\mathrm{d}^2\theta_3\cdots\mathrm{d}^2\theta_n\braket{\delta_{{1|1}}(C)(\boldsymbol{z}_1)\delta_{{1|1}}(C)(\boldsymbol{z}_2)\delta(C)(\boldsymbol{z}_3)}\,.
\end{equation}

\paragraph{Computing the ghost correlator:} The final step is to compute the ghost correlator
\begin{equation}
\braket{\delta_{{1|1}}(C)(\boldsymbol{z}_1)\delta_{{1|1}}(C)(\boldsymbol{z}_2)\delta(C)(\boldsymbol{z}_3)}\,.
\end{equation}
Expanding the delta functions
\begin{equation}
\delta(C)=c+\theta\hat\gamma\,,\quad\delta_{1|1}(C)=c\delta(\hat\gamma)-\theta c\partial c\delta'(\hat\gamma)\,
\end{equation}
and keeping in mind that only correlators with three $c$ insertions are non-vanishing, the correlator can be reduced to a sum of $c$ and $\hat{\gamma}$ correlators: 
\begin{equation}
\begin{split}
\Braket{c(z_1)c(z_2)c(z_3)}\braket{\delta(\hat\gamma)(z_1)\delta(\hat\gamma)(z_2)}&+\theta_1\theta_3\Braket{(c\partial c)(z_1)c(z_2)}\braket{\delta'(\hat\gamma)(z_1)\delta(\hat\gamma)(z_2)\hat\gamma(z_3)}\\
&-\theta_2\theta_3\braket{c(z_1)(c\partial c)(z_2)}\Braket{\delta(\hat\gamma)(z_1)\delta'(\hat\gamma)(z_2)\hat\gamma(z_3)}\,.
\end{split}
\end{equation}
The $c$ correlators are readily computed using Wick contractions, and we are left with
\begin{equation}
\begin{split}
z_{12}z_{23}z_{13}\braket{\delta(\hat\gamma)(z_1)\delta(\hat\gamma)(z_2)}&+\theta_1\theta_3z_{12}^2\braket{\delta'(\hat\gamma)(z_1)\delta(\hat\gamma)(z_2)\hat\gamma(z_3)}\\
&-\theta_2\theta_3z_{12}^2\Braket{\delta(\hat\gamma)(z_1)\delta'(\hat\gamma)(z_2)\hat\gamma(z_3)}\,.
\end{split}
\end{equation}
Finally, the correlators of the $\hat\gamma$ ghost can be computed using the methods of Section 10 of \cite{Witten:2012bh}:\footnote{Alternatively, one can bosonize the $\hat{\beta},\hat\gamma$ system and use the relations $\hat\gamma=\eta e^{\varphi}$, $\delta(\hat\gamma)=e^{-\varphi}$, and $\delta'(\hat\gamma)=\partial\xi\,e^{-2\varphi}$, following the conventions of Chapter 13 of \cite{Blumenhagen:2013fgp}.} 
\begin{equation}
\begin{split}
\braket{\delta(\hat\gamma)(z_1)\delta(\hat\gamma)(z_2)}&=\frac{1}{z_{12}}\,,\\
\braket{\delta'(\hat\gamma)(z_1)\delta(\hat\gamma)(z_2)\hat\gamma(z_3)}&=-\frac{z_{23}}{z_{12}^2}\,,\\
\Braket{\delta(\hat\gamma)(z_1)\delta'(\hat\gamma)(z_2)\hat\gamma(z_3)}&=\frac{z_{13}}{z_{12}^2}\,.
\end{split}
\end{equation}
Putting this all together, we find the superconformal ghost correlator
\begin{equation}
\braket{\delta_{{1|1}}(C)(\boldsymbol{z}_1)\delta_{{1|1}}(C)(\boldsymbol{z}_2)\delta(C)(\boldsymbol{z}_3)}=(z_1-z_3-\theta_1\theta_3)(z_2-z_3-\theta_2\theta_3)\,,
\end{equation}
which gives the measure \eqref{eq:m0n-measure} on $\mathfrak{M}_{0,n}$.

\section{The path integral measure}\label{app:measure}

In this appendix, we derive equation \eqref{eq:change of variables in the sl(2,R) measure} as well as discuss the analogous result in the heterotic theory. As a crucial step towards this goal, we will also describe the detail in going from equation \eqref{eq:superspace action} to equation \eqref{eq:superspace action with auxiliary fields}.

\subsection{Superspace actions and their path integral measures}
The action \eqref{eq:superspace action} can be written in terms of component fields as
\begin{equation}
\begin{split}
S_{\text{II}}=&\frac{L^2}{2\pi\alpha'}\int\mathrm{d}^{2|2}\boldsymbol{z}\left(D\Phi\overline{D}\Phi+e^{2\Phi}D\bar\Gamma\bar{D}\Gamma\right)\\
=&\frac{k}{2\pi}\int \mathrm{d}^2z\Bigl( (\partial\phi\bar\partial\phi-\lambda\bar\partial\lambda-\bar\lambda\partial\bar\lambda+F^2)+e^{2\phi}(-\tilde{\psi}^{\bar\gamma}\bar\partial\psi^\gamma-\tilde{\psi}^\gamma\partial\psi^{\bar\gamma}+F^\gamma F^{\bar\gamma}\\
&+\partial\bar\gamma\bar\partial\gamma+2\lambda\tilde{\psi}^{\bar\gamma}\bar\partial\gamma-2\lambda\tilde{\psi}^\gamma F^{\bar\gamma}+2\bar\lambda\tilde{\psi}^{\bar\gamma}F^\gamma+2\bar\lambda\tilde{\psi}^\gamma\partial\bar\gamma+(2F-4\lambda\bar\lambda)\tilde{\psi}^{\bar\gamma}\tilde{\psi}^\gamma) \Bigr)\,.
\label{}
\end{split}
\end{equation}
One can then integrate out the auxiliary fields $F$, $F^\gamma$, and $F^{\bar\gamma}$, which is equivalent to substituting the equations of motion
\begin{equation}
\begin{gathered}
F=e^{2\phi}\tilde{\psi}^\gamma\tilde{\psi}^{\bar\gamma},\quad F^\gamma=2\lambda\tilde{\psi}^\gamma,\quad F^{\bar\gamma}=-2\bar\lambda\tilde{\psi}^{\bar\gamma}\,,
\end{gathered}
\end{equation}
into the action yielding\footnote{Note that, after integrating out the axiliary fields, there is no fermion quartic term as is expected from a more general analysis \cite{Bergshoeff:1985qr}.}
\begin{equation}
\begin{split}
S_{\text{II}}=\frac{k}{2\pi}\int \mathrm{d}^2z\Bigl(& (\partial\phi\bar\partial\phi-\lambda\bar\partial\lambda-\bar\lambda\partial\bar\lambda)\\
&+e^{2\phi}(-\tilde{\psi}^{\bar\gamma}\bar\partial\psi^\gamma-\tilde{\psi}^\gamma\partial\psi^{\bar\gamma}+\partial\bar\gamma\bar\partial\gamma+2\lambda\tilde{\psi}^{\bar\gamma}\bar\partial\gamma+2\bar\lambda\tilde{\psi}^\gamma\partial\bar\gamma) \Bigr)\,.
\label{}
\end{split}
\end{equation}
Next, we can integrate in the auxiliary fields $\beta,\bar\beta$ and rewrite the action as\footnote{The actual (non-zero) value of the coefficient of the interaction term $e^{-2\phi}\cdots$ has no physical meaning. It is only important to distinguish whether the coefficient is vanishing or not. In particular, one can rescale the coefficient by performing the shift $\phi\to\phi+c$ where $c$ is a (real) constant. This shift also rescales $\omega$ and $\bar\omega$ as can be seen from the definition, however, this rescaling can be absorbed into the normalization of the path integral. For this reason, we have included an arbitrary parameter $\mu$ in \eqref{eq:superspace action with auxiliary fields}.}
\begin{equation}
\begin{split}
S_{\text{II}}=&\frac{1}{2\pi}\int \mathrm{d}^2z\Bigl( k(\partial\phi\bar\partial\phi-\lambda\bar\partial\lambda-\bar\lambda\partial\bar\lambda)+ke^{2\phi}(-\tilde{\psi}^{\bar\gamma}\bar\partial\psi^\gamma-\tilde{\psi}^\gamma\partial\psi^{\bar\gamma})\\
&+\beta\bar\partial\gamma+\bar\beta\partial\bar\gamma-\frac{1}{k} e^{-2\phi}(\beta-2ke^{2\phi}\lambda\tilde{\psi}^{\bar\gamma})(\bar\beta-2ke^{2\phi}\bar\lambda\tilde{\psi}^\gamma) \Bigr)\,.
\label{}
\end{split}
\end{equation}
By identifying
\begin{equation}
\omega:=ke^{2\phi}\tilde{\psi}^{\bar\gamma}\,,\quad\bar\omega:=ke^{2\phi}\tilde{\psi}^\gamma
\label{eq:definition of omega in terms of tilde fermions}
\end{equation}
(as well as $\psi:=\psi^\gamma$, $\bar\psi:=\psi^{\bar\gamma}$), rescaling the fields so that the kinetic terms are canonically normalized, and now defining the (anti-)chiral superfields
\begin{equation}
\begin{gathered}
\Gamma=\gamma+\theta\psi\,,\quad \Omega=\omega+\theta\beta\,,\quad \bar\Gamma=\bar\gamma+\bar\theta\bar\psi\,,\quad \bar\Omega=\bar\omega+\bar\theta\bar\beta\,,
\end{gathered}
\end{equation}
we see that we arrive at the action \eqref{eq:superspace action with auxiliary fields}.

\subsubsection*{The measures}
The superspace action \eqref{eq:superspace action}, after integrating out the auxiliary fields $F,F^\gamma,F^{\bar\gamma}$, comes with the natural path integral measure which explicitly reads
\begin{equation}
\begin{split}
\mathcal{D}(e^{\Phi}\Gamma)\mathcal{D}(e^{\Phi}\bar{\Gamma})\mathcal{D}\Phi=&\mathcal{D}\phi\mathcal{D}\lambda\mathcal{D}\bar\lambda\mathcal{D}(e^{\phi}\gamma)\mathcal{D}(e^{\phi}\bar{\gamma})\mathcal{D}(e^{\phi}\psi^\gamma)\mathcal{D}(e^{\phi}\psi^{\bar{\gamma}})\mathcal{D}(e^{\phi}\tilde{\psi}^\gamma)\mathcal{D}(e^{\phi}\tilde{\psi}^{\bar{\gamma}})\\
=&\mathcal{D}\phi\mathcal{D}\lambda\mathcal{D}\bar\lambda\mathcal{D}(e^{\phi}\gamma)\mathcal{D}(e^{\phi}\bar{\gamma})\mathcal{D}(e^{\phi}\psi)\mathcal{D}(e^{\phi}\bar\psi)\mathcal{D}(e^{-\phi}\omega)\mathcal{D}(e^{-\phi}\bar\omega)\,.
\end{split}
\end{equation}
However, our eventual goal is to be able to treat the fields $\gamma,\bar\gamma,\psi,\bar\psi,\omega,\bar\omega$ as free fields, hence, it is crucial to change the variables so that the path integral measure becomes that of the free field theory. More explicitly, we want to perform the following change of variables
\begin{equation}
\begin{split}
&\mathcal{D}\phi\mathcal{D}\lambda\mathcal{D}\bar\lambda\mathcal{D}(e^{\phi}\gamma)\mathcal{D}(e^{\phi}\bar{\gamma})\mathcal{D}(e^{\phi}\psi)\mathcal{D}(e^{\phi}\bar\psi)\mathcal{D}(e^{-\phi}\omega)\mathcal{D}(e^{-\phi}\bar\omega)\\
&={\cal J}\mathcal{D}\phi\mathcal{D}\lambda\mathcal{D}\bar\lambda\mathcal{D}\gamma\mathcal{D}\bar{\gamma}\mathcal{D}\psi\mathcal{D}\bar\psi\mathcal{D}\omega\mathcal{D}\bar\omega\\
&:={\cal J}\mathcal{D}\Gamma\,\mathcal{D}\bar\Gamma\,\mathcal{D}\Phi\,.
\end{split}
\end{equation}
The Jacobian for the bosonic part has long been calculated \cite{Gerasimov:1990fi,Ishibashi:2000fn,Gawedzki:1989rr} and the result is
\begin{equation}
\begin{split}
\mathcal{D}\phi\mathcal{D}(e^{\phi}\gamma)\mathcal{D}(e^{\phi}\bar{\gamma})=\mathcal{D}\phi\mathcal{D}\gamma\mathcal{D}\bar{\gamma}\exp\left( \frac{1}{4\pi}\int \mathrm{d}^2z\,(4\partial\phi\bar\partial\phi+R\phi) \right)\,.
\end{split}
\end{equation}
To calculate the Jacobian for the fermionic part, we make use of the chiral anomaly calculation. Firstly, we claim that, it is natural to package the fermions into 2 2-component spinors as follows\footnote{We take the convention that $\sigma_1=\begin{pmatrix}
0 & 1\\
1 & 0
\end{pmatrix}$, $\sigma_2=\begin{pmatrix}
0 & -i\\
i & 0
\end{pmatrix}$, and $\sigma_3=\gamma^5={\rm diag}(1,-1)$. Note that they satisfy $\sigma_i\sigma_j=\delta_{ij}+i\epsilon_{ijk}\sigma_k$.}
\begin{equation}
\boldsymbol{ \psi}=
\begin{pmatrix}
\psi\\
\bar\omega
\end{pmatrix},\quad
\boldsymbol{\bar{ \psi}}
=
\begin{pmatrix}
\bar\psi & \omega
\end{pmatrix}\,.
\end{equation}
The intuition is that, in the $\text{SL}(2,\mathbb{R})^{(1)}$ WZW model, an element $g$ transforms under the left/right actions as $g\to h_Lgh^{-1}_R$. Hence, this motivates the definition of the diagonal Cartan element to be $J^3_{\rm diag}=J^3_L-J^3_R$ instead of a sum as one may naively expect. Therefore, it is then natural to form spinors such that they transform homogeneously under the diagonal action. Rewriting the measure and the change of variables in terms of spinors, we see that
\begin{equation}
\begin{split}
\mathcal{D}(e^{\phi}\psi)\mathcal{D}(e^{\phi}\bar\psi)\mathcal{D}(e^{-\phi}\omega)\mathcal{D}(e^{-\phi}\bar\omega)=&\mathcal{D}(e^{\phi\gamma^5}\boldsymbol{\psi})\mathcal{D}(\boldsymbol{\bar\psi}e^{\phi\gamma^5})\\
=&{\cal J}_\text{chiral anomaly}\mathcal{D}\boldsymbol{\psi}\mathcal{D}\boldsymbol{\bar\psi}\,,
\end{split}
\end{equation}
where $\gamma^5={\rm diag}(1,-1)$. We want to emphasize that the chiral anomaly we need has to be valid for finite $\phi$, in contrast to the usual chiral anomaly which is normally calculated up to the first leading order. We will see that this leads to including the subleading term which is quadratic in $\phi$ and is the crucial contribution.

Let's now review the computation of the chiral anomaly in 2d, we mainly follow \cite{Fujikawa:2004cx,Weinberg:1996kr}. The action is 
\begin{equation}
S[\boldsymbol{\psi},A_i]=\frac{1}{2\pi}\int \mathrm{d}^2x\boldsymbol{\bar{\psi}}(\slashed{\partial}-\slashed{A})\boldsymbol{\psi}\,,
\end{equation}
where $\slashed{\partial}=\sigma_i\partial_i$ and similarly for $\slashed{A}$.
Writing in components, we have
\begin{equation}
S=\frac{1}{2\pi}\int \mathrm{d}^2z(\omega\overline{\partial}\psi+\bar\omega\partial\bar\psi-A\omega\psi+\bar A\bar\omega\bar{\psi})\,.
\end{equation}
Here, we use that
\begin{equation}
\begin{gathered}
2A=A_1+iA_2,\quad 2\bar A=A_1-iA_2\,,
\end{gathered}
\end{equation}
and note also that
\begin{equation}
\mathrm{d}^2z=2\mathrm{d}^2x\,.
\end{equation}
The Jacobian ${\cal J}$ from the change of variables
\begin{equation}
\mathcal{D}(e^{\phi\gamma^5}\boldsymbol{\psi})\mathcal{D}(\boldsymbol{\bar\psi}e^{\phi\gamma^5})={\cal J}\mathcal{D}\boldsymbol{\psi}\mathcal{D}\boldsymbol{\bar\psi}
\end{equation}
is
\begin{equation}
{\cal J}=\left[{\rm det}\left( e^{2\phi\gamma^5}\delta^2(x-y) \right)\right]^{-1}=\exp\left( -{\rm Tr}\left( 2\phi(x)\gamma^5\delta^2(x-y) \right) \right)\,.
\label{eq:Jacobian to first nontrivial order}
\end{equation}
We regularize the trace as
\begin{equation}
\delta^2(x-x)=\frac{1}{4\pi^2}\int \mathrm{d}^2k e^{k\cdot x}f\left(\frac{\slashed{D}^2}{M^2}\right)e^{-k\cdot x}\,,
\end{equation}
where
\begin{equation}
\slashed{D}:=\sigma_i(\partial_i-A_i)=\sigma_iD_i\,,
\end{equation}
and $M$ is the regularization scale that will be taken to infinity.
Here, we take the regulator $f(x)$ to have the properties
\begin{equation}
f(0)=1,\quad f(\infty)=f'(\infty)=0,\quad\lim_{x\to\infty}xf'(x)=0\,.
\label{eq:smoothness property}
\end{equation}
The regularized delta function becomes
\begin{equation}
\delta^2(x-x)=\frac{1}{4\pi^2}\int \mathrm{d}^2k f\left(\frac{(\slashed{D}-\slashed{k})^2}{M^2}\right)\,,
\end{equation}
again, rescaling $k\to Mk$, we get
\begin{equation}
\delta^2(x-x)=\frac{M^2}{4\pi^2}\int \mathrm{d}^2k f\left(\frac{\slashed{D}^2}{M^2}-\frac{2\slashed{k}\slashed{D}}{M}+k^2 \right)\,.
\end{equation}
Noting that the trace in equation \eqref{eq:Jacobian to first nontrivial order} involves $\gamma^5=\sigma_3$ and that any terms with odd power of $k$ integrate to zero, the only non-vanishing terms are
\begin{equation}
{\rm Tr}\left( 2\phi(x)\gamma^5\delta^2(x-y) \right)={\rm tr}\left( 2\int \mathrm{d}^2x\,\phi(x)\gamma^5\frac{1}{4\pi^2}\int d^2k(f'(k^2)+2k^2f''(k^2))\slashed{D}^2 \right)\,.
\end{equation}
Here, we use $\rm Tr$ to mean taking trace over the spinor indices \emph{and} over $x$ whereas $\rm tr$ means tracing only over the spinor indices. Doing the integration by parts and using the smoothness property \eqref{eq:smoothness property}, we get
\begin{equation}
{\cal J}(\phi,A_i)=\exp\left( -\int \mathrm{d}^2x\,\frac{\phi}{2\pi}{\rm tr}(\gamma^5\sigma_i\sigma_j)D_i D_j \right)\,.
\end{equation}
Using (note that $\epsilon^{12}=1$)
\begin{equation}
{\rm tr}(\gamma^5\sigma_i\sigma_j)=2i\epsilon^{ij}\,,
\end{equation}
we get
\begin{equation}
{\cal J}(\phi,A_i)=\exp\left( -\int \mathrm{d}^2x\frac{i\phi}{2\pi}\epsilon^{ij}[D_i, D_j] \right)=\exp\left( \int \mathrm{d}^2x\frac{i\phi}{2\pi}\epsilon^{ij}(\partial_i A_j-\partial_j A_i) \right)\,.
\end{equation}
This is the Jacobian up to first order in $\phi$. To see that the Jacobian above cannot be the complete answer, we note that we can perform the chiral transformation $\boldsymbol{\psi}\to e^{(\phi+\varphi)\gamma^5}\boldsymbol{\psi}$ in two different ways. One is to do a single chiral transformation and the other is to do $\boldsymbol{\psi}\to e^{\phi\gamma^5}\boldsymbol{\psi}\to e^{(\phi+\varphi)\gamma^5}\boldsymbol{\psi}$. However, the latter way involves two different gauge fields since
\begin{equation}
\begin{split}
\mathcal{D}\boldsymbol{\psi}''\mathcal{D}\boldsymbol{\bar\psi}''e^{-S[\boldsymbol{\psi},A_i]}=&\mathcal{D}\boldsymbol{\psi}\mathcal{D}\boldsymbol{\bar\psi}{\cal J}_{\rm full}(\phi+\varphi,A_i)e^{-S[\boldsymbol{\psi},A_i]}\\
=&\mathcal{D}(e^{\varphi\gamma^5}\boldsymbol{\psi}')\mathcal{D}(\boldsymbol{\bar\psi}'e^{\varphi\gamma^5})e^{-S[e^{-\phi\gamma^5}\boldsymbol{\psi}',A_i]}\\
=&\mathcal{D}(e^{\varphi\gamma^5}\boldsymbol{\psi}')\mathcal{D}(\boldsymbol{\bar\psi}'e^{\varphi\gamma^5})e^{-S[\boldsymbol{\psi}',A_i+i\epsilon^{ij}\partial_j\phi]}\\
=&\mathcal{D}(e^{\phi\gamma^5}\boldsymbol{\psi})\mathcal{D}(\boldsymbol{\bar\psi}e^{\phi\gamma^5}){\cal J}_{\rm full}(\varphi,A_i+i\epsilon^{ij}\partial_j\phi)e^{-S[\boldsymbol{\psi},A_i]}\\
=&\mathcal{D}\boldsymbol{\psi}\mathcal{D}\boldsymbol{\bar\psi}{\cal J}_{\rm full}(\varphi,A_i+i\epsilon^{ij}\partial_j\phi){\cal J}_{\rm full}(\phi,A_i)e^{-S[\boldsymbol{\psi},A_i]}
\end{split}
\end{equation}
where we have defined
\begin{equation}
\boldsymbol{\psi}'= e^{\phi\gamma^5}\boldsymbol{\psi},\quad \boldsymbol{\psi}''= e^{(\phi+\varphi)\gamma^5}\boldsymbol{\psi}\,.
\end{equation}
Hence, we see that the complete Jacobian, which we denote by ${\cal J}_{\rm full}$, has to satisfy the consistency condition
\begin{equation}
{\cal J}_{\rm full}(\phi+\varphi,A_i)={\cal J}_{\rm full}(\varphi,A_i+i\epsilon^{ij}\partial_j\phi){\cal J}_{\rm full}(\phi,A_i)\,.
\end{equation}
Using this consistency condition, one can deduce that the complete Jacobian is
\begin{equation}
{\cal J}_{\rm full}=\exp\left( \int \mathrm{d}^2x\frac{i\phi}{2\pi}\epsilon^{ij}(\partial_i A_j-\partial_j A_i)-\frac{1}{2\pi}\int \mathrm{d}^2x\partial_i\phi\partial^i\phi \right)\,.
\end{equation}
Since, in our case, we do not turn on any gauge field, the desired Jacobian is
\begin{equation}
{\cal J}_{\rm full}=\exp\left( -\frac{1}{2\pi}\int \mathrm{d}^2x\,\partial_i\phi\partial^i\phi \right)=\exp\left( -\frac{1}{\pi}\int \mathrm{d}^2z\,\partial\phi\bar\partial\phi \right)\,.
\end{equation}
Combining this with the bosonic contribution, we obtain
\begin{equation}
\begin{split}
&\mathcal{D}\phi\mathcal{D}(e^{\phi}\gamma)\mathcal{D}(e^{\phi}\bar{\gamma})\mathcal{D}(e^{\phi}\psi)\mathcal{D}(e^{\phi}\bar\psi)\mathcal{D}(e^{-\phi}\omega)\mathcal{D}(e^{-\phi}\bar\omega)\\
&=\mathcal{D}\phi\mathcal{D}\gamma\mathcal{D}\bar{\gamma}\mathcal{D}\psi\mathcal{D}\bar\psi\mathcal{D}\omega\mathcal{D}\bar\omega\exp\left( \frac{1}{4\pi}\int \mathrm{d}^2z\,(4\partial\phi\bar\partial\phi+R\phi)-\frac{1}{\pi}\int \mathrm{d}^2z\,\partial\phi\bar\partial\phi \right)\\
&=\mathcal{D}\phi\mathcal{D}\gamma\mathcal{D}\bar{\gamma}\mathcal{D}\psi\mathcal{D}\bar\psi\mathcal{D}\omega\mathcal{D}\bar\omega\exp\left( \frac{1}{4\pi}\int \mathrm{d}^2z\,R\phi \right)\\
&=\mathcal{D}\Phi\mathcal{D}\Gamma\mathcal{D}\bar{\Gamma}\exp\left( \frac{1}{4\pi}\int \mathrm{d}^{2|2}z\,{\cal R}\Phi \right)\,,
\end{split}
\end{equation}
as claimed in equation \eqref{eq:change of variables in the sl(2,R) measure}. Before ending this subsection, let us remark that we implicitly assume that the measure for the action \eqref{eq:superspace action with auxiliary fields} is
\begin{equation}
\begin{split}
\mathcal{D}\Phi\mathcal{D}\Gamma\mathcal{D}\bar{\Gamma}\mathcal{D}\Omega\mathcal{D}\bar{\Omega}\,,
\end{split}
\end{equation}
where now
\begin{equation}
\begin{split}
\mathcal{D}\Gamma=\mathcal{D}\gamma\mathcal{D}\psi,\quad \mathcal{D}\Omega=\mathcal{D}\beta\mathcal{D}\omega\,,
\end{split}
\end{equation}
and similarly for the anti-holomorphic fields.

\subsection{Heterotic strings}
The analysis above was for the type II strings. In this subsection, we analyze the Jacobian for the heterotic strings. Intuitively, the Jacobian in type II theory should come from the left and right moving contribution and these two contributions should be equal. Nevertheless, it will be useful to be more explicit about the Jacobian for the heterotic theory. The change of variables in this case reads
\begin{equation}
\mathcal{D}(e^{\phi\frac{\gamma^5+1}{2}}\boldsymbol{\psi})\mathcal{D}(\boldsymbol{\bar\psi}e^{\phi\frac{\gamma^5-1}{2}})={\cal J}\mathcal{D}\boldsymbol{\psi}\mathcal{D}\boldsymbol{\bar\psi}\,,
\end{equation}
which yields the Jacobian
\begin{equation}
\begin{split}
{\cal J}=&\left[{\rm det}\left( e^{\phi\frac{\gamma^5+1}{2}}\delta^2(x-y) \right){\rm det}\left( e^{\phi\frac{\gamma^5-1}{2}}\delta^2(y-z) \right)\right]^{-1}\\
=&\left[{\rm det}\left( e^{\phi\gamma^5}\delta^2(x-y) \right)\right]^{-1}\\
=&\exp\left( -{\rm Tr}\left( \phi(x)\gamma^5\delta^2(x-y) \right) \right)\,.
\end{split}
\end{equation}
Hence, we see that the Jacobian in this case is precisely the square root of the Jacobian for type II theory. Therefore, we conclude that
\begin{equation}
{\cal J}_{\rm heterotic}={\cal J}_{\text{II}}^{\frac{1}{2}}=\exp\left( -\frac{1}{2\pi}\int \mathrm{d}^2z\partial\phi\bar\partial\phi \right)\,.
\end{equation}
The total ($\rm bosonic+fermionic$) Jacobian is thus
\begin{equation}
\begin{split}
&\exp\left( \frac{1}{4\pi}\int \mathrm{d}^2z\,(4\partial\phi\bar\partial\phi+R\phi)-\frac{1}{2\pi}\int \mathrm{d}^2z\,\partial\phi\bar\partial\phi \right)\\
&=\exp\left( \frac{1}{4\pi}\int \mathrm{d}^2z\,(2\partial\phi\bar\partial\phi+R\phi) \right)\,.
\end{split}
\end{equation}

\section{\boldmath Ramond sector vertex operators for AdS\texorpdfstring{$_3$}{3}}\label{app:Ramond_sector}

In Section \ref{sec:Ramond_phys_states}, we claimed that $e^{-3\varphi/2}\mathscr{V}_{m,j}^{w,(R,\underline{\epsilon})}$, where
\begin{equation}\label{eq:Ramond_vo_superspace_full_appendix}
\mathscr{V}^{w,(R,\underline{\epsilon})}_{m,j}=e^{(w/Q-Qj)\Phi}\left(\frac{\p^{w}\Gamma}{w!}\right)^{-m-j-\frac{\epsilon_1}{2}}\delta_{w|w}^{\epsilon_1}(\Gamma)S^{\epsilon_1}_H\mathcal{S}^{\epsilon_2}_{\kappa} \mathscr{V}_C^{(R,\epsilon_3,\epsilon_4,\epsilon_5)} \,
\end{equation}
was the vertex operator that described physical states in the Ramond sector. The aim of this appendix is twofold. We will firstly verify our claim by showing that the Taylor expansion of the superfields in \eqref{eq:Ramond_vo_superspace_full_appendix} is consistent with $G_0V_{m,j}^{w,(R,\underline{\epsilon})}$ in \eqref{eq:picture_-1/2_physical_vo}. Secondly, we will verify that, of the 16 physical vertex operators that satisfy the GSO projection, only eight of them are independent in BRST cohomology.

\paragraph{Taylor expansion:} In the main text, we found expressions for NS- and R-sector vertex operators and the overall normalization of these expressions was inconsequential. Here, we will need to compare expressions written both in terms of the free fields of Section \ref{sec:quantization} and a bosonization thereof. This means we need to make consistent choices for the normalization of the vertex operator. To do this, we will slightly modify the definitions \eqref{eq:gamma_delta_function}, \eqref{eq:psi_delta_function} and \eqref{eq:Gamma_delta_function_Ramond} for the purpose of this appendix. For the bosons, we define
\begin{equation}
    \delta_w(\gamma) = \prod_{i=0}^{w-1} \delta\left( \frac{\p^i\gamma}{i!} \right)\,,
\end{equation}
whilst we must also fix the ordering for the fermions,
\begin{equation}
\delta_w(\psi) =\overleftarrow{\prod_{i=0}^{w-1}}\frac{\p^i\psi}{i!} = \frac{\p^{w-1}\psi}{(w-1)!}\dots \p\psi\psi\,.
\end{equation}
This definition agrees exactly with the OPEs of $\delta_w(\psi) = e^{iwH}$. For the superfield, the factorials cancel between odd and even delta functions, such that
\begin{equation}
    \delta_{w|w}^{\epsilon_1}(\Gamma) = \overleftarrow{\prod_{i=0}^{2w-1}} \delta\left((D_{\text{sv}}^{\epsilon_1})^i\Gamma\right) = \delta_{w|w}^{\epsilon_1}(\Gamma) = \overleftarrow{\prod_{i=0}^{w-1}}\delta(\p^i\Gamma)\delta(\p^iD_{\text{sv}}^{\epsilon_1}\Gamma)\, . 
\end{equation}

Let us now verify that the Taylor expansion of \eqref{eq:Ramond_vo_superspace_full_appendix} in terms of $\theta_{\text{sv}}^-$ is consistent with
\begin{equation}\label{eq:picture_-1/2_physical_vo_appendix}
    \begin{aligned}
        G_0 V_{m,j}^{w,(R,\underline{\epsilon})} &= (-1)^w\frac{\epsilon_1\epsilon_2}{\sqrt{2}}\left[ \frac{w}{Q}-Q\left(j-\frac{1}{2}\right)\right]V^{w,(R,\tilde{\underline{\epsilon}})}_{m,j}  \\
        &+\frac{1}{2}\left[ \left(w-\left(m+j-\frac{1}{2}\right)\right)+\epsilon_1\left(w+\left(m+j-\frac{1}{2}\right)\right) \right]V^{w,(R,\hat{\underline{\epsilon}})}_{m,j} \,.
    \end{aligned}
\end{equation}
We begin by noting that the first term of \eqref{eq:picture_-1/2_physical_vo_appendix} closely resembles the Taylor expansion in \eqref{eq:phi_lambda_superfields}, albeit with a different coefficient. The change in the coefficient is nothing more than the cocycle factor that appeared in the calculation of \eqref{eq:picture_-1/2_physical_vo_appendix} from commuting $\lambda$ past $e^{i(w+\frac{\epsilon_1}{2})H}$ (see footnote \ref{footnote:cocycle}). The same factor will also appear from the Taylor expansion.

Demonstrating consistency with the second term in \eqref{eq:picture_-1/2_physical_vo_appendix} is more computationally involved. Observe,
\begin{equation}
    \begin{aligned}
        \delta_{w|w}^{\epsilon_1}(\Gamma) &= \left[\delta\left(\frac{\p^{w-1}\hat{\psi}^{\epsilon_1} + \theta_{\text{sv}}^{\epsilon_1}\p^{w-1}\left( (z-z_i)^{-\epsilon_1}\p\gamma\right)}{(w-1)!}\right)\right]\delta_w(\gamma)\delta_{w-1}(\hat{\psi}^{\epsilon})\\
        &=\delta_w(\gamma)\delta_w(\hat{\psi}^{\epsilon_1}) + \frac{(z-z_i)^{-\epsilon_1}}{(w-1)!} \theta_{\text{sv}}^{\epsilon_1} \delta_w(\gamma)\delta_{w-1}(\hat{\psi}^{\epsilon_1})\p^w\gamma\,,
    \end{aligned}
\end{equation}
such that
\begin{equation}\label{eq:delta_function_expansions}
    \begin{aligned}
        \delta_{w|w}^{+}(\Gamma)S^+_H &= \delta_w(\gamma)\delta_w(\hat{\psi}^+)S^+_H + \frac{\p^w\gamma}{(w-1)!}(z-z_i)^{-1}\theta_{\text{sv}}^{+}\delta_w(\gamma)\delta_{w-1}(\hat{\psi}^{+}) S^+_H \\
        &= \delta_w(\gamma)\delta_w(\hat{\psi}^+) + \frac{\p^w\gamma}{(w-1)!}\theta_{\text{sv}}^{-}\delta_w(\gamma)\delta_{w}(\hat{\psi}^{-}) S^-_H\,,\\
        \delta_{w|w}^{-}(\Gamma)S^-_H &= \delta_w(\gamma)\delta_w(\hat{\psi}^-)S^-_H + \frac{\p^w\gamma}{(w-1)!}(z-z_i)\theta_{\text{sv}}^{-}\delta_w(\gamma)\delta_{w-1}(\hat{\psi}^{-})S^-_H\,.
    \end{aligned}
\end{equation}
Here, we have used that $\delta_{w-1}(\hat{\psi}^+)S^+_H = \delta_w(\hat{\psi}^-)S^-_H$ by the definition \eqref{eq:Ramond_delta_functions}. We note that, on the divisor $z=z_i$ where we insert the vertex operator, the $\epsilon_1=-$ case simplifies to $\delta_{w|w}^{-}(\Gamma)S^-_H = \delta_w(\gamma)\delta_w(\hat{\psi}^-)S^-_H$. For a similar reason, the Taylor expansion of
\begin{equation}
    \left(\frac{\partial^w\Gamma}{w!}\right)^{-m-j- \frac{1}{2}} = \left(\frac{\partial^w\gamma}{w!}\right)^{-m-j- \frac{1}{2}}
\end{equation}
for $\epsilon_1=+$ is trivial on the divisor, whilst
\begin{equation}
    \left(\frac{\partial^w\Gamma}{w!}\right)^{-m-j+ \frac{1}{2}} = \left(\frac{\partial^w\gamma}{w!}\right)^{-m-j+ \frac{1}{2}}\left[1 - \theta_{\text{sv}}^-\left(m+j-\frac{1}{2}\right)  \frac{\p^w\hat{\psi}^-}{\p^w\gamma}\right]
\end{equation}
for $\epsilon_1=-$. Combining these expressions with \eqref{eq:delta_function_expansions}, we find exact agreement with the second term of \eqref{eq:picture_-1/2_physical_vo_appendix} on the divisor $z=z_i$. For $\epsilon_1=+$,
\begin{equation}
    \begin{aligned}
        \left(\frac{\partial^w\Gamma}{w!}\right)^{-m-j- \frac{1}{2}}\delta_{w|w}^{+}(\Gamma)S^+_H &= \left(\frac{\partial^w\gamma}{w!}\right)^{-m-j- \frac{1}{2}}\delta_w(\gamma)\delta_w(\hat{\psi}^+)S^+_H \\
        &\quad+ w\left(\frac{\partial^w\gamma}{w!}\right)^{-m-j+ \frac{1}{2}}\theta_{\text{sv}}^{-}\delta_w(\gamma)\delta_{w}(\hat{\psi}^{-}) S^-_H \,,
    \end{aligned}
\end{equation}
whilst for $\epsilon_1=-$,
\begin{equation}
    \begin{aligned}
        \bigg(\frac{\partial^w\Gamma}{w!}\bigg)^{-m-j+ \frac{1}{2}}&\delta_{w|w}^{-}(\Gamma)S^-_H \\
        &=\left(\frac{\partial^w\gamma}{w!}\right)^{-m-j+ \frac{1}{2}}\delta_w(\gamma)\delta_w(\hat{\psi}^-)S^-_H \\
        &\qquad- \left(m+j-\frac{1}{2}\right)\left(\frac{\partial^w\gamma}{w!}\right)^{-m-j- \frac{1}{2}}\theta_{\text{sv}}^{-}\delta_w(\gamma)\delta_{w+1}(\hat{\psi}^{-}) S^-_H \\
        &= \left(\frac{\partial^w\gamma}{w!}\right)^{-m-j+ \frac{1}{2}}\delta_w(\gamma)\delta_w(\hat{\psi}^-)S^-_H \\
        &\qquad- \left(m+j-\frac{1}{2}\right)\left(\frac{\partial^w\gamma}{w!}\right)^{-m-j- \frac{1}{2}}\theta_{\text{sv}}^{-}\delta_w(\gamma)\delta_{w}(\hat{\psi}^{+}) S^+_H \,.
    \end{aligned}
\end{equation}
Note that a factor of $\frac{1}{w!}$ is absorbed into $\delta_{w+1}(\hat{\psi}^-) = \frac{\p^w\hat{\psi}^-}{w!}\delta_{w}(\hat{\psi}^-)$ in the first equality and once again we used that $\delta_{w}(\hat{\psi}^+)S^+_H = \delta_{w+1}(\hat{\psi}^-)S^-_H$ by \eqref{eq:Ramond_delta_functions}.

\paragraph{The BRST cohomology:} In Section \ref{sec:Ramond_phys_states} we found 16 vertex operators for each $w$ that were BRST-closed and satisfied the GSO projection. We will demonstrate here that only eight of these are independent in BRST cohomology. An analogous result was found in \cite{Yu:2024kxr} for the particular background $\rm{AdS}_3\times \rm{S}^3\times \rm{T}^4$. They showed an equality between $G_0 V_{m,j}^{w,(R,\underline{\epsilon})}(z_i)$ (up to an overall constant) for the two choices $\underline{\epsilon} =(\epsilon_1,\epsilon_2,\epsilon_3,\epsilon_4,\epsilon_5)$ and $\underline{\epsilon}=(-\epsilon_1,-\epsilon_2,\epsilon_3,\epsilon_4,\epsilon_5)$ of Ramond ground state. Here, we will find more generally that for any background $\rm{AdS}_3 \times \mathcal{N}$, the vertex operators $e^{-3\varphi/2}\mathscr{V}_{m,j}^{w,(R,\underline{\epsilon})} (z_i|\theta_i)$ with $\underline{\epsilon}=(\epsilon_1,\epsilon_2,\epsilon_3,\epsilon_4,\epsilon_5)$ are equivalent up to BRST-exact terms to a sum over operators of the same form with $\epsilon_1\mapsto -\epsilon_1$ and a similar sign flip for one of the other fermionic vacua.

First consider the case of $\theta_i=0$, for which our physical vertex operator is
\begin{equation}
         e^{-3\varphi/2}V^{w,(\text{R},\underline{\epsilon})}_{m,j} = e^{-3\varphi/2} V^{w,(R,\epsilon_1)}_{m,j} (V_S)^{w,(\text{R},\epsilon_2,\epsilon_3,\epsilon_4,\epsilon_5)}_{m,j}\,,
\end{equation}
where 
\begin{equation}
    V^{w,(R,\epsilon_1)}_{m,j} = \left(\frac{\partial^w\gamma}{w!}\right)^{-m-j- \frac{\epsilon_1}{2}}\delta_{w}(\gamma)e^{i(w+ \frac{\epsilon_1}{2})H}\,.
\end{equation}
Here, we denote by $V_S$ the part of the vertex operator that depends on $S=\mathbb{R}_{Q}^{(1)} \times \mathcal{N}$, which will ultimately be related to the seed theory of the dual CFT (see Section \ref{sec:discussion}). The heart of our argument is that
\begin{equation}
    G_0 V^{w,(R,\epsilon_1)}_{h,j} = \mu_{\epsilon_1}V^{w,(R,-\epsilon_1)}_{h,j}\,,
\end{equation}
where $\mu_+ = w$ and $\mu_- = -(m+j-\frac{1}{2})$ as in \eqref{eq:picture_-1/2_physical_vo}. It allows one to exchange vertex operators with sign $\epsilon_1$ by those with sign $-\epsilon_1$. We may rewrite
\begin{equation}\label{eq:changing_vacuum}
    \begin{aligned}
        e^{-3\varphi/2}V^{w,(\text{R},\underline{\epsilon})}_{m,j} &= e^{-3\varphi/2}\frac{1}{\mu_{-\epsilon_1}} \left(G_0V^{w,(R,-\epsilon_1)}_{m,j}\right) (V_{S})^{w,(\text{R},\epsilon_2,\epsilon_3,\epsilon_4,\epsilon_5)}_{m,j} \\
        &= e^{-3\varphi/2} \frac{1}{\mu_{-\epsilon_1}} \bigg\{ G_0\left[V^{w,(R,-\epsilon_1)}_{m,j} (V_{S})^{w,(\text{R},\epsilon_2,\epsilon_3,\epsilon_4,\epsilon_5)}_{m,j}\right] \\
        &\hspace{2cm} + \alpha V^{w,(R,-\epsilon_1)}_{m,j} \left( G_0 (V_{S})^{w,(\text{R},\epsilon_2,\epsilon_3,\epsilon_4,\epsilon_5)}_{m,j} \right)  \bigg\}\,,
    \end{aligned}
\end{equation}
where $\alpha$ represents a cocycle factor from commuting $G_0$ past $V^{w,(R,-\epsilon_1)}_{m,j}$. The first term in this expression represents a BRST-exact contribution, since $V^{w,(\text{R},-\epsilon_1,\epsilon_2,\epsilon_3,\epsilon_4,\epsilon_5)}_{m,j} = V^{w,(R,-\epsilon_1)}_{m,j} (V_{S})^{w,(\text{R},\epsilon_2,\epsilon_3,\epsilon_4,\epsilon_5)}_{m,j}$ satisfies the mass-shell condition that $L_0 = \frac{5}{8}$ whenever $V^{w,(R,\underline{\epsilon})}_{m,j}$ also satisfies the mass-shell condition. Indeed, it is a straightforward computation to show (up to cocycle factors) that
\begin{equation}
    \begin{aligned}
        ce^{-3\varphi/2}G_0V^{w,(\text{R},-\epsilon_1,\epsilon_2,\epsilon_3,\epsilon_4,\epsilon_5)}_{m,j} &= (e^{\varphi}\eta G)_0 (c e^{-5\varphi/2}\p\zeta \, V^{w,(\text{R},-\epsilon_1,\epsilon_2,\epsilon_3,\epsilon_4,\epsilon_5)}_{m,j}) \\
        &= \bigg\{ Q_{\text{BRST}} , c e^{-5\varphi/2}\p\zeta \, V^{w,(\text{R},-\epsilon_1,\epsilon_2,\epsilon_3,\epsilon_4,\epsilon_5)}_{m,j} \bigg]\,,
    \end{aligned}
\end{equation}
where $Q_{\text{BRST}}$ is the standard RNS BRST operator with superconformal ghosts $\hat\gamma = \eta e^{\varphi}$ and $\hat\beta = e^{-\varphi}\p\zeta$. Hence, up to BRST-exact terms, we can always replace a physical picture $-\frac{3}{2}$ vertex operator with sign $\epsilon_1$ by one with sign $-\epsilon_1$. The compensating factor of $G_0$ acting on the part of the vertex operator depending on $S = \mathbb{R}_{Q}^{(1)} \times \mathcal{N}$ in \eqref{eq:changing_vacuum} ensures that the GSO projection is still satisfied. Moreover, when the vertex operator describes a ground state in $\mathcal{N}$, only the sign $\epsilon_2$ will be changed by the action of $G_0$ as was observed for $\mathcal{N} = \rm{S}^3 \times \rm{T}^4$ in \cite{Yu:2024kxr}.

So far, we have only discussed the case of $\theta_i = 0$. Yet, the result naturally carries over to the general case since \eqref{eq:changing_vacuum} implies
\begin{equation}
    \begin{aligned}
        G_0 V_{m,j}^{w,(\text{R},\underline{\epsilon})} &= \frac{1}{\mu_{-\epsilon_1}} \bigg\{ G_0^2\left[V^{w,(R,-\epsilon_1)}_{m,j} (V_{S})^{w,(\text{R},\epsilon_2,\epsilon_3,\epsilon_4,\epsilon_5)}_{m,j}\right] \\
        &\quad + \alpha G_0 \left[V^{w,(R,-\epsilon_1)}_{m,j} \left( G_0 (V_{S})^{w,(\text{R},\epsilon_2,\epsilon_3,\epsilon_4,\epsilon_5)}_{m,j} \right)\right] \bigg\} \\
        &= \frac{\alpha}{\mu_{-\epsilon_1}} G_0 \left[V^{w,(R,-\epsilon_1)}_{m,j} \left( G_0 (V_{S})^{w,(\text{R},\epsilon_2,\epsilon_3,\epsilon_4,\epsilon_5)}_{m,j} \right) \right]\, .
        \end{aligned}
\end{equation}
Above, we use once again that $V^{w,(R,-\epsilon_1)}_{m,j}(V_{S})_{m,j}^{w,(R,\epsilon_2,\epsilon_3,\epsilon_4,\epsilon_5)}$ is on-shell, such that $G^2_0$ vanishes. Therefore,
\begin{equation}
    \begin{aligned}
        ce^{-3\varphi/2}\mathscr{V}^{w,(R,\underline{\epsilon})}_{m,j}(z_i|\theta_i) &= \left\{ Q_{\text{BRST}}, \frac{1}{\mu_{-\epsilon_1}}c e^{-5\varphi/2}\p\zeta \, V^{w,(\text{R},-\epsilon_1,\epsilon_2,\epsilon_3,\epsilon_4,\epsilon_5)}_{m,j}(z_i) \right] \\
        &+ \frac{\alpha}{\mu_{-\epsilon_1}} ce^{-3\varphi/2} V_{m,j}^{w,(R,-\epsilon_1)}\left( G_0(V_S)_{h,j}^{w,(R,\epsilon_2,\epsilon_3,\epsilon_4,\epsilon_5)} \right) (z_i|\theta_i) \,,
    \end{aligned}
\end{equation}
where there is no $\theta_i$-dependence in the BRST-exact term and we recall the definition \eqref{eq:P=-3/2_expansion}. This confirms that there are exactly eight independent physical vertex operators of the form \eqref{eq:Ramond_vo_superspace_full_appendix} for each $w$ that satisfy the GSO projection. In other words, one may set $\epsilon_1 = -$ without loss of generality.

\bibliography{bibliography}
\bibliographystyle{utphys.sty}

\end{document}